\documentclass[10pt]{elsarticle}

\usepackage[margin=2cm]{geometry} 	
\usepackage{amssymb,amsmath,amsthm,amsfonts}
\numberwithin{equation}{section}
\usepackage{subfigure}
\usepackage[small]{caption}
\usepackage[utf8]{inputenc} 
\usepackage{appendix}
\usepackage{braket}
\usepackage{color}
\usepackage{hyperref}

\usepackage[pagewise]{lineno}
\journal{Physics Reports}

\title{Shell-shaped atomic gases}
\author{Andrea Tononi$^{1,2,*}$, Luca Salasnich$^{3,4,*}$}
\address{
$^{1}$ICFO-Institut de Ciencies Fotoniques, The Barcelona Institute of Science and Technology, Av. Carl Friedrich Gauss 3, 08860 Castelldefels (Barcelona), Spain \\
$^{2}$Universit\'e Paris-Saclay, CNRS, LPTMS, 91405 Orsay, France \\
$^{3}$Dipartimento di Fisica e Astronomia `Galileo Galilei’ and Padua QTech, 
Universit{\`{a}} di Padova, \\ via Marzolo 8, 35131 Padova, Italy \\
$^{4}$INFN - Sezione di Padova, via Marzolo 8, 35131 Padova, Italy and CNR-INO, \\ via Carrara 1, 50019 Sesto Fiorentino, Italy}
\address{$^*$ ${\rm Corresponding \ author:}$ luca.salasnich@unipd.it}

\date{\today}
\begin{document}

\maketitle

%\vspace{2mm}
\noindent 
We review the quantum statistical properties of two-dimensional shell-shaped gases, produced by cooling and confining atomic ensembles in thin hollow shells. 
We consider both spherical and ellipsoidal shapes, discussing at zero and at finite temperature the phenomena of Bose-Einstein condensation and of superfluidity, the physics of vortices, and the crossover from the Bardeen-Cooper-Schrieffer regime to a Bose-Einstein condensate. 
The novel aspects associated to the curved geometry are elucidated in comparison with flat two-dimensional superfluids. 
We also describe the hydrodynamic excitations and their relation with the Berezinskii-Kosterlitz-Thouless transition for two-dimensional flat and curved superfluids. 
In the next years, shell-shaped atomic gases will be the leading experimental platform for investigations of quantum many-body physics in curved spatial domains. 
%%%%%%%%%%%%%%%%%%%%%%%%%%%%%%%%%%%%%%%%%%%%%%%%%%%%%%%

\tableofcontents

\section{Introduction}

\subsection{Bose-Einstein condensation and superfluidity}
Quantum mechanics was not fully established when Einstein, following a paper by Bose \cite{bose1924}, discussed the phenomenon of condensation in a series of papers in 1924 \cite{einstein1924} and 1925  \cite{einstein1925}. The exclusion principle was indeed discovered by Pauli only in 1924 \cite{pauli1924}, and the classification of the particles as bosons and fermions was still unknown. It is thus fair to say that Einstein idea was visionary and, at the same time, embryonic. Many years were necessary for reaching the full understanding of Bose-Einstein condensation and, to a certain extent, the process is still ongoing. 

When the experiments with liquid Helium of Kapitza \cite{kapitza1938} emerged in the 30's, London interpreted the superfluid properties as a manifestation of the phenomenon of Bose-Einstein condensation \cite{london1938}. Landau, on the contrary, considered Bose-Einstein condensation as a pathological condition of a noninteracting Bose gas \cite{balibar}, and described the observations only in terms of superfluidity \cite{landau1941}, developing the previous theoretical analyses by Tisza \cite{tisza1938,tisza1938bis,tisza1938tris}. 

The Bogoliubov theory \cite{bogoliubov1947} and the following theoretical works have established the existence of a Bose-Einstein condensate phase in interacting superfluid bosons, but the heredity of Landau is still present, as the relation between Bose-Einstein condensation and superfluidity is still not completely settled \cite{leggett1999}. The tension among these concepts, indeed, continues to emerge in the scientific process in which new possibilities in terms of interactions, geometry, system size, spatial dimension, etc., challenge the previous theoretical concepts, and theories are stretched to reach a better understanding. 

The key elements, which turned the study of Bose-Einstein condensation into a structured field with heterogeneous ramifications, are the tunability and the versatility of the experiments. 
The experimental milestone, from which the present diversity originates, was the discovery of Bose-Einstein condensation in 1995 \cite{anderson1995,davis1995,bradley1995}, obtained by confining alkali-metal atoms and cooling them at $\text{nK}$-range temperatures. 
Rather than being a fortuitous chance, the first observation of Bose-Einstein condensation was the outcome of progressive technical advances in cooling and trapping of neutral atoms, which were mainly obtained in the 70's and in the 80's. 
Among them, we remind the Zeeman slower \cite{phillips1982}, and the laser cooling to produce optical molasses \cite{hansch1975,wineland1975,chu1985}, combined with the trapping techniques by means of optical and magnetic potentials \cite{migdall1985,chu1986,bagnato1987}. 
With the development of sub-Doppler cooling \cite{lett1988,aspect1988} and magneto-optical traps \cite{raab1987}, most of the technical advances were available: the use of evaporative cooling \cite{masuhara1988,ketterle1996} allowed to reach a sufficiently high phase-space density and to observe the macroscopic occupation of the condensate state \cite{anderson1995,davis1995}. 

Nowadays, it is possible to tune and control experimentally all the different contributions of the many-body Hamiltonian: from kinetic and potential terms, to the interaction ones, i.~e.~engineering both weakly- and strongly-interacting systems with either short- or long-range interactions \cite{bloch2008,bloch2012,lahaye2009}. 
In perspective, these convenient features make ultracold atoms a reliable platform for the development of quantum simulators and quantum computers \cite{feynman1982,lloyd1996,georgescu2014}. 
Moreover, by strongly constraining the dynamics of an atomic gas along one (or two) spatial directions \cite{gorlitz2001}, i.~e.~decoupling the transverse dynamics from the in-plane (or in-line) degrees of freedom, ultracold atoms allow to test and develop quantum many-body physics in low spatial dimensions \cite{petrov2004,petrov2003}. 

In this review, we discuss the quantum statistical physics of systems of ultracold atoms in low dimensions, analyzing their equilibrium and nonequilibrium properties in the temperature and density regimes where quantum degeneracy occurs. We will analyze both curved geometries, such as bosonic atoms confined on spherical or on ellipsoidal surfaces, and flat configurations, such as atomic gases in box potentials. 
While most of the results regard bosonic atoms, also fermionic systems and their crossover from the Bardeen-Cooper-Schrieffer state to a Bose-Einstein condensate (BCS-BEC) will be analyzed. 
We provide in the following Sections a brief overview on the systems and on the phenomena that we will later discuss. 

\subsection{Shell-shaped atomic gases}
In the field of quantum gases, most of the experimental and of the theoretical results concerning low-dimensional systems are obtained in infinite flat geometries, or by modeling trapped three-dimensional configurations, such as pancake or cigar-like shapes \cite{dalfovo1999}. 
The idea of studying a quantum gas confined in a curved shell can be initially traced in a 2001 seminal paper by Zobay and Garraway \cite{zobay2001} (see also \cite{zobay2000,zobay2004,perrin2016}), who analyzed the magnetic confinement of atomic gases with a combination of a static field and of a radiofrequency field. 
By engineering the trap parameters, it was shown that the radiofrequency-induced adiabatic potential can confine the atoms in a shell-shaped configuration when the gravitational force can be counterbalanced or neglected. 

A new impulse to study these shell-shaped (or ``bubble-trapped'') gases follows the development of several microgravity facilities to cool and confine the atomic gases in microgravity conditions. 
Among them, we remind the NASA-JPL Cold Atom Laboratory (CAL) \cite{elliott2018,aveline2020}, on board of the International Space Station, the drop towers \cite{vanzoest2010,muntinga2013,vogt2020}, rockets \cite{becker2018}, and a free falling elevator \cite{condon2019}. 
In particular, microgravity experiments producing closed shell-shaped atomic gases were planned \cite{lundblad2019} and successfully carried on \cite{carollo2021} in CAL, and they will probably continue on BECCAL \cite{frye2021}, a future experimental facility on the International Space Station. 
Parallel to these, also several Earth-based experiments were successful in trapping quantum gases in thick shells \cite{jia2022} or in open 2D shells by using magnetic bubble traps \cite{colombe2004, white2006,merloti2013,harte2018,guo2020,guo2022,rey2022}.

To match this rising experimental interest, a great amount of theoretical research has been produced in the last five years.
Our personal contribution to this research line amounted to a wide characterization of the quantum statistical properties of shell-shaped two-dimensional bosonic gases, focusing in particular on the phenomena of Bose-Einstein condensation and of superfluidity \cite{tononi1,tononi3,tononi4,tononi2022}. 
These will be the subjects that we mainly aim to review in the present paper. 
Several authors have also considered other aspects, for instance focusing on thick shells, on long-range interactions, on the dynamics, etc., providing key contributions to these analyses. 
To underline the theoretical interest on bubble-trapped atomic gases, and to provide a synopsis of the state of the art on the subject, we now list the themes that have been studied recently and we briefly comment the pertaining publications.

\begin{itemize}
\item \textit{Quantum statistics and shell thermodynamics} --- 
Spherically-symmetric 2D shells were extensively analyzed in the last 5 years, discussing the phenomena of Bose-Einstein condensation \cite{tononi1,bereta2019}, their thermodynamics \cite{tononi4,rhyno2021,tononi2022}, and the BCS-BEC crossover of two-component fermionic gases \cite{chien2022}. 
Concerning ellipsoidal thin shells, we studied their Bose-Einstein condensation and superfluidity  \cite{tononi3}.

\item \textit{Superfluidity and vortices} --- The superfluid Berezinskii-Kosterlitz-Thouless transition in spherical shells has been studied since the 1980s with different aims and adopting various techniques \cite{kotsubo1984,kotsubo1986,wang1986,ovrut1991,mitra2008,tononi2022}.
Connected to this topic, the physics of vortices in a spherical superfluid film has been extensively analyzed \cite{vitelli2010,padavic2020,bereta2021,song2021,kanai2021,tomishiyo2022,ruban2022,xiong2023,white2023,he2023b}, and also ellipsoids and generic axially-symmetric surfaces were discussed \cite{caracanhas2022}. 
These analyses echo various mathematically-oriented works studying the motion of point vortices in fluids with the spherical geometry \cite{bogomolov1977,kimura1987,kimura1999,newton2001,kim2010}.

\item \textit{Shell dynamics} --- 
The study of the shell dynamics has mainly involved the analysis of the condensate excitations, both at zero temperature \cite{lannert2007,sun2018,padavic2018} and at finite temperature \cite{tononi4}. 
A few works discussed the emergence of equatorial modes induced by the rotation of the gas \cite{saito2023,li2023}.
The free expansion of the bubble \cite{tononi3,rhyno2021}, and the shell lensing \cite{boegel2023} were also discussed.

\item \textit{Few-body physics} --- 
Most few-body analyses on shell-shaped atomic gases were conducted for the geometry of a 2D sphere. 
In particular, we signal the one-body scattering problem \cite{zhang2018}, the two-body scattering problem in a large spherical manifold \cite{tononi2022}, and the study of emergent chiral bound states in p-wave scattering and in higher partial waves \cite{shi2017}. 
The physics of anyons on the sphere was also analyzed \cite{ouvry2019}. 
Finally, various works discussed the supersolidity of few-particle systems \cite{prestipino2019,prestipino2020} and of many-body systems with long-range interactions \cite{ciardi2023}.

\item \textit{Other related research} --- 
The anisotropic density distribution of bosons with dipolar interactions on the sphere was analyzed \cite{diniz2020,arazo2021}, also putting the topic in the context of the possible experiments.
General frameworks to model the zero-temperature dynamics of vortices on curved manifolds were developed \cite{moller2020,biral2022}. Finally, Ref.~\cite{andriati2021} discussed the stability Bose-Bose mixtures on a spherical surface, and Ref.~\cite{he2023} analyzed solitons in spinor Bose-Einstein condensates.

\end{itemize}

From our viewpoint, the research line on curved quantum gases is still partially developed, and, by solving the open research questions \cite{tononisalasnich2023,lundblad2023}, this subfield will improve our general understanding of Bose-Einstein condensation and superfluidity in ultracold quantum gases. 

\subsection{Hydrodynamic excitations and sound modes}
The two-fluid model of Landau and Tisza, developed to interpret the rich physics of superfluid Helium \cite{kapitza1938}, provides a long-wavelength hydrodynamic description of a quantum liquid \cite{landau1941}, which is modeled as a mixture of a normal fluid and of a superfluid. 
Due to the presence of two components, one of the main predictions of the model is the existence of two branches of hydrodynamic excitations in the collisional regime. 
Section \ref{section4} is devoted to reviewing the hydrodynamic modes in uniform two-dimensional superfluids which, in the box-trapped case, propagate as sound waves. 

Interestingly, the measurement of the first and second sound velocities in a two-dimensional system provides a direct evidence of the superfluid transition, which was discussed by Berezinskii \cite{berezinskii1973}, Kosterlitz and Thouless \cite{kosterlitz1972,kosterlitz1973,kosterlitz1974} (BKT). 
The BKT transition, in which the superfluid properties are suppressed by the thermal proliferation of vortices, does not lead to any discontinuity of the thermodynamic potentials \cite{desbuquois2014}, but it consists in the universal jump of the superfluid density at a critical temperature, as discussed by Kosterlitz and Nelson \cite{nelson1977}. 

While the BKT transition has already been observed in superfluid Helium \cite{bishop1978}, which is a strongly-interacting system, the first evidence in weakly-interacting bosonic superfluids was initially indirect \cite{hadzibabic2006}. 
A direct experimental proof was obtained a few years ago \cite{christodoulou2021}, by measuring the first and second sound velocities in box-trapped superfluid bosons. 
We will analyze a few theoretical results in comparison with the experiment of Ref.~\cite{christodoulou2021} in Section \ref{sectionsound2Dbosons}. 
The observation of the hydrodynamic excitations confirms the validity of the two-fluid description in weakly-interacting bosonic gases, and demonstrates the superfluid nature of the system. 

In Section \ref{sectionsound2Dfermi} we will also analyze the propagation of sound in two-dimensional uniform fermions across the BKT transition. 
In particular, we will compare the results obtained with the finite-temperature Gaussian pair fluctuation theory with the experiment of Ref.~\cite{bohlen2020}, and we will discuss the physics of the system across the crossover from weakly-bound BCS pairs to the BEC regime of composite bosons \cite{zwerger2011,strinati2018}. 

Finally, in Section \ref{sectionsoundsphere} we will consider a shell-shaped bosonic superfluid, and we will study the finite-temperature hydrodynamic excitations of the system. Their experimental characterization, indeed, can prove that the superfluid transition occurs also in topologically-nontrivial compact shells, and that it is driven by the BKT mechanism of the vortex-antivortex unbinding. 

\section{Fundamental results and formalism overview}
\label{chapter0}
We provide a brief overview of the main models and techniques that we will adopt in Section \ref{chapter1} to analyze the physics of shell-shaped quantum gases. 
We also review here a few fundamental results on Bose-Einstein condensation and on superfluidity. 

\subsection{Bose-Einstein condensation}
\label{sectionBEC}
The transition of a many-body system of identical particles to a Bose-Einstein condensate occurs when a macroscopic fraction of the particles occupies the lowest-energy single-particle state. 
The simplest realization of this transition takes place in noninteracting bosons, which is the case first analyzed by Einstein \cite{einstein1924,einstein1925}. 
Clearly, in the absence of interactions between the particles, which is the case we now discuss, the condensate phase must emerge from the quantum statistical properties of a large number of bosons that constitute the system. 

We consider a system of noninteracting particles confined in a spatial domain $V$, which we assume to be sufficiently larger with respect to the particle size. 
Working in the grand canonical ensemble, we suppose that the system is in thermal equilibrium with a bath of temperature $T$. 
In the noninteracting case, this condition can be established by turning off the interparticle interactions once that the thermalization of the interacting system has occurred. 
Moreover, we suppose that the system is in chemical equilibrium with the external reservoir of chemical potential $\mu$, so that the number of particles displays small fluctuations around its mean value $N$. 
Considering a single particle, we denote the kinetic part of the Hamiltonian with $\hat{h}_{\text{kin}}$, and we suppose that the external potential $\hat{h}_{\text{pot}}$, which acts on the particle confined in $V$, also imposes proper boundary conditions at the domain boundary $\partial V$. 
The Schr\"odinger equation of the particle reads
\begin{equation}
(\hat{h}_{\text{kin}}+\hat{h}_{\text{pot}}) \, \phi_{\alpha} = \epsilon_{\alpha} \, \phi_{\alpha}, 
\label{eigenproblem}
\end{equation}
where $\phi_{\alpha}$ are the eigenfunctions, labelled by the quantum numbers $\alpha$, and $\epsilon_{\alpha}$ are the eigenenergies. 
In the following, we suppose that the solution of this eigenproblem is known, i.~e., that $\phi_{\alpha}$ and $\epsilon_{\alpha}$ are known for each $\alpha$. After reducing the problem to this single solvable unit, we now construct the quantum statistical properties of the many-body system. 

Given the single-particle state with fixed quantum number $\alpha$, we denote with $p_0^{(\alpha)}$, $p_1^{(\alpha)}$, ... $p_r^{(\alpha)}$, ... the probabilities that it is occupied by $0$, $1$, ... $r$, ... bosons. 
Our goal is to determine, in conditions of thermal and chemical equilibrium, the average number of bosons ${N}_{\alpha}$ that occupies the state $\alpha$, given by ${N}_{\alpha} = \sum_{r=0}^{\infty} r \,  p_r^{(\alpha)}$.
According to the original derivation of Bose \cite{bose1924} and Einstein \cite{einstein1924}, i.~e.~imposing that the entropy $S$ is maximum for a fixed temperature $T$ and for a fixed chemical potential $\mu$, the probabilities $p_r^{(\alpha)}$ are given by $p_r^{(\alpha)} = B \, e^{-\beta (\epsilon_{\alpha}-\mu)\, r}$, where $\beta = 1/(k_{\text{B}} T)$, with $k_{\text{B}}$ the Boltzmann constant, and $B$ is a positive constant. 
To determine $B$, one imposes that the total probability of occupying the state $\alpha$ is equal to one, i.~e.~$\sum_{r=0}^{\infty} p_r^{(\alpha)} = 1$, getting $B = 1 - e^{-\beta (\epsilon_{\alpha}-\mu)}$. 
We then calculate the sum over $r$ in the definition of ${N}_{\alpha}$, obtaining 
\begin{equation}
{N}_{\alpha} = \frac{1}{e^{\beta \,(\epsilon_{\alpha}-\mu)}-1}, 
\end{equation} 
which is the Bose-Einstein distribution. 
We emphasize that, to have positive probabilities $p_r^{(\alpha)}$ and positive occupation numbers ${N}_{\alpha}$, the chemical potential must satisfy the inequality $\mu < \epsilon_{\bar{\alpha}}$, with $\epsilon_{\bar{\alpha}} = \text{min}_{\alpha} ( \epsilon_{\alpha})$ the single-particle ground-state energy. 
In particular, when $\epsilon_{\bar{\alpha}} = 0$, as in the cases discussed in the next subsections \ref{nonintbos3Dbox}, \ref{nonintbos2Dbox} and \ref{spherenonint}, the chemical potential can assume only negative values: $\mu <0$. 

The phenomenon of Bose-Einstein condensation consists in the macroscopic occupation of the lowest-energy single-particle state $\bar{\alpha}$ by a macroscopic fraction of the atoms in the system \cite{einstein1925}. 
In general, this condition can be achieved by following two slightly different procedures:
either fixing the particle number and decreasing the temperature, or fixing the temperature and increasing the particle number. 
Actually, when working in the grand canonical ensemble the particle number is determined by the chemical potential,
and the relevant thermodynamic variables are therefore $T$ and $\mu$. 
Let us rewrite the total number of bosons as 
\begin{align}
\begin{split}
N = N_{\bar{\alpha}} + \tilde{N}, 
\qquad \qquad 
N_{\bar{\alpha}} = \frac{1}{e^{\beta \,(\epsilon_{\bar{\alpha}}-\mu)}-1}, \qquad \tilde{N} = \sum_{\alpha \neq \bar{\alpha}} \frac{1}{e^{\beta \,(\epsilon_{\alpha}-\mu)}-1}, 
\label{becnoninteracting}
\end{split}
\end{align}
where $N_{\bar{\alpha}}$ are the particles in the lowest-energy single particle state $\bar{\alpha}$, and $\tilde{N}$ is the number of particles in the excited states $\alpha \neq \bar{\alpha}$. 

If $\mu  \to {\epsilon_{\bar{\alpha}}}^{-}$ at a fixed temperature $T$, the number of particles $\tilde{N}$ tends to the critical atom number $N_c$ from below.
Strictly speaking, depending on the dimensionality of the system and on its (infinite or finite) size, $N_c$ could be infinite when $\mu = \epsilon_{\bar{\alpha}}$. Thus, the following discussion applies only to the case in which $N_c$ is finite. 
When this occurs, since the chemical potential $\mu$ controls the total number of particles $N$, for $\mu$ larger than a critical value $\mu_c$ we find that $N > N_c$. 
Therefore, in the regime of $\mu_c < \mu < \epsilon_{\bar{\alpha}}$ a macroscopic number of particles $N_{\bar{\alpha}}$ occupies the condensate state $\bar{\alpha}$. 
{A commonly-used approximation to calculate $\tilde{N}$ consists in setting $\mu  = \epsilon_{\bar{\alpha}}$. By so doing, since the blunt application of Eq. \eqref{becnoninteracting} would yield an infinite occupation number $N_{\bar{\alpha}}$ of the condensate state $\bar{\alpha}$, this quantity must be interpreted as an unknown parameter, to determine a posteriori as a function of $N$ and $T$.}

\subsubsection{Thermodynamics of the noninteracting Bose gas}
In the grand canonical ensemble, the thermodynamics of a noninteracting Bose gas can be derived from the grand canonical partition function $\mathcal{Z}$, which reads \cite{huang1987}
\begin{equation}
\mathcal{Z} = \sum_{N=0}^{\infty} z^N \, \mathcal{Q}_N(V,T),
\end{equation}
where $z=e^{\beta \mu}$ is the fugacity.
In this expression, we introduce the canonical partition function of the $N$-particle system as $\mathcal{Q}_N(V,T) =  \sum_{ \{N_{\alpha}\} } e^{-\beta E_{ \{ N_{\alpha} \} }}$,
which is calculated as the sum, over all possible occupation numbers $ N_{\alpha} $ satisfying the constraint $N= \sum_{\alpha} N_{\alpha}$, of the Boltzmann factors $e^{-\beta E_{ \{ N_{\alpha} \}}}$, where $E_{ \{ N_{\alpha} \}} = \sum_{\alpha} N_{\alpha} \epsilon_{\alpha}$. 
After a few steps, the grand canonical partition function can be rewritten as \cite{huang1987}
\begin{equation}
\mathcal{Z} = \prod_{\alpha} \sum_{N_{\alpha}=0}^{\infty} \big[e^{-\beta (\epsilon_{\alpha}-\mu)}\big]^{N_{\alpha}} = \prod_{\alpha} \frac{1}{1-e^{-\beta (\epsilon_{\alpha}-\mu)}}, 
\end{equation}
which, assuming that the solution of eigenproblem of Eq.~\eqref{eigenproblem} is known, is a {given} function of $T$ and $\mu$.
The grand canonical potential $\Omega = U -TS-\mu N$, with $U$ the internal energy, can be calculated from the grand canonical partition function as $\Omega = -\beta^{-1} \ln(\mathcal{Z})$, obtaining 
\begin{equation}
\Omega = \frac{1}{\beta} \sum_{\alpha} \ln \big[1-e^{-\beta (\epsilon_{\alpha}-\mu)} \big],  
\end{equation}
and, using standard thermodynamic relations, also the other thermodynamic functions can be obtained. 
For instance, the number of atoms is given by 
\begin{equation}
N = - \bigg( \frac{\partial \Omega}{\partial \mu}  \bigg)_{V,T} = \sum_{\alpha} N_{\alpha},
\end{equation}
which coincides with the result of Eq.~\eqref{becnoninteracting}. 
Moreover, the entropy reads
\begin{equation}
S = - \bigg( \frac{\partial \Omega}{\partial T}  \bigg)_{\mu,V} = k_{\text{B}} \sum_{\alpha} \bigg\{  \frac{\beta \,(\epsilon_{\alpha}-\mu)}{e^{\beta \,(\epsilon_{\alpha}-\mu)}-1} - \ln \big[1-e^{-\beta (\epsilon_{\alpha}-\mu)} \big] \bigg\},
\end{equation}
while the internal energy $U$ can be calculated from $U = \Omega + TS + \mu N$, and reads
\begin{equation}
U =  \sum_{\alpha} \epsilon_{\alpha} N_{\alpha}. 
%\frac{\epsilon_{\alpha}}{e^{\beta \,(\epsilon_{\alpha}-\mu)}-1}.
\end{equation}
Finally, the pressure is defined as 
\begin{equation}
P = - \bigg( \frac{\partial \Omega}{\partial V}  \bigg)_{\mu,T},
\end{equation}
which, for a uniform system, coincides with the simple relation $P=-\Omega/V$. 

\subsubsection{Noninteracting bosons in a uniform 3D box} 
\label{nonintbos3Dbox}
We consider a noninteracting Bose gas confined in a cubic box of volume $V=L^3$. The solution of the single-particle eigenproblem of Eq.~\eqref{eigenproblem} yields the eigenergies $\epsilon_{\mathbf{k}} = \hbar^2 k^2/(2m)$,
where, due to the imposition of periodic boundary conditions on the eigenfunctions $\phi_{\mathbf{k}} = e^{i \mathbf{k} \cdot \mathbf{x}} / \sqrt{V}$, the wave vector is given by $\mathbf{k} = (2\pi/L) (n_x,n_y,n_z)$, with $n_x, n_y, n_z \in \mathbb{Z}$. 
In this configuration, Bose-Einstein condensation occurs in the state $\bar{\alpha} = \mathbf{0}$, in which the wave vector $\mathbf{k}$ is zero. To get an analytical insight on this problem we set the chemical potential to $\mu = \epsilon_{\mathbf{0}} = 0$.

Implementing Eq.~\eqref{becnoninteracting} for $N_{\mathbf{0}} = 0$, we obtain a relation between the critical temperature of Bose-Einstein condensation $T_{\text{BEC}}^{(0)}$ and the total number of atoms $N = \tilde{N} = N_c$, over which the condensate state is macroscopically occupied. 
In particular
\begin{equation}
N = 8 \sum_{n_x=1}^{\infty} \sum_{n_y=1}^{\infty} \sum_{n_z=1}^{\infty} \frac{1}{e^{\epsilon_{\mathbf{k}}/(k_{\text{B}}T_{\text{BEC}}^{(0)})}-1},
\label{Ncubicbox}
\end{equation}
where, in the sum, we are neglecting the terms with one null quantum number and the others nonzero, and the terms with two null quantum numbers and the other nonzero. 
For sufficiently large $V$ these contributions are irrelevant, and the discrete wave vectors $\mathbf{k}$ can be thought, in this limit, as a continuum of values. 
In this case, we can substitute the sums with integrals, $\sum_{n_i=1}^{\infty} \to L/(2\pi) \int_{2\pi/L}^{\infty} \text{d}k_i$, 
and the lower bounds $2\pi / L$ can actually be approximated with $0$. 
Before doing this substitution, we briefly discuss the case of a large but finite volume: 
since the distance between the energy levels $\epsilon_{\mathbf{k}}$ increases with the quantum numbers $n_x, n_y, n_z$, the continuum approximation seems to become invalid. 
However, the Bose-Einstein distribution cuts off the higher-energy states, and the semiclassical approximation will work well. 

If we evaluate Eq.~\eqref{Ncubicbox} analytically, by performing the wave vector integrals in the thermodynamic limit and using spherical coordinates, we find the critical density 
\begin{equation}
\frac{N}{L^3} = \zeta(3/2) \bigg( \frac{m k_{\text{B}} T_{\text{BEC}}^{(0)}}{2 \pi \hbar^2} \bigg)^{3/2},
\label{tbec03D}
\end{equation}
where $\zeta(x)$ is Riemann's zeta function. 
Inverting Eq.~\eqref{tbec03D}, we finally obtain 
\begin{equation}
T_{\text{BEC}}^{(0)} = \frac{2\pi}{[\zeta(3/2)]^{2/3}} \frac{\hbar^2 n^{2/3}}{m k_{\text{B}}},
\end{equation}
which is the critical temperature of a noninteracting Bose gas with number density $n=N/L^3$ in a uniform box.
We stress again that this relation between the critical temperature $T_{\text{BEC}}^{(0)}$ and $n$ holds in the thermodynamic limit of $N,L^3 \to \infty$, with $n$ fixed. 

\subsubsection{Noninteracting bosons in a uniform 2D box}
\label{nonintbos2Dbox}
Let us now calculate the critical temperature of a bosonic gas confined in a uniform square box with area $V=L^2$. Similarly to the cubic case, the eigenergies are given by $\epsilon_{k} = \hbar^2 k^2 / (2m)$, where $\mathbf{k} = (2\pi/L) (n_x,n_y)$ is the two-dimensional wave vector, with $n_x, n_y \in \mathbb{Z}$. 
In this context, the analogous of Eq.~\eqref{Ncubicbox} reads 
\begin{equation}
N = 4 \sum_{n_x=1}^{\infty} \sum_{n_y=1}^{\infty} \frac{1}{e^{\epsilon_{k}/(k_{\text{B}}T_{\text{BEC}}^{(0)})}-1}, 
\label{tbec02Dsum}
\end{equation}
and, assuming that the energy levels are finely spaced in the region where the Bose-Einstein distribution is nonzero, we write
\begin{equation}
\frac{N}{L^2} = \int_{2\pi/L}^{\infty} \frac{\text{d} k_x}{(2\pi)} \int_{2\pi/L}^{\infty} \frac{\text{d} k_y}{(2\pi)} \frac{1}{e^{\epsilon_{k}/(k_{\text{B}}T_{\text{BEC}}^{(0)})}-1}, 
\label{tbec02D}
\end{equation}
where, as in the three-dimensional case, we have substituted the sum with an integral. 
However, in contrast with the three-dimensional case, here we cannot consider the thermodynamic limit and set the lower bound $2\pi/L$ of the integrals to $0$. 
Indeed, a simple numerical test shows that, for a nonzero critical temperature $T_{\text{BEC}}^{(0)}$, the critical density $n=N/L^2$ is finite only if the system size $L^2$ is finite. 
More quantitatively, in the thermodynamic limit of $L^2 \to \infty$, and assuming $T_{\text{BEC}}^{(0)}>0$, the critical density of Eq.~\eqref{tbec02D} diverges in the infrared as $\int_0^{\infty} \text{d}k /k$. 
This divergence is a manifestation of the Hohenberg-Mermin-Wagner theorem \cite{mermin1966, hohenberg1967}, which, in this context, states that there cannot be Bose-Einstein condensation at finite temperature in a two-dimensional system at the thermodynamic limit. 

For a finite-size system, both the critical density and the critical temperature are finite, and the precise calculation of these quantities requires the numerical evaluation of the sum in Eq.~\eqref{tbec02Dsum}. 
However, an approximated result, which neglects subleading corrections scaling with the inverse system size, can be obtained expressing the integral of Eq.~\eqref{tbec02D} in polar coordinates, in which the radial component of the wave vector is integrated in the interval $[4 \sqrt{\pi}/L,\infty[$\footnote{The integral in Eq.~\eqref{tbec02D} cannot be expressed in polar coordinates in a straightforward way, due to the squared shape of the cutoffed area $(-2\pi/L,2\pi/L) \times (-2\pi/L,2\pi/L)$. Our choice of the infrared cutoff for the radial wave vector coordinate, i.~e.~$k_{\text{c}}=4 \sqrt{\pi}/L$, allows to keep the cutoffed area constant.}. 
In this case, repeating the same steps of subsection \ref{nonintbos3Dbox}, we find
\begin{equation}
k_B T_{\text{BEC}}^{(0)} \approx 2\pi  \frac{\hbar^2 n}{m}
\bigg\{ 
\frac{8 \pi \hbar^2}{ m L^2 k_{\text{B}} T_{\text{BEC}}^{(0)}} 
- \ln \big[ 
e^{8 \pi \hbar^2/( m L^2 k_{\text{B}} T_{\text{BEC}}^{(0)})} -1 \big] \bigg\}^{-1},
\label{Tbec0flat}
\end{equation}
which is an implicit equation relating the critical temperature $T_{\text{BEC}}^{(0)}$ with the two-dimensional number density $n$.
The analysis of this result confirms our previous considerations, namely, that $n$ cannot be finite in the limit of infinite system size $L^2 \to \infty$ unless $T_{\text{BEC}}^{(0)} \to 0$. 

\subsubsection{Noninteracting bosons on the surface of a sphere}
\label{spherenonint}
We now describe Bose-Einstein condensation of noninteracting bosons confined on the surface of a sphere of radius $R$. 
We parametrize the surface of the sphere, whose area is given by $V = 4 \pi R^2$, with the spherical coordinates $\{\theta,\varphi\} \in [0,\pi]\times [0,2\pi]$.
For this configuration, the single-particle Schr\"odinger equation of Eq.~\eqref{eigenproblem} can be written as 
\begin{equation}
\frac{\hat{L}^2}{2mR^2} \, \mathcal{Y}_l^{m_l} (\theta,\varphi) = \epsilon_l \, \mathcal{Y}_l^{m_l} (\theta,\varphi), 
\end{equation}
where
\begin{equation}
\hat{L}^2 = -\hbar^2 \bigg[  \frac{1}{\sin \theta} \frac{\partial}{\partial \theta} \bigg( \sin \theta \frac{\partial}{\partial \theta}  \bigg) + \frac{1}{\sin^2 \theta} \frac{\partial^2}{\partial \varphi^2}\bigg],
\label{angularmomentum}
\end{equation}
is the angular momentum operator in spherical coordinates, $\mathcal{Y}_l^{m_l} (\theta,\varphi)$ are the spherical harmonics, labelled by the main quantum number of the angular momentum $l$, and by the magnetic quantum number $m_l = -l, -l+1, ... l-1, l$. 
The eigenenergies $\epsilon_l$, which are degenerate in $m_l$, read
\begin{equation}
\epsilon_l = \frac{\hbar^2 l(l+1)}{2mR^2}, 
\end{equation}
and the lowest-energy condensate state $\bar{\alpha}$ corresponds to $l=0$, $m_l = 0$, so that $\epsilon_0 = 0$. 

\begin{figure}
\centering
\includegraphics[width=0.518\columnwidth]{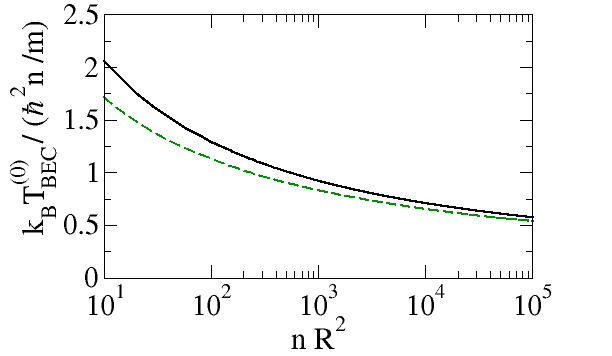}
\caption{Critical temperature of noninteracting bosons on the surface of a sphere (black line), plotted as a function of the parameter $n R^2$. The green dashed line represents the direct evaluation of the sum of Eq.~\eqref{Ncspheresum}, showing that the analytical result improves for larger values of $n R^2$. 
Note that the critical temperature tends logarithmically to zero in the thermodynamic limit. 
From Ref.~\cite{tononi1}.
}
\label{figTbec0sphere}
\end{figure}

The relation between the critical temperature of Bose-Einstein condensation $T_{\text{BEC}}^{(0)}$ and the number of atoms $N$ is given by 
\begin{equation}
N = \sum_{l=1}^{\infty} \sum_{m_l=-l}^{l} \frac{1}{e^{\epsilon_{l}/(k_{\text{B}}T_{\text{BEC}}^{(0)})}-1}, 
\label{Ncspheresum}
\end{equation}
where we set the chemical potential to $\mu = \epsilon_0 = 0$. To obtain an analytical result for $T_{\text{BEC}}^{(0)}$, in analogy to the previous cases, we substitute the sum with an integral, i.~e.~$\sum_{l=1}^{\infty} \sum_{m_l = -l}^{l} \to \int_{1}^{\infty} \text{d}l \, (2l+1)$ (see \cite{bereta2019} for detailed analyses of the density of states), finding
\begin{equation}
\frac{N}{4\pi R^2} =  \frac{m k_{\text{B}} T_{\text{BEC}}^{(0)} }{ 2\pi\hbar^2} \bigg\{ 
\frac{\hbar^2}{ mR^2 k_{\text{B}} T_{\text{BEC}}^{(0)}} - 
\ln{\big[ e^{\hbar^2/(mR^2 k_{\text{B}} T_{\text{BEC}}^{(0)})} -1 \big]} \bigg\}, 
\end{equation}
which allows to calculate the critical number of atoms for a given critical temperature. 
The critical temperature can be obtained by solving the following equation \cite{tononi1}
\begin{equation}
k_B T_{\text{BEC}}^{(0)} = 2\pi  \frac{\hbar^2 n}{m}
\bigg\{ 
\frac{\hbar^2}{ mR^2 k_{\text{B}} T_{\text{BEC}}^{(0)}} 
- \ln \big[ 
e^{\hbar^2/( mR^2 k_{\text{B}} T_{\text{BEC}}^{(0)})} -1 \big] \bigg\}^{-1},
\label{Tbec0sphere}
\end{equation}
where we define the two-dimensional number density of the spherical Bose gas as $n=N/(4\pi R^2)$. 
In Fig.~\ref{figTbec0sphere} we plot the dimensionless critical temperature as a function of the parameter $n R^2$, obtained by solving numerically Eq.~\eqref{Tbec0sphere}. 
The critical temperature, for a fixed density $n$, is finite for a finite radius of the sphere $R$, and it tends to $0$ in the limit of $R \to \infty$. 
As seen in the two-dimensional flat case, this behavior is consistent with the prescription of the Hohenberg-Mermin-Wagner theorem \cite{mermin1966, hohenberg1967}. 

\begin{figure}
\centering
\includegraphics[width=0.518\columnwidth]{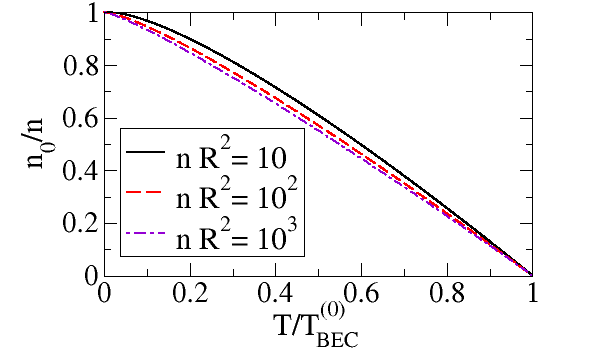}
\caption{
Condensate fraction of noninteracting bosons on the surface of a sphere plotted as a function of the adimensional temperature $T/T_{\text{BEC}}^{(0)}$ for different values of $n R^2$. 
From Ref.~\cite{tononi1}.}
\label{fign0overnsphere}
\end{figure}

Note that it is not possible to identify the critical temperature in the spherical case with Eq.~\eqref{Tbec0flat}, i.~e.~the analogous result for the two-dimensional box, by simply substituting $4 \pi R^2 = L^2$ into Eq.~\eqref{Tbec0sphere}. Indeed, even by doing this, a factor of $2$ remains in the expressions and prevents their identification. {Indeed, while the high-energy spectra of the two geometries are essentially equivalent, the low-energy part of the spectrum is sensitive to the specific geometry. The different critical temperatures signal} the different value of the infrared energy cutoff in the two cases: while in the spherical case the first state above the condensate has the energy $\hbar^2/(m R^2)$, in the spherical one we assumed $\hbar^2 k_{\text{c}}^2/(2m)=8 \pi \hbar^2/(m L^2)$, and these, even setting $4 \pi R^2 = L^2$, differ by a factor $2$.

At temperatures lower than $T_{\text{BEC}}^{(0)}$, determined by Eq.~\eqref{Tbec0sphere}, the condensate state is occupied by a macroscopic fraction of bosons $n_0/n$, with $n_0 = N_0 / (4\pi R^2)$. 
The condensate fraction reads \cite{tononi1}
\begin{equation}
\frac{n_0}{n} = 1 - \frac{ {1 - k_{\text{B}} T \frac{mR^2}{\hbar^2} 
\ln{\big[ e^{\hbar^2/(mR^2 k_{\text{B}} T)} -1 \big]}} 
}{
{1 - k_{\text{B}} T_{\text{BEC}}^{(0)} \frac{mR^2}{\hbar^2} 
\ln{\big[ e^{\hbar^2/(mR^2 k_{\text{B}} T_{\text{BEC}}^{(0)})} -1 \big]}} } \; , 
\label{condfracnonint}
\end{equation}
which can be calculated evaluating the sum in Eq.~\eqref{becnoninteracting} and following the analogous of the procedure adopted for the critical temperature. In Fig.~\ref{fign0overnsphere}, we plot $n_0/n$ for different values of the parameter $n R^2$.

\subsection{Functional integral of a many-body quantum system}
\subsubsection{The bosonic case}
\label{sectionfunctionalint}
The quantum statistical properties of a many-body system, due to the interaction between the particles, cannot be determined in a straightforward manner from the solution of a single-particle eigenvalue equation. 
In the formalism of first quantization, the solution of the $N$-body bosonic problem requires the proper symmetrization of the wave function of the system, and the complexity of the problem increases exponentially with $N$. 
Second quantization offers an elegant way to reformulate the problem, including the quantum statistics in the properties of operators, and focusing only on the occupation numbers of the different states. 
Here, starting from the second-quantized Hamiltonian of a system of bosons, we express the grand canonical partition function as a coherent-state functional integral of a bosonic complex field. 

We consider a many-body system of interacting bosons, whose grand canonical partition function $\mathcal{Z}$ is given by 
\begin{equation}
\mathcal{Z} = \text{Tr} \big[ e^{-\beta (\hat{H} - \mu \hat{N})} \big], 
\label{ZTr}
\end{equation}
where $\hat{H}$ is the many-body Hamiltonian of the system, and $\hat{N}$ is the number operator.
We express $\hat{H}$ and $\hat{N}$ in terms of the field operator $\hat{\psi}(\mathbf{r})$ in second quantization, which annihilates a boson at the spatial coordinate $\mathbf{r}$. 
In particular, considering a generic $D$-dimensional system of bosons in the hypervolume $V$, the Hamiltonian reads 
\begin{align}
\hat{H} = \int_{V} \text{d}\mathbf{r} \, \hat{\psi}^{\dagger}(\mathbf{r}) \bigg( - \frac{\hbar^2 \nabla^2}{2m} \bigg) \hat{\psi}(\mathbf{r}) 
+ \frac{1}{2} \int_{V} \text{d}\mathbf{r} \int_{V} \text{d}\mathbf{r'} \, \hat{\psi}^{\dagger}(\mathbf{r}) \hat{\psi}^{\dagger}(\mathbf{r'}) V(\mathbf{r}-\mathbf{r'}) \hat{\psi}(\mathbf{r'}) \hat{\psi}(\mathbf{r}),
\label{Hamiltonianfieldoperator}
\end{align}
where $V(\mathbf{r}-\mathbf{r}')$ is the two-body interaction potential between the particles. 
Moreover, we write 
\begin{equation}
\hat{N} = \int_{V} \text{d}\mathbf{r} \, \hat{\psi}^{\dagger}(\mathbf{r}) \hat{\psi}(\mathbf{r}), 
\label{Nfieldoperator}
\end{equation}
which is the number operator in second quantization. 

Since the grand canonical partition function is the trace of the operator $e^{-\beta (\hat{H} - \mu \hat{N})}$, it can be calculated by summing the expectation values of this operator over all the possible states of a proper basis. 
For this scope, we choose the basis of generalized coherent states $\ket{\psi}$, which are defined as the eigenstates of the field operator $\hat{\psi}(\mathbf{r}) \ket{\psi} = \psi(\mathbf{r}) \ket{\psi}$. 
These states are normalized by construction, but are not orthogonal, and, calculating the trace in Eq.~\eqref{ZTr} in this basis \cite{salasnich2017}, we get 
\begin{equation}
\mathcal{Z} = \int \prod_{\mathbf{r}} \frac{\text{d}\bar{\psi}_0(\mathbf{r}) \text{d}\psi_0(\mathbf{r}) }{2 \pi i} \, \bra{\psi_0} e^{-\beta (\hat{H} - \mu \hat{N})} \ket{\psi_0},
\label{Ztrace0} 
\end{equation}
where, in the following, we will express the measure of the integral in a compact form as $\text{d}[\bar{\psi}_0,\psi_0]$. 
To proceed further, we reinterpret the expectation value $\bra{\psi_0} e^{-\beta (\hat{H} - \mu \hat{N})} \ket{\psi_0}$ as the probability amplitude of a system with Hamiltonian $\hat{H}-\mu \hat{N}$ that propagates in the imaginary time interval $\tau \in [0,\beta\hbar]$. 
Specifically: the many-body system starts at time $0$ in the initial state $\ket{\psi_0}$, evolves under the action of the operator $e^{-\, (\tau/\hbar) (\hat{H} - \mu \hat{N})}$, and ends at time $\beta\hbar$ in the same state $\ket{\psi_0}$. 
Exploring this fruitful analogy, the partition function of a many-body system can be calculated as a Feynman path integral \cite{feynman1948}. 
Thus, we split $\beta$ into many imaginary time intervals $\Delta \tau$, by writing $\beta = M \Delta \tau /\hbar$, with $M \gg 1$. 
In this way, the exponential $ e^{-\frac{M \Delta \tau}{\hbar} (\hat{H} - \mu \hat{N})} $ can be split into the product of $M$ exponentials, committing for each subdivision an error of $\text{o}(\Delta \tau^2)$, which is negligible for sufficiently large $M$. 
Inserting $M-1$ coherent-state identities between the $M$ exponentials, the field operators inside $\hat{H}-\mu\hat{N}$ are substituted by the classical fields $\psi(\mathbf{r},j \Delta \tau) =: \psi(\mathbf{r},\tau)$, and this occurs for each time step $j \Delta \tau$, with $j=0,...,M$.  
These calculations are developed in detail in Refs.~\cite{negele1998,salasnich2017,tononithesis}, let us now report the final results. 

The grand canonical partition function of a many-body system of interacting bosons is given by 
\begin{equation}
\mathcal{Z}=\int{\mathcal{D} [\bar{\psi},\psi] \; 
e^{-\frac{S[\bar{\psi},\psi]}{\hbar}}},
\label{partfunction}
\end{equation}
where the Euclidean action $S$, i.~e.~the action in imaginary time, is defined as  
\begin{equation}
S[\bar{\psi},\psi] = \int_{0}^{\beta\hbar} \text{d}\tau \, 
\int_{V} \text{d}\mathbf{r} \, \mathcal{L}(\bar{\psi},\psi) ,
\label{action}
\end{equation}
and the Euclidean Lagrangian $\mathcal{L}$ reads 
\begin{equation}
{ \cal L } = \bar{\psi}(\mathbf{r},\tau) 
\bigg(\hbar\partial_{\tau}-\frac{\hbar^2 \nabla^2}{2m}-\mu\bigg) 
\psi(\mathbf{r},\tau) + 
\frac{1}{2} \int_{V} \text{d}\mathbf{r'} \, |\psi(\mathbf{r'},\tau)|^2 V(\mathbf{r}-\mathbf{r'}) |\psi(\mathbf{r},\tau)|^2 .
\label{lagr} 
\end{equation}
Note that the measure of the functional integral is defined as $\mathcal{D} [\bar{\psi},\psi] = \prod_{j=0}^{M} \text{d}[\bar{\psi}_j,\psi_j]$, 
namely, as the product of the integration measures for all the intermediate imaginary-time steps. 

Let us interpret this result for the partition function. 
Within this formalism, we are calculating $\mathcal{Z}$ as a sum over all possible configurations of the classical field $\psi(\mathbf{r},\tau)$ which describes a specific trajectory of the system in the imaginary time interval $[0,\hbar\beta]$. 
The factor $e^{-S[\bar{\psi},\psi]/\hbar}$ assigns a different weight, throughout the value of the action $S[\bar{\psi},\psi]$, to the different configurations. 
The quantum nature of the system, embodied by the bosonic statistics, is included through the periodic boundary conditions in imaginary time, which determine the specific form of the trace in Eq.~\eqref{Ztrace0}. 
The functional integral representation of the partition function is equivalent to the general definition of $\mathcal{Z}$, and thus it does not involve any approximation. At the same time, it constitutes an intuitive description in terms of a complex bosonic field and allows for a convenient implementation of approximated calculations of the system thermodynamics. 

\subsubsection{The fermionic case}
In analogy with the bosonic case, we now implement a coherent-state functional integral formulation of the grand canonical partition function of a fermionic system. 
For this scope, we need to implement explicitly Eq.~\eqref{ZTr}, specifying the second-quantized Hamiltonian and the number operators of the fermions. 

Let us describe a system of uniform fermions confined in the $D$-dimensional hypervolume $V$. 
If we would consider a single fermionic species, i.~e.~fermions in a single hyperfine state, the Pauli principle would prevent $s$-wave interactions to occur. 
To avoid this limitation, we suppose to have a mixture of fermionic atoms in two different hyperfine states, labelled with the index $\sigma=\{\uparrow, \downarrow \}$, and we assume an attractive contact interaction between fermions with opposite spins.
The grand-canonical Hamiltonian, written in terms of the second-quantized field operator $\hat{\psi}_\sigma(\mathbf{r})$, is given by 
\begin{equation}
\hat{H} -\mu \hat{N} = \sum_{\sigma=\uparrow, \downarrow} \int_V 
\mathrm{d} \mathbf{r} \bigg[ \hat{\psi}^\dagger_\sigma (\mathbf{r}) 
\bigg( - \frac{\hbar^2\nabla^2}{2m} -\mu\bigg) 
\hat{\psi}_\sigma(\mathbf{r})
+ g \ \hat{\psi}^{\dagger}_\uparrow (\textbf{r}) 
\hat{\psi}^{\dagger}_\downarrow (\textbf{r}) 
\hat{\psi}_\downarrow (\textbf{r}) 
\hat{\psi}_\uparrow (\textbf{r}) \bigg] \; ,
\label{bcs hamiltonian}
\end{equation}
where $m$ is the atomic mass, $\mu$ is the chemical potential, and $g$ is the $s$-wave contact interaction strength between fermions with opposite spins. 
Note that, by tuning the Feshbach resonance with the application of a static magnetic field \cite{chin2010}, it is possible to change the strength $g$ and study the crossover from BCS pairs of weakly-interacting fermions to a bosonic BEC made of tightly-bound composite molecules \cite{strinati2018}. 
The grand canonical partition function $\mathcal{Z}$, according to the definition of Eq.~\eqref{ZTr}, is given by
\begin{equation}
\mathcal{Z} = \int \prod_{\sigma=\uparrow, \downarrow} \prod_{\mathbf{r}} \text{d}\bar{\psi}_{\sigma}(\mathbf{r}) \text{d}\psi_{\sigma}(\mathbf{r})  \, \bra{-\psi} e^{-\beta (\hat{H} - \mu \hat{N})} \ket{\psi},
\label{Ztrace0fermions} 
\end{equation}
where we are calculating the trace in the basis of generalized fermionic coherent states \cite{negele1998,salasnich2017}. 
The latter are the eigenstates of the fermionic field operator with eigenvalue $\psi_{\sigma}(\mathbf{r})$, namely $\hat{\psi}_\sigma (\textbf{r}) \ket{\psi} =  \psi_{\sigma}(\mathbf{r}) \ket{\psi}$,
and the fermionic statistics results in the non commutativity of the complex fermionic field $\psi_{\sigma}(\mathbf{r})$, which assumes the values of Grassmann anticommuting numbers \cite{negele1998}. 
The construction of the functional integral in the fermionic case is similar to that of the bosonic case, with the crucial difference that the fermionic Grassmann field satisfies antiperiodic conditions at the boundaries of the imaginary time interval\footnote{{This fact is encoded in Eq. \eqref{Ztrace0fermions} by using the notation $\bra{-\psi}$.}}, i.~e.~$\psi_{\sigma}(\mathbf{r},0) = - \psi_{\sigma}(\mathbf{r},\beta\hbar)$. 
Keeping in mind this distinction, which stems from the fermionic quantum statistics, we now report the results (the full derivation can be found in Ref.~\cite{negele1998}). 

The grand canonical partition function of a system of fermions across the BCS-BEC crossover reads
\cite{nagaosa1999}
\begin{equation}
\mathcal{Z}=\int{\mathcal{D} [\bar{\psi}_{\sigma},\psi_{\sigma}] \; 
e^{-\frac{S[\bar{\psi}_{\sigma},\psi_{\sigma}]}{\hbar}}},
\label{Zfermions0}
\end{equation}
where the Euclidean action is defined as 
\begin{equation}
S[\bar{\psi_{\sigma}},\psi_{\sigma}] = \int_{0}^{\beta\hbar} \text{d}\tau \, 
\int_{V} \text{d}\mathbf{r} \, \mathcal{L}(\bar{\psi_{\sigma}},\psi_{\sigma}) ,
\end{equation}
and the Euclidean Lagrangian reads
\begin{equation}
\mathcal{L} = \sum_{\sigma=\uparrow, \downarrow} \bar{\psi}_\sigma (\mathbf{r},\tau)
\bigg(\hbar\partial_{\tau}-\frac{\hbar^2 \nabla^2}{2m}-\mu\bigg) 
\psi_{\sigma}(\mathbf{r},\tau) 
+
g \, \bar{\psi}_\uparrow (\textbf{r},\tau) \,
\bar{\psi}_\downarrow (\textbf{r},\tau) \,
\psi_\downarrow (\textbf{r},\tau) \,
\psi_\uparrow (\textbf{r},\tau),
\end{equation}
where the imaginary-time dependence of the fermionic field emerges from the construction of the functional integral, and $\mathcal{D} [\bar{\psi},\psi] = \prod_{\sigma=\uparrow, \downarrow} \prod_{\mathbf{r}} \prod_{j=0}^{M} \text{d}\bar{\psi}_{\sigma}(\mathbf{r},j\beta\hbar/M) \, \text{d}\psi_{\sigma}(\mathbf{r},j\beta\hbar/M)$,
with $M \to \infty$, is the measure of the integral. 

The grand canonical partition function of Eq.~\eqref{Zfermions0} has been obtained without further approximations with respect to the Hamiltonian in second quantization. 
As a natural consequence, due to the quartic dependence of the Lagrangian on the fermionic field, it is not possible to calculate exactly the functional integral. 
To facilitate the development of an approximated theory, we introduce the pairing field $\Delta(\mathbf{r},\tau)$, which pairs fermions with opposite spins and, therefore, represents the Cooper pairs in the system \cite{cooper1956}. {Since this field is bosonic, it obeys periodic boundary conditions in imaginary time. 
We define it as}
\begin{equation}
\Delta(\mathbf{r},\tau) = - g \psi_\downarrow (\textbf{r},\tau) \,
\psi_\uparrow (\textbf{r},\tau),
\end{equation}
and analogously for the complex conjugate field $\bar{\Delta}$, and we perform the following Hubbard-Stratonovich transformation \cite{nagaosa1999, tempere2012}
\begin{align}
\begin{split}
 &\exp \bigg[- \int_{0}^{\beta\hbar} \text{d}\tau \, 
\int_{V} \text{d}\mathbf{r} \, g \, \bar{\psi}_\uparrow (\textbf{r},\tau) 
\bar{\psi}_\downarrow (\textbf{r},\tau) 
\psi_\downarrow (\textbf{r},\tau) 
\psi_\uparrow (\textbf{r},\tau)\bigg] = \int \mathcal{D} [\bar{\Delta},\Delta]
\\ 
& \exp \bigg\{  \int_{0}^{\beta\hbar} \text{d}\tau \, 
\int_{V} \text{d}\mathbf{r} \, \bigg[ \frac{\bar{\Delta}(\mathbf{r},\tau)\Delta(\mathbf{r},\tau)}{g} 
+ \bar{\psi}_\uparrow (\textbf{r},\tau) 
\bar{\psi}_\downarrow (\textbf{r},\tau) \Delta(\mathbf{r},\tau) 
 + \bar{\Delta}(\mathbf{r},\tau)
\psi_\downarrow (\textbf{r},\tau) 
\psi_\uparrow (\textbf{r},\tau)  \bigg] \bigg\}
\end{split}
\end{align}
which, despite the disadvantage of introducing an additional field, allows us to get a Gaussian integral in the fermionic fields. 
Indeed, the partition function has been transformed in this way as \cite{diener2008}
\begin{align}
\begin{split}
\mathcal{Z}=& \int \mathcal{D} [\bar{\psi}_{\sigma},\psi_{\sigma}] \int \mathcal{D} [\bar{\Delta},\Delta] \exp \bigg\{-\int_{0}^{\beta\hbar} \text{d}\tau \, 
\int_{V} \text{d}\mathbf{r} 
\bigg[
\sum_{\sigma=\uparrow, \downarrow} \bar{\psi}_\sigma (\mathbf{r},\tau)
\bigg(\hbar\partial_{\tau}-\frac{\hbar^2 \nabla^2}{2m}-\mu\bigg) 
\psi_{\sigma}(\mathbf{r},\tau)
\label{Zfermions}
\\
&-\frac{\bar{\Delta}(\mathbf{r},\tau)\Delta(\mathbf{r},\tau)}{g} - \bar{\psi}_\uparrow (\textbf{r},\tau) 
\bar{\psi}_\downarrow (\textbf{r},\tau) \Delta(\mathbf{r},\tau) - \bar{\Delta}(\mathbf{r},\tau)
\psi_\downarrow (\textbf{r},\tau) 
\psi_\uparrow (\textbf{r},\tau)\bigg]\bigg\},
\end{split}
\end{align}
which is an alternative form with respect to Eq.~\eqref{Zfermions0} that does not involve any further approximation.
Later on, we will describe how to derive the system thermodynamics, by developing an approximated calculation of $\mathcal{Z}$.

\subsection{Magnetic trapping of atomic gases in a shell-shaped geometry}
\label{sectionmagnetictrapping}
Ultracold atoms undergo the phase transition of Bose-Einstein condensation when the phase-space density exceeds a critical value. 
This phenomenon was observed for the first time in 1995 \cite{anderson1995,davis1995,bradley1995}, by cooling atomic gases with the combined use of laser and evaporative cooling techniques. 
These experiments relied on the magnetic and optical confinement of the sample, which could not be cooled in the $\text{nK}$ temperature range if held in physical containers. 
In this section,  we discuss briefly the basic ideas to confine atoms with magnetic traps. 
We avoid analyzing the pre-cooling stage in magneto-optical traps and the evaporative cooling (for details on this topic see \cite{ketterle1999}), and the short analysis implemented here aims to provide a background for discussing radiofrequency-induced adiabatic potentials. 

Let us discuss how weakly-interacting bosonic gases are magnetically confined. In these systems, the interactions are so weak and the gas is so dilute that, to describe the trapping mechanism, it is sufficient to understand how a single atom interacts with the magnetic field. 
Given a static space-dependent magnetic field $\mathbf{B}_0 (\mathbf{r})$, the potential energy of an atom interacting with the field reads 
\begin{equation}
U(\mathbf{r}) = - \hat{\boldsymbol{\mu}} \cdot \mathbf{B}_0 (\mathbf{r}), 
\end{equation} 
where $\hat{\boldsymbol{\mu}}$ is the magnetic dipole moment of the atom. 
This dipole moment is proportional to the total angular momentum operator $\hat{\mathbf{F}} = \hat{\mathbf{I}} + \hat{\mathbf{J}}$, which is given by the sum of the nuclear spin $\hat{\mathbf{I}}$ and of the angular momentum of the electrons $\hat{\mathbf{J}}$. 
In particular, we write $\hat{\boldsymbol{\mu}} = - g_F \mu_{\text{B}} \hat{\mathbf{F}}/\hbar$, where $\mu_{\text{B}}$ is Bohr's magneton and $g_F$ is the Landé factor \cite{perrin2016}. The projection of $\hat{\mathbf{F}}$ on the local direction of the magnetic field is quantized as $\hbar m_F$, where $m_F = -F, ...,+F$ is the magnetic quantum number which, for bosonic atoms, assumes only integer values. 
Therefore, the space-dependent energy levels of the atom, which result from the interaction of its magnetic dipole moment with the magnetic field, are given by 
\begin{equation}
u_{m_F}(\mathbf{r}) =  m_F g_F \, \mu_{\text{B}} |\mathbf{B}_0 (\mathbf{r})|.
\label{umF}
\end{equation}

By engineering the static magnetic field to have a minimum at a certain spatial position, the atoms with $m_F g_F >0$, usually called low-field-seeking states, will be subject to the force $-\nabla u_{m_F}(\mathbf{r})$ directed towards the trap minimum. To confine the atoms, a possible magnetic field configuration is the quadrupole field $\mathbf{B}_0 (\mathbf{r}) \propto (x,y,-2z)$, which however suffers from losses of atoms due to the likely occurrence of Majorana spin-flips at the trap minimum \cite{ketterle1999}. Other static magnetic field configurations include the Ioffe-Pritchard trap and cloverleaf traps (see, for more details, Refs.~\cite{majorana1932,ketterle1999} and the references therein).

The atoms are typically confined in magnetic conservative traps after the stage of laser cooling, in which the optical molasses reach a temperature in the $\text{mK}$ range and densities of $10^{12} \, \text{cm}^{-3}$. 
To produce a Bose-Einstein condensate, it is however necessary to further decrease the temperature: this is typically done by letting the most energetic atoms escape from the trap, allowing the system to thermalize at a lower temperature. 
This technique is called evaporative cooling \cite{masuhara1988,ketterle1996}, and the loss of energetic atoms is realized by coupling, via a radiofrequency magnetic field, the potential of low-field seeking states with repulsive potentials in the regions far from the trap minimum. 

But adopting the setup used to perform the evaporative cooling, it is also possible to engineer radiofrequency-induced adiabatic potentials that trap the atoms in a spatially-confined superposition of their hyperfine states \cite{zobay2001}. 
Let us consider an atom in a region of space where both a static magnetic field $\mathbf{B}_0 (\mathbf{r})$ and a time-dependent magnetic field  $\mathbf{B}_{\text{rf}} (\mathbf{r},t)$ are nonzero. 
Due to the static magnetic field, the atom precesses around the local direction of the static field with the Larmor frequency 
\begin{equation}
\omega_{\text{L}}(\mathbf{r}) = \frac{|g_F| \mu_{\text{B}} |\mathbf{B}_0 (\mathbf{r})|}{\hbar},
\end{equation}
which depends on the spatial position. 
Due to the radiofrequency field $\mathbf{B}_{\text{rf}} (\mathbf{r},t)$, whose frequency is given by $\omega_{\text{rf}}$, tunneling between different magnetic sublevels can occur, and it is more likely to happen in the regions where $\omega_{\text{L}}(\mathbf{r})\approx\omega_{\text{rf}}$. Thus, when both fields are present, the atoms can be confined or repelled in \textit{dressed} magnetic levels which correspond to a superposition of the \textit{bare} energy levels of Eq.~\eqref{umF}. By writing the Hamiltonian of the interaction between the magnetic fields in the rotating wave approximation, and moving to the frame rotating at $\omega_{\text{rf}}$, the radiofrequency-induced adiabatic potentials read (see Ref.~\cite{perrin2016} for the details)
\begin{equation}
U_{M_F}(\mathbf{r}) = M_{\text{F}} \sqrt{[ \hbar \omega_{\text{L}}(\mathbf{r}) -\hbar \omega_{\text{rf}} ]^2 + [\hbar \Omega_{\text{r}}(\mathbf{r})]^2}, 
\label{Ububble}
\end{equation}
where $M_F$ labels the dressed magnetic state, while 
\begin{equation}
\Omega_{\text{r}}(\mathbf{r}) = \frac{|g_F| \mu_{\text{B}} |\mathbf{B}_{\text{rf}}^{\perp} (\mathbf{r})|}{2\hbar}
\end{equation}
is the Rabi frequency among the bare levels, with $\mathbf{B}_{\text{rf}}^{\perp} (\mathbf{r})$ the component of $\mathbf{B}_{\text{rf}} (\mathbf{r},t)$ perpendicular to $\mathbf{B}_0 (\mathbf{r})$ in the position $\mathbf{r}$. 

Considering a system confined optically along two spatial directions, the potential $U_{M_F}(\mathbf{r})$ consists of a double well potential, as it is illustrated in Ref.~\cite{schumm2005}. 
In two dimensions, $U_{M_F}(\mathbf{r})$ confines the atoms on a ring, while in three-dimensions, the atoms will be confined around a two-dimensional shell-shaped surface \cite{zobay2001}. 

\section{Quantum physics of shell-shaped gases}
\label{chapter1}
In this section, which discusses the central results of the review, we analyze the physics of two-dimensional shell-shaped Bose gases. 
To investigate experimentally the properties of this atomic configuration, it is necessary to implement a magnetic confinement with radiofrequency-induced adiabatic potentials, whose essential details are discussed in Section \ref{sectionmagnetictrapping}.
We model these external potentials as \cite{zobay2001}
\begin{equation}
U_{\text{bubble}}(\mathbf{r}) = M_{\text{F}} \sqrt{[ u(\mathbf{r}) -\hbar \Delta]^2 + (\hbar \Omega_{\text{r}})^2}, 
\label{bubble}
\end{equation}
where $M_{\text{F}}$ is the magnetic quantum number of the dressed state populated by the atoms, $\Delta$ and $\Omega_{\text{r}}$ are tunable frequencies, and $u(\mathbf{r}) = m (\omega_x^2 x^2 + \omega_y^2 y^2 + \omega_z^2 z^2)/2$ is the bare harmonic trap with frequencies $\omega_{x,y,z}$. 
The set of points which minimize the potential $U_{\text{bubble}}(\mathbf{r})$ corresponds to the surface of a triaxial ellipsoid, whose equation reads $\omega_x^2 x^2 + \omega_y^2 y^2 + \omega_z^2 z^2 =2\hbar \Delta/m $. When the energy contribution associated to the trapping potential is sufficiently stronger than both the mean kinetic and the interaction energy, the particles will be confined across the surface of this ellipsoid. 

{
Actually, the first experiments with radiofrequency-induced adiabatic potentials \cite{colombe2004} featured an additional gravitational potential contribution $m g_{\text{grav}} z$, with $g_{\text{grav}}$ the acceleration of gravity.  
In the presence of gravity, thus, it was only possible to produce condensates with weak curvature \cite{colombe2004,white2006}, or to engineer ring-shaped traps \cite{sherlock2011,guo2020,deherve2021}. 
In the last years, various gravity-compensation mechanisms were devised in Earth-based laboratories to achieve coverings of larger portions of the shell \cite{guo2022}, or to analyze the dynamics of the atoms along the shell surface \cite{guo2020}: these efforts led thus to the realization of partially-open Bose-Einstein condensate shells. 
Complementary to these studies, the confinement of atomic clouds in fully-closed two-dimensional shells is currently possible only by carrying on the experiments in microgravity facilities such as the Cold Atom Lab \cite{carollo2021,elliott2018,lundblad2019}, or, potentially, in free-falling experiments conducted in a drop tower \cite{vanzoest2010} or in a falling elevator \cite{condon2019}. 
Moreover, a setup using phase-separated mixture in a harmonic confinement was also successful in producing closed shells \cite{jia2022}, whose three-dimensional atomic cloud is {a} Bose-Einstein condensate. 
Most of the results analyzed in the following concern the physics of shell-shaped condensates in the absence of gravity and, therefore, will neglect the gravitational potential energy $m g_{\text{grav}} z$. 
}

When all the harmonic frequencies are equal, i.~e.~$\omega_{x,y,z} = \omega_{\text{r}}$, the trapping configuration is spherically symmetric. 
In this case, and considering the limit of $\Delta \gg \Omega_{\text{r}}$, the potential of Eq.~\eqref{bubble} can be approximated as the radially-shifted harmonic trap 
\begin{equation}
U_{\text{thin}} = \frac{m}{2} \omega_{\perp}^2 (r - R)^2,
\label{shellradial}
\end{equation}
where $\omega_{\perp}= \omega_{\text{r}} (2 M_F \Delta/\Omega_{\text{r}})^{1/2}$ is the transverse frequency, and $R=[2\hbar\Delta/(m \omega_r^2)]^{1/2}$ is the radius of the sphere. 
In the next subsections, we will review the thermodynamic properties of a system of interacting bosonic atoms confined in microgravity conditions on a spherically-symmetric thin shell. 
Instead of describing a three-dimensional system of bosonic particles confined in the external potential of Eq.~\eqref{shellradial}, we will adopt the formalism of functional integration (see subsection \ref{sectionfunctionalint}) to model a uniform Bose gas on the surface of a sphere. 
The explicit discussion of the three-dimensional trapping potential is useful to check if a purely two-dimensional formalism is adequate. 
Indeed, the typical energies of the 2D phenomena we will analyze must always be lower than the energy of the transverse confinement $\hbar \omega_{\perp}$. 

The properties of ellipsoidal shells will be discussed only in Section \ref{sectionellipsoidalshells}, which aims to model the microgravity experiments on shell-shaped condensates \cite{carollo2021,lundblad2019}. 

\subsection{Bose-Einstein condensation and thermodynamics}
Let us consider a spherically-symmetric two-dimensional Bose gas, obtained by confining a system of atomic bosons on the surface of a thin spherical shell. 
We now implement a two-dimensional description of the thermodynamic properties of the system, based on the coherent state functional integral formulation of quantum field theory developed in the subsection \ref{sectionfunctionalint}. 

\subsubsection{Derivation of the grand potential}
\label{subsectionfunctionalintegralsphere}
At the equilibrium, the quantum statistical properties of the Bose gas can be derived from the grand canonical partition function $\mathcal{Z}$, which reads
\begin{equation}
\label{partfunctionsphere}
\mathcal{Z}=\int{{ \cal D} [\bar{\psi},\psi] \; 
e^{-\frac{S[\bar{\psi},\psi]}{\hbar}}},
\end{equation}
where we define the Euclidean action $S$ of a spherical gas as 
\begin{equation}
S[\bar{\psi},\psi] = \int_{0}^{\beta\hbar} \text{d}\tau \, 
\int_{0}^{2 \pi} \text{d} \varphi \, \int_{0}^{\pi} \text{d} \theta \, \sin\theta 
\, R^2 \,{ \cal L }(\bar{\psi},\psi).
\label{actionsphere}
\end{equation}
and $\mathcal{L}$ is the Euclidean Lagrangian. 
We limit ourselves to the description of bosons with a zero-range interaction of strength $g_0$, for which $\mathcal{L}$ reads
\begin{equation}
{ \cal L } = \bar{\psi}(\theta,\varphi,\tau) 
\bigg(\hbar\partial_{\tau}+\frac{\hat{L}^2}{2mR^2}-\mu\bigg) 
\psi(\theta,\varphi,\tau) + \frac{g_0}{2} |\psi(\theta,\varphi,\tau)|^{4},
\label{lagrsphere} 
\end{equation}
where the kinetic part contains the angular momentum operator in spherical coordinates, see Eq.~\eqref{angularmomentum},
and the radius of the sphere $R$ is considered a fixed constant. 

Note that, to avoid discussing the details of the external potential, we are directly implementing the description of a two-dimensional Bose gas on a spherical manifold. 
The connection between theory and the experiment, and the discussion of the trapping parameters requires the modeling in this curved geometry of the bare contact interaction strength $g_0$ between the bosons. {
Through a scattering-theory calculation, we will show in subsection \ref{sectionscatteringth} that $g_0$ depends logarithmically on a high-momentum cutoff which is crucial to get the correct renormalized equation of state. 
To favor a clear presentation of the following material, we now derive a general theory in which $g_0$ is simply treated as a generic input parameter. }

The standard Bogoliubov-Popov theory of a two-dimensional Bose gas \cite{popov1972}, and particularly its implementation with the functional integral, can be extended to the spherical case. The main technical differences concern the different geometry, which produces different quantum numbers in the implementation of the Bogoliubov transformations. 
We decompose the bosonic field as 
\begin{equation}
\psi (\theta, \varphi, \tau) = \psi_0 + \eta (\theta, \varphi, \tau), 
\label{bogoliubovsphere}
\end{equation}
where, as in the noninteracting problem studied in subsection \ref{spherenonint}, the condensate state $\psi_0$ is represented by the $l=0$, $m_l = 0$ mode of the field. The complex fluctuation field $\eta (\theta, \varphi, \tau)$ contains all the components $\{l,m_l\} \neq \{0,0\}$, and can therefore be written as $\eta(\theta,\varphi,\tau) = R^{-1} \sum_{\omega_{n}} \sum_{l=1}^{\infty} 
\sum_{m_{l}=-l}^{l} \, \eta(l,m_{l},\omega_{n}) \, e^{-i\omega_{n}\tau} \, \mathcal{ Y}_{m_{l}}^{l}(\theta,\varphi) $,
where $\omega_n = 2 \pi n / (\beta \hbar)$ are the Matsubara frequencies, and where the factor $R$ is introduced to adimensionalize the components. 

We substitute the field decomposition \eqref{bogoliubovsphere} into the Lagrangian of Eq.~\eqref{lagrsphere}, and we neglect the contributions containing cubic and quartic powers of the fluctuation field, finding \cite{tononi1}
\begin{equation}
\mathcal{L} = \mathcal{L}_0 + \mathcal{L}_{\text{g}},
\end{equation}
where the mean-field Lagrangian is given by
\begin{equation}
\mathcal{ L}_0 = - \mu \psi_0^2 + \frac{g_0}{2} \psi_0^4,
\end{equation}
and where the Gaussian Lagrangian reads 
\begin{align}
\begin{split}
\mathcal{ L}_g = \, \bar{\eta} (\theta,\varphi,\tau) \bigg( \hbar \partial_{\tau} + \frac{\hat{L}^2}{2mR^2} - \mu + 2 g_0 \psi_0^2 \bigg) \eta (\theta,\varphi,\tau)
+{\frac{g_0}{2}} \psi_0^2 \, [\bar{\eta} (\theta,\varphi,\tau) \bar{\eta} (\theta,\varphi,\tau) + \eta (\theta,\varphi,\tau) \eta (\theta,\varphi,\tau) ].
\label{Lagrgsphere}
\end{split}
\end{align}

By expanding the fluctuation field, we express the action of Eq.~\eqref{actionsphere} as a sum over $\omega_n$, $l$ and $m_l$, and, using the property of orthonormality of the spherical harmonics $ \delta_{l l'} \delta_{m_l m_l'} = \int_0^{2\pi} \text{d}\varphi \int_{0}^{\pi} \text{d} \theta \sin \theta \, \mathcal{Y}_{m_{l}'}^{l' *}(\theta,\varphi) \mathcal{ Y}_{m_{l}}^{l }(\theta,\varphi)$, we obtain 
\begin{equation}
S = S_0 + S_{\text{g}},
\end{equation}
where $S_0 = 4\pi R^2 \beta \hbar \, (-\mu \psi_0^2 + g \psi_0^4/2)$ is the mean-field action. The Gaussian action $S_{\text{g}}$ can be written in the following matrix form
\begin{equation}
S_{\text{g}}[\bar{\eta},\eta]=\frac{\beta\hbar}{2} \sum_{\omega_{n}} \sum_{l=1}^{\infty} 
\sum_{m_{l}=-l}^{l}
\begin{bmatrix}
\bar{\eta}(l,m_{l},\omega_{n}) \\ \eta(l,-m_{l},-\omega_{n})
\end{bmatrix}^{T}\textbf{M}
\begin{bmatrix}
\eta(l,m_{l},\omega_{n}) \\
\bar{\eta}(l,-m_{l},-\omega_{n})
\end{bmatrix},
\end{equation}
where the elements of the matrix $\textbf{M}$ are defined as 
\begin{align}
\begin{split}
&\textbf{M}_{11} \; = - i\hbar\omega_{n} + \epsilon_l 
- \mu + 2 g_0 \psi_{0}^2,  \\
&\textbf{M}_{22} \; = + i\hbar\omega_{n} + \epsilon_l 
- \mu + 2 g_0 \psi_{0}^2, \\
&\textbf{M}_{12} = \, \textbf{M}_{21} = (-1)^{m_l} g_0 \psi_{0}^2,
\end{split}
\end{align}
and $\epsilon_l = \hbar^2 l(l+1)/(2 m R^2)$ are the energy levels of a free particle on the sphere. 
Having neglected the non-Gaussian terms in the Lagrangian, it is possible to calculate the partition function by performing the Gaussian functional integral of the action $S_{\text{g}}$. 
Here we simply report the final result for the grand potential 
\begin{equation}
\Omega = -\beta^{-1} \ln \mathcal{Z} = \Omega_0 + \Omega_{\text{g}},
\end{equation}
where 
\begin{equation}
\Omega_0 = 4\pi R^2 \, \bigg(- \mu \psi_0^2 + \frac{g_0}{2} \psi_0^4 \bigg)
\label{omega0sphere}
\end{equation} 
is the mean-field grand potential, and with 
\begin{equation}
\Omega_{\text{g}}(\mu,\psi_0^2) = \frac{1}{2 \beta} \sum_{\omega_{n}} 
\sum_{l=1}^{\infty} 
\sum_{m_{l}=-l}^{l} \ln\{\beta^2[\hbar^2\omega_{n}^2+E_{l}^2(\mu,\psi_0^2)]\},
\label{omegagsphere}
\end{equation}
the Gaussian beyond-mean-field grand potential. 
In the previous expression, we define $E_{l}(\mu,\psi_0^2)$ as 
\begin{equation}
E_{l}(\mu,\psi_0^2)=\sqrt[]{(\epsilon_l
-\mu + 2 g_0 \psi_{0}^2 )^{2}-g_0^2 \psi_{0}^4},
\label{effectivespectrumsphere}
\end{equation}
which represents the excitation spectrum of the quasiparticles. 

The last steps consist in calculating the sum over the Matsubara frequencies in Eq.~\eqref{omegagsphere}, namely, the sum of the discrete frequencies $\omega_n = 2 \pi n /(\beta \hbar)$ over all $n \in \mathbb{Z}$. To perform this operation, we multiply the logarithm by the convergence factor $e^{i \omega_n \Delta \tau}$, with $\Delta \tau \to 0^{+}$. 
The reason for including this term lies in the construction of the functional integral, where the field $\bar{\psi}$ is evaluated {at a time infinitesimally later} than the field $\psi$ (infinitesimally for $M \to \infty$ imaginary time slices) \cite{diener2008,altland2010,salasnich2016}. 
With this operation, the sum in Eq.~\eqref{omegagsphere} converges, producing two Gaussian contributions, one is temperature-independent, the other depends on temperature \cite{salasnich2016}. 
We report the result for the effective grand potential $\Omega(\mu,\psi_0^2)$, which reads \cite{tononi1}
\begin{align}
\begin{split}
\Omega(\mu,\psi_0^2) &= 4 \pi R^2 \big( -\mu \psi_{0}^2 +g_0 \psi_{0}^4 /2 \big) 
+  \frac{1}{2} \sum_{l=1}^{\infty} \sum_{m_{l}=-l}^{l} \, 
[E_{l}(\mu,\psi_0^2)-\epsilon_l - \mu] 
+ \frac{1}{\beta}  \sum_{l=1}^{\infty} 
\sum_{m_{l}=-l}^{l} \, \ln\big[1-e^{-\beta E_{l}(\mu,\psi_0^2)}\big],
\label{grandpotential2}
\end{split}
\end{align}
and where the counterterms at the first line appear due to the convergence-factor regularization. The classical field $\psi_0$ can be determined imposing that it extremizes the grand canonical potential, i.~e.~$\partial \Omega / \partial \psi_0 = 0$. This condition, defining the condensate density as $n_0 = \psi_0^2$, leads to the following relation 
\begin{equation}
n_0(\mu) = \frac{\mu}{g_0} - \frac{1}{4 \pi R^2} \sum_{l=1}^{\infty} 
\sum_{m_{l}=-l}^{l} \, \frac{2\epsilon_l + \mu}{E_{l}^{\text{B}}} 
\, \bigg(\frac{1}{2} + \frac{1}{e^{\beta E_{l}^{\text{B}} }-1}  \bigg),
\label{conddenssphere}
\end{equation}
where we treat the Gaussian contributions perturbatively, considering them as small contributions with respect to the mean-field term \cite{kleinert2004,kleinert2005}. 
In the previous relation, we define the Bogoliubov spectrum as  
\begin{equation}
E_{l}^{\text{B}}=\sqrt[]{\epsilon_l 
(\epsilon_l +  2 \mu ) },
\label{Bogoliubovspectrumsphere}
\end{equation}
which is obtained from \eqref{effectivespectrumsphere} neglecting the higher-order beyond-mean-field corrections of Eq.~\eqref{conddenssphere}.

Substituting the previous equation into the effective grand potential, we finally obtain the grand canonical potential $\Omega$ as a function of the chemical potential
\begin{equation}
\Omega[\mu,n_0(\mu)]  = \Omega_0[\mu,n_0(\mu)] + \Omega_{\text{g}}^{(0)}[\mu,n_0(\mu)] + \Omega_{\text{g}}^{(T)}[\mu,n_0(\mu)], 
\label{grandpotentialsphere}
\end{equation}
where we define the mean-field grand potential as
\begin{equation}
\Omega_0[\mu,n_0(\mu)] = - (4 \pi R^2) \, \frac{\mu^2}{2 g_0},
\end{equation}
and the beyond-mean-field Gaussian contributions at zero and at finite temperature, respectively, as
\begin{align}
\Omega_{\text{g}}^{(0)}[\mu,n_0(\mu)] &=
\frac{1}{2} \sum_{l=1}^{\infty} 
\sum_{m_{l}=-l}^{l} \, (E_{l}^{\text{B}} -\epsilon_l - \mu),
\\ 
\Omega_{\text{g}}^{(T)}[\mu,n_0(\mu)] &= 
\frac{1}{\beta}  
\sum_{l=1}^{\infty} \sum_{m_{l}=-l}^{l} \, \ln(1-e^{-\beta E_{l}^{\text{B}}}).
\end{align}

{
We will compute $\Omega$ explicitly in subsection \ref{sectionthermodynamicssphere}, but, before that, let us discuss the approximations adopted insofar.
The Bogoliubov approach for deriving the thermodynamics is expected to be quantitatively reliable
in temperature regimes where the condensate density is large. At higher temperatures, in which the non-condensate density and the anomalous density become large, more refined methods such as Hartree-Fock Bogoliubov \cite{griffin1996,giorgini1997} give a more accurate description of the thermodynamics. We postpone the presentation of this method to Section \ref{sectionellipsoidalshells}. 
Relying instead on the Bogoliubov framework developed insofar, we calculate in the next section the critical temperature of Bose-Einstein condensation of an interacting Bose gas on the surface of a sphere.
}

\subsubsection{Critical temperature and condensate fraction}
\label{sectioncriticalTcondfrac}
\begin{figure}
\centering
\includegraphics[width=0.518\columnwidth]{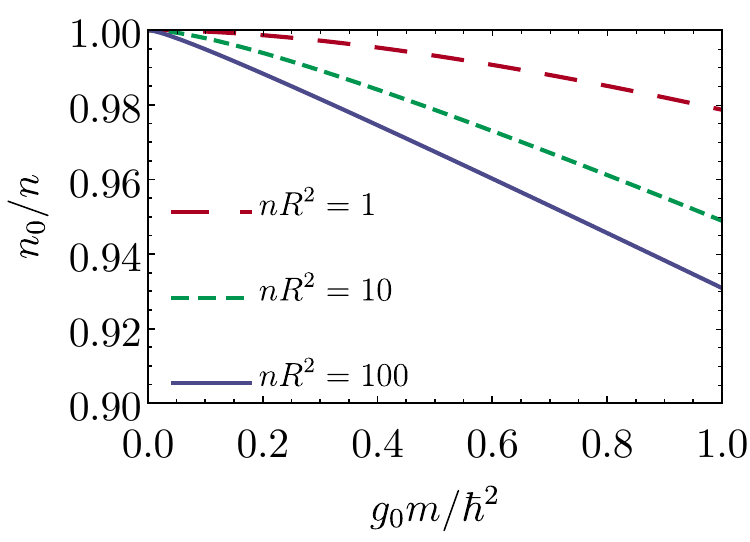}
\caption{
Condensate fraction of interacting bosons on the surface of a sphere at zero temperature, plotted as a function of the adimensional interaction strength $g_0 m /\hbar^2$ and for different $n R^2$.}
\label{fign0overnsphereinteracting}
\end{figure}

Let us calculate the critical temperature and the condensate fraction in terms of the bare interaction strength $g_0$. 
According to standard thermodynamic relations, we can calculate the number density as 
\begin{equation}
n(\mu) = - \frac{1}{4 \pi R^2} \frac{\partial \Omega[\mu,n_0(\mu)]}
{\partial \mu}, 
\end{equation}
where the grand potential is given by Eq.~\eqref{grandpotentialsphere}. To calculate the condensate fraction $n_0 / n$ we aim to obtain a perturbative expression of the density as a function of the condensate density, i.~e.~$n(n_0)$. For this scope, we express the chemical potential in the previous equation as $\mu = g_0 n_0 + ...$ by inverting Eq.~\eqref{conddenssphere}, and $n(n_0)$ reads
\begin{equation}
n(n_0) = n_{0} + f_{g}^{(0)}(n_0) + f_{g}^{(T)}(n_0),
\label{nofn0sphere}
\end{equation}
where we define the beyond-mean-field Gaussian contribution to the density at zero and at finite temperature respectively, as 
\begin{align}
f_{g}^{(0)}(n_0) &= \frac{1}{4 \pi R^2} \frac{1}{2} \sum_{l=1}^{\infty} 
\sum_{m_{l}=-l}^{l} \, \bigg\{ \frac{\epsilon_l + g_0 n_0}{ E_{l}[\mu(n_0),n_0]} - 1\bigg\},
\label{fg0spheren0}
\\
f_{g}^{(T)}(n_0) &= \frac{1}{4 \pi R^2} \sum_{l=1}^{\infty} 
\sum_{m_{l}=-l}^{l}  \, \frac{\epsilon_l + g_0 n_0}{ E_{l}[\mu(n_0),n_0]} 
\, \frac{1}{e^{\beta E_{l}[\mu(n_0),n_0]}-1}.
\label{fgTspheren0}
\end{align}
The Ref.~\cite{tononi1} provides an analytical expression of $n_0 / n$, and, motivated by some applications of variational perturbation theory at the lowest order \cite{kleinert2004,kleinert2005}, makes use of the approximation $n_0 \approx n$ in Eqs.~\eqref{fg0spheren0} and \eqref{fgTspheren0}. 
By doing so and performing as in the noninteracting case of subsection \ref{spherenonint} the integral over $l$ instead of the sum {(equivalent to the integral for $nR^2 \gg 1$, see Fig. \ref{figTbec0sphere})}, we get
\begin{align}
f_{g}^{(0)}(n) &= \frac{m g_0 n}{4 \pi \hbar^2} + \frac{1}{4 \pi R^2} \bigg( 
1 - \sqrt{1+\frac{2 g_0 m n R^2}{\hbar^2}} \bigg),
\label{fg0spheren}
\\
f_{g}^{(T)}(n) &= \frac{1}{2 \pi R^2} 
\sqrt{1+\frac{2 g_0 m n R^2}{\hbar^2}} - \frac{m k_{\text{B}} T}{2 \pi \hbar^2} 
\ln \bigg( e^{\frac{\hbar^2}{m R^2 k_{\text{B}} T} \sqrt{1+\frac{2 g_0 m n R^2}
{\hbar^2}}}-1\bigg), 
\label{fgTspheren}
\end{align}
and putting all these contributions together into Eq.~\eqref{nofn0sphere}, we simply divide by the density $n$ to calculate the condensate fraction of a Bose gas on the surface of a sphere. We find 
\begin{align}
\begin{split}
\frac{n_0}{n} = &1 - \frac{m g_0}{4 \pi \hbar^2} - \frac{1}{4 \pi R^2 n} 
\bigg( 1 + \sqrt{1+\frac{2 g_0 m n R^2}{\hbar^2}} \bigg)
+ \frac{m k_{\text{B}} T} {2 \pi \hbar^2n} 
\ln \bigg( e^{\frac{\hbar^2}{m R^2 k_{\text{B}} T} 
\sqrt{1+\frac{2 g_0 m n R^2}{\hbar^2}}}-1\bigg),
\label{condfracsphere}
\end{split}
\end{align}
which is a valid approximation of the condensate fraction for sufficiently low interactions. 
At zero temperature, taking into account that $x \ln (e^{b/x}-1) \overset{x\to0}{\longrightarrow} \, b$ for $b>0$, the condensate fraction is given by 
\begin{equation}
\frac{n_0}{n}(T=0) = 1 - \frac{m g_0}{4 \pi \hbar^2} - \frac{1}{4 \pi R^2 n} 
\bigg( 1 - \sqrt{1+\frac{2 g_0 m n R^2}{\hbar^2}} \bigg),
\label{condfracspheret0}
\end{equation}
which is shown in Fig.~\ref{fign0overnsphereinteracting}, plotted as a function of $m g_0 / \hbar^2$, for different choices of the parameter $n R^2$. 

In the thermodynamic limit in which $N,R \to \infty$ with $n$ fixed, the quantum statistical properties of an infinite sphere must be equivalent to those of a flat two-dimensional system. In that limit (and working at zero temperature) we obtain $n_0/n = 1 - m g_0/(4 \pi \hbar^2)$, a result equivalent to the condensate fraction calculated by Schick for a 2D Bose gas \cite{schick1971}. 
When considering a finite radius of the sphere, instead, the zero-temperature condensate fraction, at the leading order in the small interaction parameter $m g_0/\hbar^2$, is given by 
\begin{equation}
\frac{n_0}{n}(T=0) \sim 1 - \frac{nR^2}{8\pi} \bigg(\frac{m g_0}{\hbar^2}\bigg)^2.
\end{equation}
so that the finite-radius quantum depletion scales quadratically with the interaction strength. 

\begin{figure}
\centering
\includegraphics[width=0.518\columnwidth]{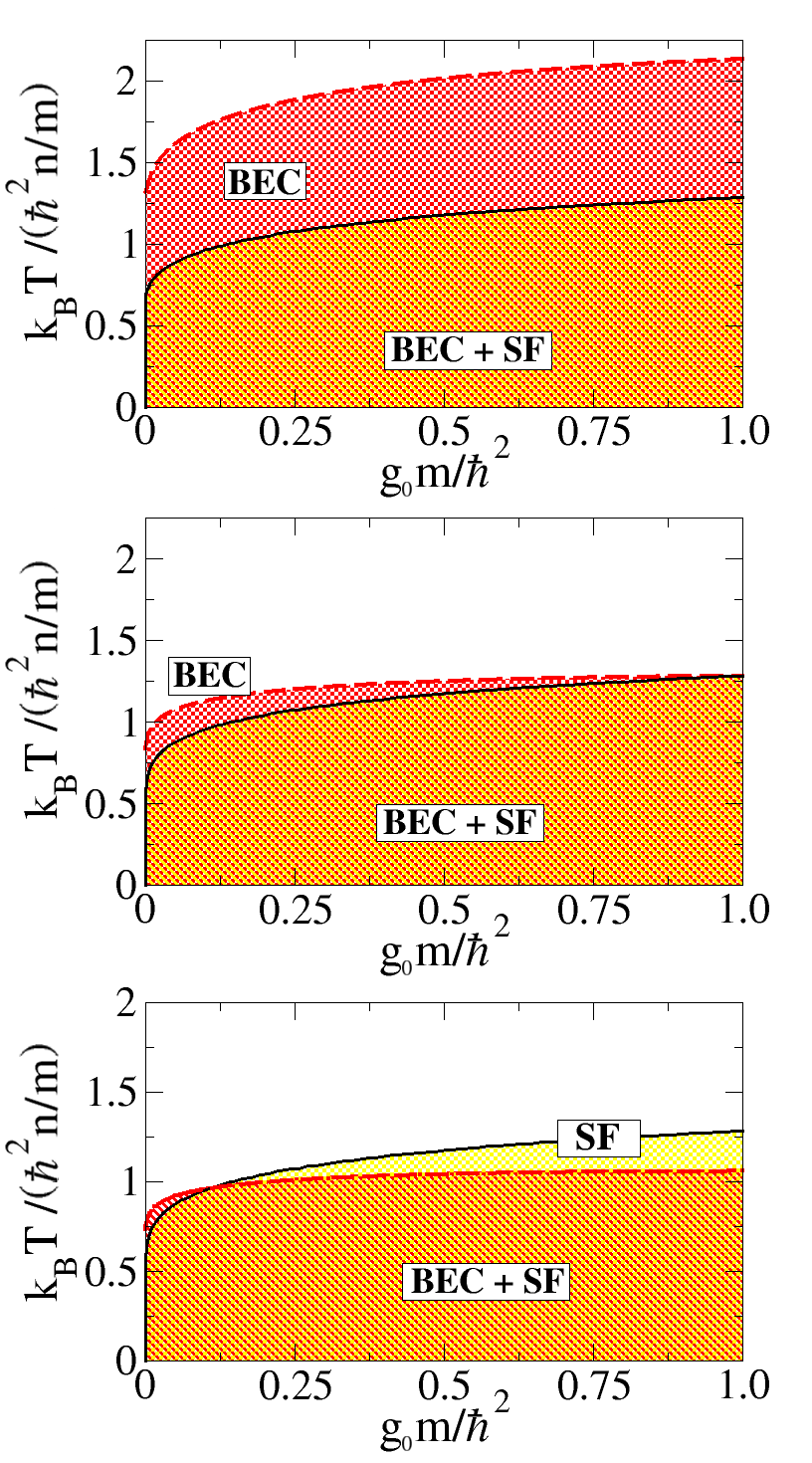}
\caption{
Phase diagram of a Bose gas confined on the surface of a sphere, considering $nR^2=10^2$ in the top panel, $nR^2=10^4$ in the middle one, and $nR^2=10^5$ in the bottom one. 
The red dashed line represents the critical temperature of Eq.~\eqref{Tcinteracting}, properly rescaled with physical constants, under which the condensate fraction of the system is nonzero and the system is a Bose-Einstein condensate (BEC). 
Under the black solid line, determined by the temperature at which the superfluid density satisfies the Kosterlitz-Nelson criterion \cite{tononi1}, the system is also superfluid (SF). 
The calculations based on Bogoliubov theory suggest that a sufficiently small sphere displays a superfluid phase in which the system is condensate but not superfluid. 
This prediction holds also for the more refined modeling of Section \ref{sectionBKTsphere}. 
At the larger value of $nR^2$, which, for a fixed density $n$, corresponds to a larger sphere, the phase diagram is more similar to that of a conventional infinite 2D superfluid. 
In that case, even if the system is superfluid below the BKT critical temperature, there is no Bose-Einstein condensation at finite temperature. 
Adapted from Ref.~\cite{tononi1}.
}
\label{figcondsphere-f2}
\end{figure}

The critical temperature of the interacting Bose gas on a sphere $T_{\text{BEC}}$ can be calculated imposing that $n_0/n=0$ into Eq.~\eqref{condfracsphere}, and we obtain 
\begin{equation}
k_{\text{B}} T_{\text{BEC}} = \frac{\frac{2 \pi \hbar^2 n}{m} - \frac{g_0 n}{2} }{ 
\frac{\hbar^2 }{2 m R^2 k_{\text{B}} T_{\text{BEC}}} \bigg( 1 + \sqrt{1+\frac{2 g_0 m n R^2}{\hbar^2}} \bigg) 
- \ln \bigg( e^{\frac{\hbar^2}{m R^2 k_{\text{B}} T_{\text{BEC}}} 
\sqrt{1+\frac{2 g_0 m n R^2}{\hbar^2}}}-1\bigg)}.
\label{Tcinteracting}
\end{equation}
which is an implicit analytical expression for $T_{\text{BEC}}$. Note that the critical temperature of the noninteracting case $T_{\text{BEC}}^{(0)}$, reported at Eq.~\eqref{Tbec0sphere}, is reproduced by putting $g_0 = 0$ in this expression. 
{We emphasize that this result for $T_{\text{BEC}}$ is obtained within the framework of the Bogoliubov theory improved by the variational perturbation theory at the lowest order \cite{kleinert2004,kleinert2005}, which has the advantage of providing an analytical result for a two-dimensional finite-size system of interacting bosons. In the weakly-interacting regime, the comparison between Monte Carlo simulations and the method illustrated above (see Ref. \cite{kleinert2005}) shows that the relative error on the estimate of $T_{\text{BEC}}$ is always $<10\%$. 
Since in the last decades very precise methods were developed to calculate the critical temperature of Bose-Einstein condensation in three-dimensional free space, both analytical (see Ref. \cite{yukalov2017} for a recent review, as well as Refs. \cite{baym2000,holzmann2001}) and numerical (among them, a few examples include Refs. \cite{holzmann1999,kashurnikov2001}), future theoretical analyses could focus on the extension of these methods to the curved geometry of the sphere, with the goal of improving the precision of Eq. \eqref{Tcinteracting}.
}

In Fig.~\ref{figcondsphere-f2} we plot the adimensional critical temperature (red dashed line) as a function of the adimensional contact interaction strength $m g_0/\hbar^2$. 
Due to the finite system size, Bose-Einstein condensation occurs at finite temperature and, due to the peculiar form of the superfluid density, we find that for small enough radii, a shell-shaped condensate displays a phase of Bose-Einstein condensation without superfluidity. 
The superfluid transition, modeled for Fig.~\ref{figcondsphere-f2} in a simplified way with the Kosterlitz-Nelson criterion \cite{nelson1977}, will be discussed more in detail in Section \ref{sectionBKTsphere}, but the phase diagram presented here will be qualitatively unchanged. 
{
This delicate interplay of Bose-Einstein condensation and superfluidity in the finite-size spherical geometry requires a careful interpretation, and a reflection on the phenomena of Bose-Einstein condensation and superfluidity themselves.

Condensation is a static and purely statistical phenomenon which consists of the macroscopic occupation of the same lowest-energy single-particle state. 
It involves off-diagonal long-range order and it can occur in both interacting and noninteracting systems. Superfluidity is instead a collective dynamical phenomenon and therefore needs the interaction between bosons (a system of non-interacting bosons at $T=0$ is condensed but not superfluid, even in the thermodynamic limit). As soon as a weak interaction between the bosons is turned on, even though the phenomenon of Bose-Einstein condensation is practically unaffected, the infinite system becomes superfluid. 
Operationally, we can identify the critical temperature for BEC as the maximum temperature at which the condensate fraction is finite and non-negative, and the BKT temperature as the one which satisfies the Kosterlitz-Nelson criterion, and these yield that $T_{BEC} = 0$ and $T_{BKT} > 0$ in the 2D case at the thermodynamic limit. 

The situation is however different in a finite-size system like a gas confined on the surface of a sphere. 
For sufficiently small values of $nR^2$ and of $g_0 m/\hbar^2$ the gas turns out to be too weakly-interacting to show a collective superfluid behavior, but, if the system reaches a large enough density in the phase space, there can still be condensation: this region has been marked in Fig.~\ref{figcondsphere-f2} as BEC (without superfluidity). 
In addition, in a finite-size system there is always a residual fraction of the particles occupying the condensate state and more refined theories could show that the BEC region extends even more. 
Interestingly, Monte Carlo simulations of a two-dimensional finite-size gas have shown the existence of a phase of nonzero ``quasicondensate'' density well above the temperature of the superfluid BKT transition \cite{prokofev2002}.
This phase of BEC without superfluidity should be the object of more sophisticated future analyses. 
One could indeed devise, in a matter consistent with the renomalization group framework presented in Section \ref{sectionBKTsphere}, a calculation of the renormalized superfluid density $n_s$ from the non-negative residual $n_0$, to judge quantitatively whether the superfluid phase extends significantly into the region of BEC of Fig.~\ref{figcondsphere-f2}.
}

\subsubsection{Scattering theory on a sphere}
\label{sectionscatteringth}
With the goal of discussing the thermodynamics of spherical bubble-trapped condensates, we now analyze the scattering theory on a spherical surface. 
In particular, we will restrict our discussion to a large sphere with $R \gg a_s$, where $a_s$ is the two-dimensional scattering length of the system, which we will later define precisely. 
To analyze the scattering of two identical particles on a large sphere, interacting with the interaction operator $\hat{V}$, we can consider the motion of a single particle with reduced mass $m/2$ in the potential $\hat{V}$. 
In the absence of scattering, the dimensionless Hamiltonian of the particle with reduced mass would be given by $\hat{H}_0=\hat{L}^2/\hbar^2$, with $\hat{L}^2$ the angular momentum operator, and the dimensionless energy eigenstates read $\mathcal{E}_{l_0}=l_0 (l_0+1)$, with $l_0$ a quantized positive integer. The free eigenfunctions are the spherical harmonics $\mathcal{ Y}_l^{m_l}(\theta,\varphi) = \braket{\theta,\varphi|\, l,m_l}$, where the brakets satisfy the following relations
\begin{equation}
\sum_{l=0}^{\infty} \sum_{m_l=-l}^{l} \, \ket{l,m_l} \bra{l,m_l} = 1,
\qquad
\int_0^{2\pi} \text{d}\varphi \int_{0}^{\pi} \text{d} \theta \, \sin \theta \, \ket{\theta,\varphi} \bra{\theta,\varphi} = 1.
\label{identitysphericalharmonics}
\end{equation}

The full Hamiltonian $\hat{H}_0 + \hat{V}$ has the eigenstates $\ket{\Psi_{l_0}}$, which are in principle unknown, and we define the transition operator $\hat{\mathcal{T}}$ through the relation $\hat{\mathcal{T}} \ket{l,m_l} = \hat{V} \ket{\Psi_{l_0}}$,
namely, as the operator whose action on the eigenstates of the free Hamiltonian is equivalent to the action of the interaction operator on the eigenstates of the full Hamiltonian. 
The Schr\"odinger equation $(\hat{H}_0 + \hat{V})\ket{\Psi_{l_0}} = \mathcal{E}_{l_0} \ket{\Psi_{l_0}}$ can be formally solved as 
\begin{equation}
\ket{\Psi_{l_0}} = \ket{l_0,m_{l_0}} + \frac{1}{\mathcal{E}_{l_0} -\hat{H}_0 +i \eta} \, \hat{V} \ket{\Psi_{l_0}}, 
\end{equation}
where $\eta \to 0^{+}$ is a small vanishing parameter included to regularize the calculations, since the eigenvalues of $\hat{H}_0$ coincide with $\mathcal{E}_{l_0}$. 
Acting with $\hat{V}$ on the left, we find the Lippmann-Schwinger equation \cite{lippmann1950,stoof2009}
\begin{equation}
\hat{\mathcal{T}} = \hat{V} + \hat{V} \, \frac{1}{\mathcal{E}_{l_0}-\hat{H}_0 +i \eta} \, \hat{\mathcal{T}}, 
\label{Tsphere}
\end{equation}
where we used the definition of the transition operator. 

Considering the s-wave interaction between bosons $\hat{V} \approx \hat{V}_0 = \tilde{g}_0 \, \delta(1 -\cos \theta)\, \delta(\varphi)$, where {$\tilde{g}_0$ is dimensionless and} $g_0 = \hbar^2 \tilde{g}_0/m$ is precisely the interaction strength introduced in the functional integral calculations, we calculate the matrix element for s-wave scattering $\mathcal{T}_{l',l_0}=\bra{l',m_l'=0}\hat{\mathcal{T}} \ket{l_0,m_{l_0}=0}$ using Eq.~\eqref{Tsphere}.
After a few steps, which include neglecting the matrix elements in partial waves higher than the s-wave, we obtain \cite{tononi2022}
\begin{align}
\mathcal{T}_{l',l_{0}} =& \tilde{g}_0 \, \frac{\sqrt{(2l'+1)(2l_0+1)}}{4\pi} 
\bigg[ 1 +\sum_{l=0}^{\infty} \frac{\sqrt{2l+1}}{\sqrt{2l_0+1}}
\frac{\mathcal{T}_{l,l_{0}}}{\mathcal{E}_{l_0}-\mathcal{E}_{l}+i\eta}\bigg],
\label{bornTll0}
\end{align}
whose iterative solution yields
\begin{equation}
\frac{1}{\tilde{g}_e (\mathcal{E}_{l_0}+i\eta)} = \frac{1}{\tilde{g}_0} + \frac{1}{4\pi}\sum_{l=0}^{\infty} \frac{2l+1}{\mathcal{E}_l-\mathcal{E}_{l_0}-i\eta},
\label{TEnocutoff}
\end{equation}
where we define the renormalized interaction strength $\tilde{g}_e $ as $\tilde{g}_e (\mathcal{E}_{l_0}+i\eta) = 4\pi \, \mathcal{T}_{l',l_{0}}/\sqrt{(2l'+1)(2l_0+1)}$.

Following Refs.~\cite{mora2003,mora2009}, instead of the summation at the right-hand side of the previous equation we integrate over $l$ up to an ultraviolet cutoff $l_c$ {(the corrections with respect to the sum exceed the accuracy sought in the calculation)}. Thus, calling $\tilde{g}_{e}$ as  $\tilde{g}_{e,l_c}$ to remind ourselves of the cutoff, we get
\begin{equation}
\tilde{g}_{e,l_c}(\mathcal{E}_{l_0}) = - \frac{2\pi}{-\frac{2\pi}{\tilde{g}_0} + \frac{1}{2}\ln\big[ \frac{l_0(l_0+1)}{l_c(l_c+1)} \big] -i\frac{\pi}{2} },
\label{Tlc}
\end{equation}
where the limit of $\eta \to 0^+$ has been taken after the integration.

Our goal is to find a relation between the contact interaction strength $g_0$, and the cutoff $l_c$. 
In particular, we will also include this cutoff in the zero temperature grand potential and, by using the relation between $g_0$ and $l_c$, we will be able to get a renormalized cutoff-independent grand potential $\Omega$. 
For low-energy scattering, the renormalized interaction strength is equal to the $s$-wave scattering amplitude $f$, which is given by \cite{zhang2018}
\begin{equation}
f(\mathcal{E}_{l_0}) = - \frac{4}{\cot \delta_0(E_{l_0})-i},
\label{flogeneral}
\end{equation}
where, differently from Ref.~\cite{zhang2018}, we have multiplied by a factor $-\sqrt{i/(8\pi k)}$ to get the flat-case scaling of $f(\mathcal{E}_{l_0})$ at large distances \cite{tononi2022}. 
In the scattering amplitude, the phase shift $\delta_0 (\mathcal{E}_{l_0})$ of the partial $s$-wave reads 
\begin{equation}
\cot \delta_0(\mathcal{E}_{l_0}) = -\frac{2}{\pi} [Q_{l_0}(\cos\theta_0)+B^{-1}], 
\end{equation}
where $Q_{l_0}(\cos\theta_0)$ is the Legendre associated function of second kind, $\theta_0$ is the range of the potential, and $B^{-1}$ is a constant which depends on high-energy scattering properties. 
As in Ref.~\cite{landauquantum}, the parameter $B^{-1}$ can be fixed introducing the $s$-wave scattering length, and we set $B^{-1}=\ln(\theta_0/\theta_s)$, with $\theta_s$ the $s$-wave scattering angle. Note that we can define $a_s$, i.~e.~the two-dimensional $s$-wave scattering length \textit{on the sphere}, as $a_s = R\, \theta_s$.
For a sufficiently large spherical surface, such that the potential range $\theta_0$ is very small and $l_0$ is much larger than the zero-point motion, the scattering amplitude can be expanded as \cite{zhang2018}
\begin{equation}
f(\mathcal{E}_{l_0}) = -\frac{2\pi}{\ln\big( \frac{l_0 \theta_s e^{\gamma}}{2} \big) -i\frac{\pi}{2} },
\label{fsphere}
\end{equation}
{where $\gamma$ is the Euler-Mascheroni constant.}
We impose that $\tilde{g}_{e,l_c}(\mathcal{E}_{l_0}) = f(\mathcal{E}_{l_0})$ at the leading order in $l_0$, and thus, from Eqs.~\eqref{Tlc}, \eqref{fsphere}, we find the contact interaction strength
\begin{equation}
g_0 = -\frac{2\pi\hbar^2}{m} \frac{1}{\ln[\sqrt{l_c(l_c+1)}\, a_s e^{\gamma}/(2R)]}, 
\label{g0rensphere}
\end{equation}
expressed as a function of the cutoff $l_c$, and of the $s$-wave scattering length on the sphere. 

Before of concluding the discussion of scattering on a spherical surface, we remark that the results of Eqs.~\eqref{fsphere} and \eqref{g0rensphere} are obtained assuming a large radius of the sphere. 
From a quantitative point of view, we thus suppose that the radius is much larger than the healing length $\xi$, which can be modeled as 
\begin{equation}
\xi = \sqrt{\frac{\hbar^2}{2m g_{2D} n}},
\label{healinglength}
\end{equation}
where we can estimate $g_{2D}$ as the mean-field contact interaction strength in two-dimensional weakly-interacting condensates, namely $g_{2D} = -\frac{4\pi\hbar^2}{m} \frac{1}{\ln(n a_s^2)}$.
Strictly speaking, the two-dimensional $s$-wave scattering length in $g_{2D}$ refers to a flat two-dimensional system, while the length appearing in Eq.~\eqref{g0rensphere} refers to scattering on the spherical surface. 
These quantities are in principle different, but the hypothesis of working in the large-radius regime justifies the use of relations obtained for flat condensates.
{
We stress that we cannot simply use $g_0$ to define the healing length because it depends on the cutoff $l_c$, which is unknown\footnote{{Actually, whatever choice of the interaction strength is fine due to its logarithmic dependence on the typical system scales, and any factor appearing inside the logarithm is negligible in the weakly-interacting regime.}}. 
We therefore proceed by assuming} for $a_s$ the value calculated in works on quasi two-dimensional condensates \cite{petrov2001}, namely $a_s = 2 \sqrt{\pi/C} \exp( -\sqrt{\pi/2} \, l_{\perp}/a_{3D} -\gamma_{\text{E}}) l_{\perp}$,
where $\gamma_{\text{E}}$ is the Euler-Mascheroni constant, $C=0.915$ is calculated solving numerically the two-body problem, $l_{\perp}$ is the shell thickness, and $a_{3D}$ is the $s$-wave scattering length in three-dimensions. 
We emphasize that all these quantities are known and, in particular, that $l_{\perp} = \sqrt{\hbar/(m\omega_{\perp})}$, with $\omega_{\perp}$ defined in Eq.~\eqref{shellradial} in terms of the trap frequencies. 

\subsubsection{Thermodynamics}
\label{sectionthermodynamicssphere}
After discussing the scattering properties of a spherical Bose gas, we are ready to derive the regularized grand potential. 
The zero-temperature contributions to the grand potential of Eq.~\eqref{grandpotentialsphere} can be expressed as
\begin{equation}
\frac{\Omega^{(0)}}{4\pi R^2} = -\frac{\mu^2}{2g_0} + \frac{1}{2} \int_{1}^{l_c} \text{d} l 
\, (2l+1) \, (E_{l}^{\text{B}} -\epsilon_l - \mu),
\end{equation} 
where we integrate instead of summing\footnote{{See Ref. \cite{tononi2022} for subleading corrections that stem from evaluating the sum instead of the integral.}}, and where we include an ultraviolet cutoff $l_c$. 
The integral can be performed analytically, and its logarithmic divergence in the parameter $l_c$ is balanced exactly by the same scaling of the bare interaction $g_0$, as can be seen in Eq.~\eqref{g0rensphere}.
Including also the finite-temperature contribution of Eq.~\eqref{grandpotentialsphere}, we get

\begin{align}
\frac{\Omega}{4 \pi R^2}  = &-\frac{m\mu^2}{8\pi\hbar^2} \bigg\{ \ln \bigg[ \frac{4\hbar^2}{m(\mu + E_1^{\text{B}} +\epsilon_1) a_s^2 \, e^{2\gamma+1}}\bigg] +\frac{1}{2}\bigg\}
+\frac{m E_1^{\text{B}}}{8\pi\hbar^2} (E_1^{\text{B}} -\epsilon_1 - \mu) 
+ \label{omegasphere} \frac{1}{4\pi R^2}\frac{1}{\beta} \sum_{l=1}^{\infty} \sum_{m_l=-l}^{l} \, \ln (1 - e^{-\beta E_l^{\text{B}}}),
\end{align}
which is the grand potential per unit of area of a spherically symmetric Bose gas. 
Note that $E_1^B=\sqrt{\epsilon_1[\epsilon_1 + 2\mu]}$, with $\epsilon_1=\hbar^2/(m R^2)$, and in the thermodynamic limit in which $R \to \infty$ we have $E_1^{\text{B}}, \epsilon_1 \to 0$. 
In this limit, and at a one-loop level, our grand potential coincides with the one obtained in Refs.~\cite{mora2003,mora2009} where an infinite and uniform Bose gas is studied.

We calculate the number of atoms in the condensate deriving the grand potential with respect to the chemical potential $\mu$. We obtain \cite{tononi2022}
\begin{align}
n = 
\frac{m \mu}{4\pi\hbar^2} \, \ln \bigg\{ \frac{4\hbar^2 [1-\alpha(\mu)]}{m\mu\, a_s^2 \, e^{2\gamma+1+\alpha(\mu)}}\bigg\} 
+  \frac{1}{4\pi R^2} \, \sum_{l=1}^{\infty} \sum_{m_l=-l}^{l} \, \frac{\epsilon_l}{E_l^B} \frac{1}{e^{\beta E_l^B}-1},
\label{Nsphere} 
\end{align}
where $\alpha(\mu) = 1-\mu/(\mu+E_1^B +\epsilon_1)$ encodes the finite-size contributions, and it therefore vanishes in the thermodynamic limit. 

The typical experiments with Bose-Einstein condensates are done with a fixed number of particles and are in principle not compatible with a description in the grand canonical ensemble: the systems are not exchanging particles with an external reservoir. 
It is however simpler to calculate the partition function in this ensemble, and the spurious fluctuations in the number of atoms do not usually prevent the correct description of the experiments, provided that the number of particles is sufficiently large. 
Despite these considerations, it is formally inconsistent to fix a temperature-independent value of the chemical potential when, on the contrary, it is the number of atoms that is kept fixed. 
The correct procedure is, in this case, to perform the following Legendre transformation: 
\begin{equation}
F(T,V,N) = \mu(T,V,N) N + \Omega[T,V,\mu(T,V,N)],
\end{equation}
in which $F$, the free energy of the system that is determined by fixing $T$, $V$, $N$, is obtained from the grand potential. 
For this operation, it is necessary to know the chemical potential as a function of $T$, $V$, $N$. To obtain it numerically, we calculate the number of atoms $N$ of Eq.~\eqref{Nsphere} for a fixed volume $V$ and on a grid of values of $T$ and $\mu$. After that, we fit and invert numerically the function $N(T,V,\mu)$, obtaining $\mu(T,V,N)$. 

\begin{figure}
\centering
\includegraphics[width=0.518\columnwidth]{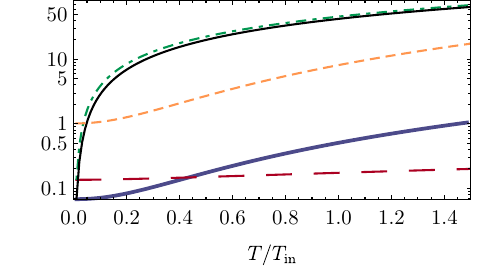}
\caption{
Thermodynamic functions of a spherically-symmetric bosonic gas plotted as a function of the temperature $T/T_{\text{in}}$, where $T_{\text{in}} = 35 \, \text{nK}$. 
These curves represent: the specific heat per unit of mass and of atom number $\tilde{c}_v$ (green dot-dashed), the entropy per unit of mass and atom number $\tilde{s}$ (black), the ratio $\kappa_T/\kappa_S$ between the isothermal and adiabatic compressibilities (orange dashed), the adimensional chemical potential $\mu/E_{l_{\perp}}$ (red long-dashed), and the grand potential $\Omega/E_{l_{\perp}}$ (blue thick). 
For this plot, we consider a spherically-symmetric gas of $^{87}$Rb atoms trapped with the external potential of Eq.~\eqref{bubble}, where we set $M_F=1$, $\omega_{r} = 2\pi \times 173 \, \text{Hz}$, $\Delta = 2\pi \times 30 \, \text{kHz}$, and $\Omega_r = 2\pi \times 3 \, \text{kHz}$ \cite{lundblad2019}, and we have $E_{l_{\perp}} = \hbar^2/(2ml_{\perp}^2)$, with $l_{\perp} = 0.4 \, \mu \text{m}$ the shell thickness. Moreover, the scattering length is modeled as discussed after Eq.~\eqref{healinglength}.  From \cite{tononi4}.}
\label{figpelster1}
\end{figure}

Once that the free energy is known, all the other thermodynamic functions can be derived with the usual thermodynamic identities. 
We show in Fig.~\ref{figpelster1} some relevant thermodynamic functions of a spherically-symmetric Bose gas. 
{We also point out that} in microgravity experiments the bubble shape is produced by inflating atomic gases initially confined in harmonic potentials, and the thermodynamics of noninteracting bosons in spherically-symmetric shells and its evolution during the expansion of the bubble was modeled in Ref.~\cite{rhyno2021}.

\subsection{Bare superfluid density}
We derive the superfluid density of a two-dimensional spherical superfluid, by extending the functional integral calculation implemented in subsection \ref{subsectionfunctionalintegralsphere}. 
Inspired by Landau, i.~e.~Ref.~\cite{landau1941}, we suppose that the trapping potential rotates along a fixed axis with a constant angular velocity, so that the superfluid part of the fluid remains unperturbed while the normal fluid rotates with the trap. 

Thus, when a spherical superfluid is rotating, the angular momentum of the system is proportional to the nonclassical moment of inertia and, therefore, it is also proportional to the density of the normal fluid. 
Given the normal density, it is then possible to derive the superfluid density as the total density minus the normal one. 
To implement quantitatively these concepts, we impose a rotation of the normal fluid along $z$, with angular velocity $\Omega_z$, by shifting the imaginary time derivative in Eq.~\eqref{lagrsphere} as $\hbar \partial_{\tau} \rightarrow \hbar \partial_{\tau} + \Omega_z \hat{L}_z$,
where $\hat{L}_z = -i \hbar\, \partial_{\varphi}$. 
Taking into account this modified term, the steps in subsection \ref{subsectionfunctionalintegralsphere} can be essentially repeated up to the Lagrangian of Eq.~\eqref{Lagrgsphere}, which is shifted as $\mathcal{L}_{\text{g}} \to \mathcal{L}_{\text{g}} - i \hbar \, \bar{\eta} (\theta,\varphi,\tau) \, \partial_{\varphi} \eta (\theta,\varphi,\tau)$. Then, besides including a few additional terms, all the other calculations can be done in the same formal order, and the grand potential contribution at Eq.~\eqref{omegagsphere} reads
\begin{equation}
\Omega_{\text{g}}(\mu,\psi_0^2) = \frac{1}{2 \beta} \sum_{\omega_{n}} 
\sum_{l=1}^{\infty} 
\sum_{m_{l}=-l}^{l} \ln\{\beta^2[\hbar^2\omega_{n}^2+\xi_{l}^2(\mu,\psi_0^2)]\},
\end{equation}
where $\xi_l (\mu,\psi_0^2) = E_l(\mu,\psi_0^2) + m_l \hbar \Omega_z$, with $E_l(\mu,\psi_0^2)$ given by Eq.~\eqref{effectivespectrumsphere}. 
As before, we perform the sum over the Matsubara frequencies with standard techniques, and the total grand potential becomes
\begin{equation}
\begin{split}
\Omega (\mu, \psi_0^2) = 
(4 \pi R^2) \bigg( - \mu \psi_0^2 + \frac{g_0}{2} \psi_0^4 \bigg) 
+ \frac{1}{2} \sum_{l=1}^{\infty} \sum_{m_l = -l}^l E_l (\mu, \psi_0^2)
+ \frac{1}{\beta} \sum_{l=1}^{\infty} \sum_{m_l = -l}^l \ln \{1-e^{-\beta[E_l (\mu, \psi_0^2) + m_l \hbar \Omega_z]}\},
\end{split}
\end{equation}
where the zero-temperature counterterms, included in the previous Eq.~\eqref{grandpotential2}, are inessential for the derivation of the present section. 

Given the grand potential of the rotating fluid, and considering the analogy with a similar calculation done in flat geometries \cite{tononi7}, the angular momentum of the normal fluid can be calculated as 
\begin{equation}
L_n = -  \frac{\partial \Omega (\mu, \psi_0^2)}{\partial \Omega_z}  \bigg\rvert_{\psi_0^2 = n_0(\mu)}
\overset{\Omega_z \sim 0}{\longrightarrow} \,
\beta \sum_{l=1}^{\infty} \sum_{m_l = -l}^l \hbar^2 m_l^2 \, \frac{e^{\beta E_l^{\text{B}} }}{(e^{\beta E_l^{\text{B}} }-1)^2} \,  \Omega_z,
\end{equation}
where the right-hand side is obtained by expanding the resulting angular momentum for a small angular velocity $\Omega_z$. 
Taking into account the known identity $\sum_{m_l = -l}^l m_l^2 = (2l+1) (l^2+l)/3$,
the angular momentum of the normal fluid, which is dragged by the rotating trap, reads
\begin{equation}
L_n = 
\frac{\beta}{3} \sum_{l=1}^{\infty} (2l+1) \, \hbar^2 (l^2+l) \, \frac{e^{\beta E_l^{\text{B}} }}{(e^{\beta E_l^{\text{B}} }-1)^2} \,  \Omega_z, 
\end{equation}
which results from a microscopic calculation. But the angular momentum can also be expressed as $L_n = I_n \Omega_z$, where $I_n = (2/3) M_n R^2$
is the moment of inertia of a hollow sphere with mass $M_n = m n_n (4\pi R^2)$, and $n_n^{(0)}$ is the (bare) number density of the normal fluid. 
A simple comparison of the previous relations yields the bare normal density of the spherical superfluid
\begin{equation}
n_n^{(0)} = \beta \sum_{l=1}^{\infty} \frac{(2l+1)}{4\pi R^2} \, \frac{\hbar^2 l (l+1)}{2 m R^2} \, \frac{e^{\beta E_l^{\text{B}} }}{(e^{\beta E_l^{\text{B}} }-1)^2},
\label{barenn}
\end{equation}
and, consequently, the bare superfluid density 
\begin{equation}
n_s^{(0)} = n - n_n^{(0)},
\label{barens}
\end{equation}
which coincides with the result postulated in Ref.~\cite{tononi1}. 

The superfluid density derived here is denoted as \textit{bare}, meaning that it takes into account only the Bogoliubov excitations of the system and that it neglects the vortex-antivortex excitations. 
In two-dimensional systems, including spherical superfluids, these topological excitations proliferate with the temperature, renormalizing the superfluid density. 
In Section \ref{sectionvortices}, we will include the physics of vortices in the theoretical description, by modeling explicitly their contribution to the energy of the superfluid. 

\subsection{Vortices in a spherical superfluid shell}
\label{sectionvortices}
In the calculations of the previous sections we obtained the grand potential $\Omega$ of a spherical bosonic gas as a sum of a mean-field part and of a beyond-mean-field part, the latter obtained at a one-loop level. 
In particular, the beyond-mean-field terms describe the Bogoliubov excitations of the system on top of the mean-field condensate state. 
In ultracold bosonic gases, however, the Bogoliubov quasiparticles are not the only excitations that the system may possess. 
Indeed, even at zero temperature, a superfluid can host quantized vortices, namely, configurations of the macroscopic field in which the fluid rotates around a single point (the core of the vortex) with quantized angular momentum. 
In line with the current experimental and theoretical understanding, the physics of vortices and their eventual thermal proliferation cannot be detected through discontinuities of any of the free energy derivatives \cite{desbuquois2014}. Even though this justifies the exclusion of vortex excitations in the modeling of the system thermodynamics (cf. Section \ref{sectionthermodynamicssphere}), it must be stressed that there are currently no analytical methods allowing to evaluate their eventual role. {Therefore, including the vortices within a microscopic calculation of the grand potential, for instance by evaluating directly how they renormalize the grand canonical partition function, is an interesting topic for future research.}

In this section, we construct an effective model to calculate the energy of a system of vortices on a spherical superfluid film. 
For this scope, we suppose that the superfluid at a finite temperature is described by the order parameter\footnote{ 
Note that the square modulus of the order parameter \eqref{superfluidorderpar} is the superfluid density, whereas the usual Madelung transformation of the zero-temperature condensate wave function (see, for instance, Ref. \cite{stringari1996}) contains the condensate density. 
While these quantities coincide at zero temperature and for weak interactions, they can differ substantially outside these regimes. 
Our order parameter enables us to describe effectively the superfluid transition, even in the absence of a microscopic theory that includes the vortices in the grand canonical partition function. Interestingly, such a theory would allow to obtain the renormalized superfluid density with a direct microscopic calculation.
}
\begin{equation}
\psi(\theta,\varphi) = \big[n_s^{(0)}(T)\big]^{1/2}  \, e^{i \Phi(\theta,\varphi)},
\label{superfluidorderpar}
\end{equation}
where $n_s^{(0)}$ is the uniform (bare) superfluid density given by Eq.~\eqref{barens}, and where the field $\Phi(\theta,\varphi)$ represents the phase field of the bosonic fluid. 
The bare superfluid density, which contains only the contribution due to the Bogoliubov excitations vanishes at a temperature $T^{*}$, and therefore, when this superfluid transition occurs, the order parameter becomes zero. 
Actually, it turns out that the vortical configurations of the superfluid can be thermally excited, and renormalize the superfluid density. 
Here we will only discuss the calculation of the vortex energy, and the issue of renormalization will be analyzed in the Section \ref{sectionBKTsphere}. 

We calculate the free energy associated to the order parameter $\psi(\theta,\varphi)$ as \cite{vitelli2010}
\begin{equation}
E = \int_{0}^{2 \pi} \text{d} \varphi \, \int_{0}^{\pi} \text{d} \theta \, \sin\theta
\, R^2 \, \psi^{*}(\theta,\varphi) \bigg( \frac{\hat{L}^2}{2mR^2} \bigg) \psi(\theta,\varphi), 
\end{equation}
where the energy contributions associated to the radial motion, as done insofar, are also not included here. 
After some simple steps, the energy can be rewritten as
\begin{equation}
E = \frac{1}{2}m n_s^{(0)} \int_{0}^{2 \pi} \text{d} \varphi \, \int_{0}^{\pi} \text{d} \theta \, \sin\theta
\, \mathbf{v}(\theta,\varphi) \cdot \mathbf{v}(\theta,\varphi),
\label{energyvorticesgeneral}
\end{equation}
in which we define the velocity field $\mathbf{v}(\theta,\varphi)$ as
\begin{equation}
\mathbf{v}(\theta,\varphi) = \frac{\hbar}{m R} \tilde{\nabla}_{R} \Phi(\theta,\varphi), 
\end{equation}
where $\tilde{\nabla}_{R} = \mathbf{e}_{\theta} \, \partial_{\theta} +  (\sin\theta)^{-1}\, \mathbf{e}_{\varphi} \,\partial_{\varphi}$ is the dimensionless gradient in spherical coordinates for $R$ fixed, with $\mathbf{e}_{\theta}$ and $\mathbf{e}_{\varphi}$ the unitary vectors along $\theta$ and $\varphi$. 
Due to its definition, the velocity field is irrotational in all the spatial coordinates where the phase field is defined, namely $\nabla \times \mathbf{v} = 0$,
which can be verified calculating the curl in spherical coordinates and considering that the velocity field has a zero radial component. 
However, a superfluid can have some phase defects, namely, point singularities where the phase field is not defined and the curl of the velocity is nonzero. 
In general, we can express $\mathbf{v}(\theta,\varphi)$ as
\begin{equation}
\mathbf{v}(\theta,\varphi) = \mathbf{v}_0(\theta,\varphi) + \mathbf{v}_{\text{v}}(\theta,\varphi),
\end{equation}
decomposing it in an irrotational part $\mathbf{v}_0(\theta,\varphi)$ without phase defects, and in a part with nonzero curl $\mathbf{v}_{\text{v}}(\theta,\varphi)$ that describes the velocity field of the vortices. 
For the vortical part, we write the Feynman-Onsager condition of quantized circulation \cite{onsager1949,feynman1955}, namely
\begin{equation}
\oint_{\partial\Sigma} \mathbf{v}_{\text{v}} \cdot \text{d}\mathbf{l} = 2 \pi \frac{\hbar}{m} \sum_i q_i, 
\label{feynmanonsager}
\end{equation}
where $q_i$ are the integer charges of the vortices inside the region $\Sigma$, with border $\partial\Sigma$.

The fields $\mathbf{v}_0$ and $\mathbf{v}_{\text{v}}$ are orthogonal, and the free energy splits into the sum of the kinetic energy of the vortical fluid, and of the free energy of the (everywhere) irrotational fluid. 
The energy of the part of the fluid without phase defects is assumed equal to the grand potential derived previously in Eq.~\eqref{grandpotentialsphere}. 
This analogy is motivated by works on two-dimensional superfluid fermionic systems \cite{babaev1999}, where the kinetic energy contribution without vortices, in the form of Eq.~\eqref{energyvorticesgeneral}, is obtained from a microscopic calculation analogous to our Bogoliubov-Popov theory. 
To analyze, as stated, the energy of the vortical part of the fluid, here we focus only on the kinetic energy associated to $\mathbf{v}_{\text{v}}(\theta,\varphi)$, which reads
\begin{equation}
E^{\text{(vor)}} = \frac{1}{2}m n_s^{(0)} \int_{0}^{2 \pi} \text{d} \varphi \, \int_{0}^{\pi} \text{d} \theta \, \sin\theta \, R^2 
\, \mathbf{v}_{\text{v}}(\theta,\varphi) \cdot \mathbf{v}_{\text{v}}(\theta,\varphi), 
\label{vortexenergy0}
\end{equation}
and, once that the velocity field is known, can be calculated analytically. 

To obtain $\mathbf{v}_{\text{v}}(\theta,\varphi)$, we consider a system of $M_{\text{v}}$ vortices with charges $q_i$, where $i=1,...,M_{\text{v}}$. 
Due to topological constraints, the net vortex charge of a spherical superfluid must be zero \cite{bogomolov1977}. 
Indeed, a path $\partial\Sigma$ on the sphere corresponds to two complementary spherical caps $\Sigma_1$ and $\Sigma_2$. 
If the path $\partial\Sigma$ is chosen in a such a way that $\Sigma_1$ does not contain vortices, and that $\Sigma_2$ contains all the vortices, one finds that $\sum_{i=1}^{M_{\text{v}}} q_i = 0$
by applying to both caps the condition of quantized circulation of Eq.~\eqref{feynmanonsager}. 
We now assume that the flow associated to the vortical velocity field $\mathbf{v}_{\text{v}}(\theta,\varphi)$ is incompressible, namely, that $\nabla \cdot \mathbf{v}_{\text{v}} = 0$. 
From this condition, it follows that
\begin{equation}
\mathbf{v}_{\text{v}}(\theta,\varphi) = 2 \pi \, \frac{\hbar}{mR} \, \mathbf{e}_{r} \times \big[\tilde{\nabla}_{R} \, \chi(\theta,\varphi)\big],
\label{velocityvorticalchi}
\end{equation}
where $\mathbf{e}_{r}$ is the unitary vector along the radial direction, and where $\chi(\theta,\varphi)$ is the stream function, which is constant along the streamlines of the fluid. 
For point vortices, the stream function can be calculated analytically, and the velocity field is therefore known. 
Indeed, introducing the vortex charge density 
\begin{equation}
n_{\text{v}}(\theta,\varphi) = \sum_{i=1}^{M_{\text{v}}} q_i \, \bigg[\delta (\cos\theta-\cos\theta_i) \, \delta(\varphi-\varphi_i)-\frac{1}{4\pi}\bigg], 
\end{equation}
where the factor $1/(4\pi)$ is introduced for regularization purposes by using the condition of charge neutrality, the stream function is determined by 
\begin{equation}
- \frac{\hat{L}^2}{\hbar^2} \chi(\theta,\varphi) = n_{\text{v}}(\theta,\varphi),
\label{greeeqsphere}
\end{equation}
where the angular momentum operator is defined as in Eq.~\eqref{angularmomentum}. The general form of the stream function reads 
\begin{equation}
\chi(\theta,\varphi) = \sum_{i=1}^{M_{\text{v}}} \chi_i(\theta,\varphi), \qquad \chi_i(\theta,\varphi) = \frac{q_i}{2\pi} \, \ln \bigg[ \sin \bigg(\frac{\gamma_i}{2}\bigg) \bigg],
\label{chisphere}
\end{equation}
where $\gamma_i$ is the angular distance between $(\theta,\varphi)$ and $(\theta_i,\varphi_i)$.  
The detailed steps to solve Eq.~\eqref{greeeqsphere}, which is essentially the Green's function of the Laplace equation in spherical coordinates, are shown in the \ref{appendixgreen}. 
We stress that, by using the bisection formula of the sine and writing explicitly the cosine of the angular distance, one finds $\sin (\gamma_i/2) = \sqrt{[1- \cos \theta \cos \theta_i - \sin \theta \sin \theta_i \cos(\varphi-\varphi_i)]/2}, $
which allows us to determine the single-vortex stream function $\chi_i(\theta,\varphi)$ once that the position of its core $(\theta_i,\varphi_i)$ is fixed. 

For a given configuration of the vortices, the velocity field $\mathbf{v}_{\text{v}}(\theta,\varphi)$ can be calculated with Eq.~\eqref{velocityvorticalchi}, and the energy of Eq.~\eqref{vortexenergy0} can be expressed as 
\begin{equation}
E^{\text{(vor)}} = \sum_{i=1}^{M_{\text{v}}} E_{i}^{\text{(vor)}} + \sum_{ \substack{i,j=1 \\ i \neq j}}^{M_{\text{v}}} E_{ij}^{\text{(vor)}},
\end{equation}
where the self-energy contribution reads 
\begin{equation}
E_{i}^{\text{(vor)}} = \frac{\hbar^2 n_s^{(0)}}{2m}  \int_0^{2\pi} \text{d}\varphi \int_\epsilon^\pi \text{d}\theta \sin \theta \, \big[2\pi \, (\tilde{\nabla}_{\text{R}} \chi_i)\big]^2,
\end{equation}
with $\epsilon$ a small angular cutoff included to regularize the self-energy, and where 
\begin{equation}
E_{ij}^{\text{(vor)}} = \frac{\hbar^2 n_s^{(0)}}{2m} \int_0^{2\pi} \text{d}\varphi \int_0^\pi \text{d}\theta \sin \theta \\ \, \big[2\pi \, (\tilde{\nabla}_{\text{R}} \chi_i)\big] \cdot \big[2\pi \, (\tilde{\nabla}_{\text{R}} \chi_j)\big]
\label{inten}
\end{equation}
is the ``interaction''-energy contribution among the vortices. 
Due to symmetry considerations, the self-energy integrals can be evaluated for $\theta_i = 0$, $\varphi_i = 0$ without loss of generality, leading to
\begin{equation}
E_{i}^{\text{(vor)}} = \frac{\hbar^2 n_s^{(0)}}{m} \, \pi q_i^2 \, \bigg{\{}  \ln \bigg[ \frac{1}{\sin (\epsilon/2)} \bigg] - \frac{1 +\cos \epsilon}{4}  \bigg{\}},
\label{e1}
\end{equation}
where we simply integrated $[2\pi \, (\partial_{\theta} \chi_i)]^2 = [(q_i/2) \cot(\theta/2)]^2$ over the spherical coordinates. 
The terms $E_{ij}^{\text{(vor)}}$ can be calculated integrating Eq.~\eqref{inten} by parts and using the property of Eq.~\eqref{poissoneqappendix} of the Green's function, obtaining 
\begin{equation}
E_{ij}^{\text{(vor)}} = - \frac{\hbar^2 n_s^{(0)}}{m} \, \pi \, q_i q_j \, \bigg{\{}  \ln \bigg[ \sin \bigg(\frac{\gamma_{ij}}{2} \bigg) \bigg] + \frac{1}{2}  \bigg{\}},
\label{eij}
\end{equation}
where
\begin{equation}
\sin \bigg(\frac{\gamma_{12}}{2} \bigg) = \sqrt{\frac{1- \cos \theta_1 \cos \theta_2 - \sin \theta_1 \sin \theta_2 \cos(\varphi_1-\varphi_2)}{2}}
\end{equation}
is the angular distance between the couple of vortices. 
Putting everything together, the general expression of $E^{\text{(vor)}}$ is given by 
\begin{equation}
\begin{split}
E^{\text{(vor)}} = \sum_{i=1}^{M_{\text{v}}} \frac{\hbar^2 n_s^{(0)}}{m} \, \pi q_i^2 \, \bigg(  \ln \bigg\{ \frac{\sin[\xi/(2R)]}{\sin (\epsilon/2)} \bigg\}  + \frac{1 -\cos \epsilon}{4}  \bigg) - \sum_{i,j=1}^{M_{\text{v}}} \frac{\hbar^2 n_s^{(0)}}{m} \, \pi \, q_i q_j \,  \ln \bigg\{ \frac{\sin (\gamma_{ij}/2)}{\sin[\xi/(2R)]}  \bigg\}, 
\label{EwithxiRgeneral}
\end{split}
\end{equation}
where we used the conditions of charge neutrality and the properties of the logarithms to include $M_{\text{v}}$ terms of the form of $\ln\{\sin[\xi/(2R)]\}$, where $\xi$ is the healing length of the superfluid. 

The kinetic energy of the vorticous superfluid depends on the cutoff $\epsilon$, which, in general, is unknown and arbitrary. 
In particular, the first line of Eq.~\eqref{EwithxiRgeneral} represents the energy necessary to create a system of $M_{\text{v}}$ vortices with charges $q_i$, with $i=1,...,M_{\text{v}}$. 
We expect that, at least for a sphere with a radius $R \gg \xi$, these self-energy terms coincide with those obtained by Kosterlitz and Thouless in the flat case \cite{kosterlitz1973}. 
Thus, denoting with $2 q^2 \mu_{\text{v}}$ the energy necessary to create a vortex-antivortex dipole with charges $\pm q$ at the minimal distance of $\epsilon \propto \xi/R$, we assume that the vortex chemical potential $\mu_{\text{v}}$ is given by 
\begin{equation}
\mu_{\text{v}} \approx \frac{\hbar^2 n_s^{(0)}}{m} \,  \pi\, \ln \bigg[ \frac{(\xi/R)}{\epsilon} \bigg] = \frac{\hbar^2 n_s^{(0)}}{m} \, \pi \, [\ln (2 \sqrt{2}) + \gamma_{\text{E}}], 
\label{muv}
\end{equation}
where the last expression coincides with the value obtained in Ref.~\cite{kosterlitz1973}\footnote{{A different choice will be made for a fermionic system in subsection \ref{sectionsound2Dfermions}, see the discussion therein.}}.
In conclusion, we obtain the kinetic energy of a spherical vorticous superfluid, namely
\begin{equation}
E^{\text{(vor)}} = \sum_{i=1}^{M_{\text{v}}} q_i^2 \mu_{\text{v}} - \pi \, \frac{\hbar^2 n_s^{(0)}}{m} \sum_{ \substack{i,j=1 \\ i \neq j}}^{M_{\text{v}}}  q_i q_j \,  \ln \bigg[ \frac{2R \, \sin (\gamma_{ij}/2)}{\xi}  \bigg] ,
\label{EwithxiR}
\end{equation}
which holds in the large-sphere regime. 
{Above, we have thus provided a detailed derivation of the vortex-antivortex interaction in which we identify the self-energy terms with a suitable value of the chemical potential $\mu_{\text{v}}$, on which the calculations of the following section rely.}
Note that the energy derived here reproduces the result reported in Ref.~\cite{vitelli2010}, and the same result was also obtained in Ref.~\cite{bereta2021}, where it was employed to analyze the dynamics of vortices in a spherical superfluid film.

\subsection{Renormalization of the superfluid density}
\label{sectionBKTsphere}
At zero temperature, in a uniform system of weakly-interacting bosons, the superfluid density coincides with the density itself. 
When the temperature is increased, however, thermal excitations appear spontaneously in the system, decreasing the portion of the fluid which displays superfluid properties. 
From a microscopic point of view, this ``normal'' fluid component that appears at finite temperature is composed by two kinds of excitations: the Bogoliubov quasiparticles, and the vortices. 
Actually, in a nonzero but low temperature regime, the production of free vortices at large distances from each other requires a large amount free energy, and is therefore highly unfavored. 
In this case, one may assume that the Bogoliubov excitations are the only quasiparticles in the system, and that the Landau superfluid density, obtained in Eq.~\eqref{barens}, is a good approximation of the real superfluid density $n_{\text{s}}$. 

Concerning its temperature dependence, the Landau superfluid density $n_{\text{s}}^{(0)}$ goes to zero smoothly at a temperature $T^*$, but, in two-dimensional systems, this simple behavior does not represent what occurs in the experiments. 
Indeed, while the free vortices have a high free-energy cost at low temperatures, at which they can only exist as vortex-antivortex dipoles, they actually unbind and exist as thermal excitations when a critical temperature is reached. 
At this ``BKT'' transition, which is named after Berezinskii, Kosterlitz and Thouless, the vortices proliferate, and the vortical velocity field disrupts any underlying superfluid flow. 

This superfluid transition has been qualitatively and quantitatively analyzed in Refs.\cite{berezinskii1973,kosterlitz1973,kosterlitz1974}, and the specific analysis for two-dimensional superfluid Helium was done by Nelson and Kosterlitz in Ref.~\cite{nelson1977}. 
In the infinite-size case, Bose-Einstein condensation cannot occur due to the Hohenberg-Mermin-Wagner theorem \cite{hohenberg1967,mermin1966}, but superfluidity, associated to quasi-long-range order and to a power-law decay of the phase correlations, does occur. 
It was shown that the bare superfluid density $n_s^{(0)}$ is renormalized to $n_s$ by the thermal excitation of vortices, and that $n_s$ jumps abruptly to zero at a temperature given by the Kosterlitz-Nelson criterion \cite{nelson1977}
\begin{equation}
\frac{n_s(T_{\text{BKT}}^{-})}{T_{\text{BKT}}} = \frac{2}{\pi} \frac{m k_{\text{B}}}{\hbar^2}, 
\end{equation}
so that the size of the jump is a universal constant which does not depend on the interatomic interactions. 
In finite-size two-dimensional superfluids, Bose-Einstein condensation takes place, and the BKT transition typically occurs as a smooth nonuniversal crossover, rather than a universal system-independent jump. 
The main underlying mechanism is however the same: the unbinding of vortex-antivortex dipoles and the proliferation of free vortices.

Let us model the BKT transition in bubble-trapped bosonic superfluids. 
In particular, to employ the analytical relations obtained before, we will limit here to the spherically-symmetric case. 
The Kosterlitz-Thouless analysis is based on the analogy between a system of electric charges in two dimensions, whose interaction scales logarithmically with their distance, and the physics of vortices in a superfluid. 
In extending this analogy to the spherical case, we consider a vortex-antivortex dipole with quantized unitary charges, whose energy, due to Eq.~\eqref{EwithxiR}, reads
\begin{equation}
\beta U_0 (\theta) = 2 \beta \mu_{\text{v}} + 2\pi K_0 \ln \bigg[ \frac{2R \, }{\xi} \sin \bigg(\frac{\theta}{2}\bigg) \bigg],
\end{equation}
where, without loss of generality, we consider $\varphi_1=\varphi_2=0$ and $\theta_1=\theta$, $\theta_2=0$. 
In the previous equation, we choose the symbol $U_0$ for the bare interaction energy between the vortices instead of $E^{\text{(vor)}}$, and we define the parameter $K_0 = \hbar^2 n_s^{(0)}/(m k_{\text{B}} T)$,
which is essentially the adimensionalized bare superfluid density. 
Due to the presence of other vortex-antivortex dipoles in the angular interval $\theta$ among the charges, the ``Coulomb'' force is screened as 
\begin{equation}
\frac{\text{d} U}{\text{d} \theta} = \frac{1}{\varepsilon(\theta)} \frac{\text{d} U_0}{\text{d} \theta},
\end{equation}
and $U$, the renormalized interaction potential of the vortices, reads 
\begin{equation}
\beta U(\theta) = \int_{\xi/R}^{\theta} \text{d}\theta' \, \frac{\pi K(\theta')}{\tan (\theta'/2)} = 2\pi  \int_{\ell(\xi/R)}^{\ell(\theta)} K(\theta') \ \text{d}[\ell(\theta')],
\label{renoU}
\end{equation}
where {we define the renormalization group scale} $\ell(\theta) = \ln[(2R/\xi)\sin(\theta/2)]$ and the renormalized parameter $K(\theta) =  \hbar^2 /(m k_B T) \, [n_s^{(0)}/\varepsilon(\theta)]$,
which is essentially the renormalized superfluid density $n_s=n_s^{(0)}/\varepsilon(\theta)$ in an adimensional form. 
Once that the relative dielectric function $\varepsilon(\theta)$ is known, it is then possible to calculate the renormalized superfluid density, and the renormalized interaction $U(\theta)$. 

Let us develop a perturbative calculation of the dielectric function $\varepsilon(\theta)$. 
We express $\varepsilon(\theta)$ in terms of the electric susceptibility of the dipoles, i.~e.~$\chi_{\text{e}}(\theta)$, as $\varepsilon(\theta) = 1 + 4\pi \chi_{\text{e}}(\theta)$, and we calculate $\chi_{\text{e}}(\theta)$ as
\begin{equation}
\chi_{\text{e}}(\theta) = \int_{\xi/R}^{\theta} \text{d}\theta' \, n_{\text{d}}(\theta') \, \alpha(\theta'), 
\end{equation}
namely, as the product between the polarizability of the superfluid medium $\alpha(\theta')$ and the density of dipoles $n_{\text{d}}(\theta')$ at a distance $\theta'$, integrated for all $\theta' < \theta$. 
In this way, the renormalized superfluid parameter $K(\theta)$ can be expressed as 
\begin{equation}
K^{-1}(\theta) = K_0^{-1} + 4\pi K_0^{-1} \int_{\xi/R}^{\theta} \text{d}\theta' \, n_{\text{d}}(\theta') \, \alpha(\theta'),
\label{Kgeneric}
\end{equation}
as follows from its definition. 
We now implement a perturbative calculation of $n_{\text{d}}(\theta')$ and of $\alpha(\theta')$. 
We express the density of dipoles as \cite{kotsubo1984,kotsubo1986}
\begin{equation}
n_{\text{d}}(\theta') =  2 \pi \,\frac{ R^2 }{\xi^4} \sin \theta'  \, y_0^2 \, e^{-\beta U(\theta')}   + o(y_0^4), 
\label{ndfinal}
\end{equation}
namely, as the product between the integral measure $2\pi \sin \theta' (R/\xi)^2$ and the dipole density along $\theta'$, which, at the lowest order in the vortex fugacity $y_0=e^{-\beta \mu_{\text{v}}}$, is proportional to the Boltzmann factor $y_0^2 \, e^{-\beta U(\theta')}$.

To calculate $\alpha(\theta')$, we consider a vortex-antivortex dipole with an angular distance $\theta'$ in an external superfluid flow with velocity $v_{\text{ext}}$. For small $\theta'$, the external flow is practically uniform, and it is the analogous of an ``electric'' field $|\mathbf{e}|=2mv_{\text{ext}}/\hbar$ that polarizes the medium between the vortex and the antivortex. 
At a finite temperature, the dipole moment of the vortex antivortex dipole is given by $\langle\mathbf{d}(\theta')  \rangle=\langle 2 R \sin(\theta'/2) \cos \varphi  \rangle$,
where we assume that the flow forms and angle $\varphi$ with the dipole moment, and where the thermal average is performed with the Boltzmann factor $\exp \{-[\beta U_0(\theta'') -\pi K_0 \, \mathbf{d}(\theta'') \cdot \mathbf{e}]\}$,
and integrating in the annulus $\theta' < \theta'' < \theta' + d\theta$.
The polarizability is then given by \cite{kosterlitz1973,kotsubo1986}
\begin{equation}
\alpha(\theta') = \frac{\partial}{\partial |\mathbf{e}|}\langle 2 R \sin(\theta'/2) \cos \varphi  \rangle \bigg|_{|\mathbf{e}|=0}, 
\end{equation}
from which we find that
\begin{equation}
\alpha (\theta') = 2\pi K_0 \, R^2 \sin^2(\theta'/2). 
\label{alphafinal}
\end{equation}

Putting Eq.~\eqref{ndfinal} and \eqref{alphafinal} inside Eq.~\eqref{Kgeneric}, we can write the inverse of the renormalized superfluid parameter as 
\begin{equation}
K^{-1}(\theta) = K_0^{-1} + 4\pi^3  \int_{\ell(\xi/R)}^{\ell(\theta)} y^2(\theta') \ \text{d}[\ell(\theta')],
\label{Kreno3}
\end{equation}
where 
\begin{equation}
y^2(\theta) = y_0^2 \, \frac{\sin^4(\theta/2)}{[\xi/(2R)]^4} \,  e^{-\beta U(\theta)} 
\label{yrenosimple}
\end{equation}
is the renormalized fugacity of a spherical superfluid. 
Considering the different equivalent forms of the renormalized interaction $U(\theta)$ of Eq.~\eqref{renoU}, the renormalized fugacity can also be written as 
\begin{equation}
y^2(\theta) = y_0^2  \, \exp \bigg\{  4 \ell(\theta)
-2\pi  \int_{\ell(\xi/R)}^{\ell(\theta)} K(\theta') \ \text{d}[\ell(\theta')]
 \bigg\},  
\label{yreno}
\end{equation}
where the dependence from the renormalization group scale $\ell(\theta)$, which contains the vortex-antivortex chordal distance, is made explicit. 
We emphasize that Eq.~\eqref{yreno} leads to an expression of $y(\theta)$ in the spherical case which is the analogous of the flat-case one obtained by Young in 1978, and avoids making further approximations \cite{young1978,kosterlitz2016}. 
Finally, we derive Eqs.~\eqref{Kreno3} and \eqref{yreno} with respect to $\ell(\theta)$, to obtain the renormalization group differential equations of a spherical superfluid, namely
\begin{equation}
\begin{aligned}
\frac{\partial K^{-1}(\theta)}{\partial \ell(\theta)} &= 4 \pi^3 y^2 (\theta),
\label{Kfinal}
\\
\frac{\partial y(\theta)}{\partial \ell(\theta)} &= [2 - \pi K(\theta)] \, y(\theta),
%\nonumber
%\label{yfinal}
\end{aligned}
\end{equation}
which allow to calculate the renormalized superfluid parameters using the bare ones as initial conditions. 
Note that this two-step renormalization group procedure yields a renormalized superfluid density which takes into account both the Bogoliubov quasiparticles and the vortices. 
The validity of this approach has been carefully analyzed in Ref.~\cite{maccari2020}, where the critical temperature obtained through the two-step procedure is found to be in good agreement with that of Quantum Monte Carlo simulations and of functional renormalization group ones. 

The numerical solution of the renormalization group equations \eqref{Kfinal} in the interval $\ell(\theta) \in [\ell(\xi/R),\ell(\pi)]$ yields the solutions $\{y(\pi),K(\pi)\}$ as a function of temperature and, therefore, allows to calculate the renormalized superfluid density $n_s = m k_{\text{B}} T K(\pi)/\hbar^2$. 
In particular, as initial condition at $\ell(\theta)=\ell(\xi/R)$, we consider the bare values of $y_0$ and of $K_0$. 
These parameters, which are calculated numerically according to their definitions, depend on the bare Landau superfluid density $n_s^{(0)}$ of Eq.~\eqref{barens}. 
In turn, the calculation of $n_s^{(0)}$ at a fixed temperature, number of atoms, and area,  requires the knowledge of the equation of state, i.~e. of the chemical potential $\mu(T,V,N)$ discussed in subsection \ref{sectionthermodynamicssphere}. 
By putting all these elements together, it is possible to study how the BKT transition occurs in bubble-trapped superfluids. 

\begin{figure}
\centering
\includegraphics[width=0.508\columnwidth]{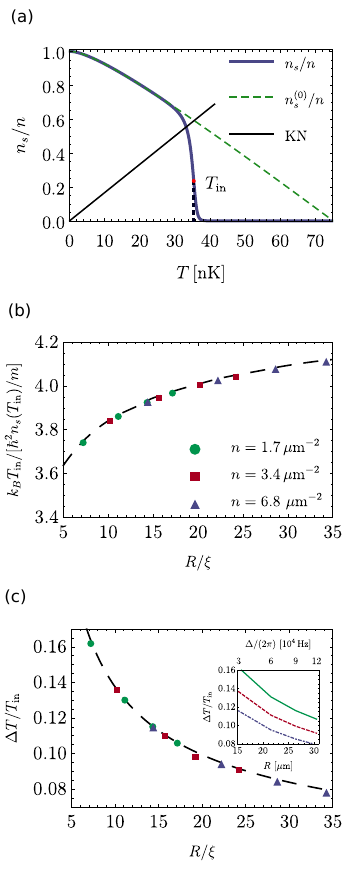}
\caption{
Superfluid BKT transition in bubble-trapped condensates. 
Panel (a) shows the superfluid fraction as a function of temperature, with $T_{\text{in}}$ being the inflection point. 
In panel (b) we plot $m k_{\text{B}} T_{\text{in}}/[\hbar^2 n_s(T_{\text{in}})]$ as a function of $R/\xi$, showing how, for different values of the two-dimensional density $n$, all the data collapse on the same line. 
Panel (c) highlights the logarithmic scaling of the transition width as $\Delta T/T_{\text{in}}= 2.1/\ln^2(5 R/\xi)$ (see Refs.~\cite{szeto1985,bramwell1994,komura2012}), which is justified by the correlation length of BKT theory \cite{kosterlitz1974}. 
This plot uses the same parameters of Fig.~\ref{figpelster1}. 
From Ref.~\cite{tononi4}.
}
\label{figpelsterfig2}
\end{figure}

Implementing this description for the typical number of atoms, temperature, and interaction regimes, we analyzed the superfluid BKT transition for experimentally relevant configurations in Ref.~\cite{tononi4}. 
Due to the finite size of the system, the renormalization group equations are solved up to the finite scale $\ell(\pi)$, which implies that the superfluid density vanishes smoothly when increasing the temperature. 
This behavior can be seen in Fig.~\ref{figpelsterfig2}(a), where we plot the superfluid fraction as a function of temperature. 

We identify the typical temperature at which the finite-size BKT transition occurs with $T_{\text{in}}$, which is the temperature of the inflection point of $n_s/n$ \cite{foster2010}. 
{Note that $T_{\text{in}}$ is very close to the critical temperature identified by the Kosterlitz-Nelson criterion, which is given by the intersection between the KN curve and the blue line in Fig.~\ref{figpelsterfig2}(a), and also corresponds to the black line in Fig.~\ref{figcondsphere-f2}. Therefore, the interplay of Bose-Einstein condensation and superfluidity in the spherical geometry, discussed at the end of Section \ref{sectioncriticalTcondfrac}, is essentially unchanged by the renormalization-group analysis, which broadens the superfluid transition into a smooth crossover.
The width of the crossover is identified} as the temperature interval corresponding to the intersection of the tangent in $n_s(T_{\text{in}})$ with the constants $n_s/n=0$, $n_s/n=1$. 
The other panels of Fig.~\ref{figpelsterfig2} show the universal scaling of finite-size BKT physics, depicting in Fig.~\ref{figpelsterfig2}(b) the values of $m k_{\text{B}} T_{\text{in}}/[\hbar^2 n_s(T_{\text{in}})]$, and showing in Fig.~\ref{figpelsterfig2}(c) how the finite width of the transition vanishes when increasing the system size. 
These universal, i.~e.~interaction independent, scalings emerge when the horizontal axis is rescaled as $R/\xi$, where the vortex core size $\xi$ is identified with the healing length of Eq.~\eqref{healinglength}. 

\subsection{Ellipsoidal shells}
\label{sectionellipsoidalshells}
We insofar reviewed the quantum statistics and the superfluid properties of bubble-trapped condensates, discussing various analytical results for a spherically-symmetric configuration. 
Relaxing this symmetry constraint, the study of these systems requires the adoption of numerical techniques, and the phenomenology becomes richer and more involved. 
In this section, we discuss prolate axially-symmetric shells using the typical experimental parameters of the microgravity experiments, illustrated in Refs.~\cite{lundblad2019,carollo2021}.

\subsubsection{Critical condensation temperature and ground state}
To calculate the critical temperature of ellipsoidal shells, we use the Hartree-Fock-Bogoliubov theory developed in Refs.~\cite{griffin1996,giorgini1997} within the Hartree-Fock approximation. 
Referencing to these works, we now review the essential details of this approach. 
The general idea is to start from the Heisenberg equation for the field operator $\hat{\psi}(\mathbf{r})$, which evolves under the action of the grand canonical Hamiltonian $\hat{H} - \mu \hat{N}$, with $\hat{H}$ given by Eq.~\eqref{Hamiltonianfieldoperator} and $\hat{N}$ by Eq.~\eqref{Nfieldoperator}. In particular, the field operator satisfies the equation
\begin{equation}
i \hbar \frac{\partial \hat{\psi}(\mathbf{r},t)}{\partial t} = \bigg[ - \frac{\hbar^2 \nabla^2}{2m} + U(\mathbf{r}) - \mu + g_0 \, \hat{\psi}^{\dagger}(\mathbf{r},t) \hat{\psi}(\mathbf{r},t) \bigg] \hat{\psi}(\mathbf{r},t), 
\label{Heisenbergequationpsi}
\end{equation}
where we limit ourselves to considering a contact interaction between bosons, i.~e.~$V(\mathbf{r}-\mathbf{r'}) = g_0 \, \delta^{(3)}(\mathbf{r}-\mathbf{r'})$. 
We then decompose the field operator as $\hat{\psi}(\mathbf{r},t) = \Phi (\mathbf{r}) + \hat{\eta}(\mathbf{r},t)$,
where $\Phi (\mathbf{r}) = \braket{\hat{\psi}(\mathbf{r})}$ is the condensate field, with $\braket{...}$ denoting the statistical average, and $\hat{\eta}(\mathbf{r},t)$ is the fluctuations operator, which has zero thermal average $\braket{\hat{\eta}(\mathbf{r},t)} = 0$. 
Inserting this decomposition into Eq.~\eqref{Heisenbergequationpsi} and calculating the average, we get \cite{griffin1996}
\begin{equation}
\bigg\{ - \frac{\hbar^2 \nabla^2}{2m} + U(\mathbf{r}) - \mu + g_0 \,[n_c(\mathbf{r}) + 2\tilde{n}(\mathbf{r})] \bigg\} \Phi(\mathbf{r}) + \tilde{m}(\mathbf{r}) \Phi^{*}(\mathbf{r}) = 0,
\label{HFBPhi}
\end{equation}
with $n_c(\mathbf{r}) = |\Phi(\mathbf{r})|^2$, and where we define the noncondensed density $\tilde{n}(\mathbf{r}) = \braket{\hat{\eta}^{\dagger}(\mathbf{r},t) \hat{\eta}(\mathbf{r},t)}$, and the anomalous density $\tilde{m}(\mathbf{r}) = \braket{\hat{\eta}(\mathbf{r},t) \hat{\eta}(\mathbf{r},t)}$. Note that, to obtain Eq.~\eqref{HFBPhi}, it is necessary to evaluate the average of cubic terms in the fluctuation field within a mean-field approximation \cite{griffin1996}.  

The equation of motion of the fluctuation operator $\hat{\eta}(\mathbf{r},t)$ is obtained subtracting Eq.~\eqref{HFBPhi} from Eq.~\eqref{Heisenbergequationpsi}, getting
\begin{equation}
i \hbar \frac{\partial \hat{\eta}(\mathbf{r},t)}{\partial t} = \bigg\{ - \frac{\hbar^2 \nabla^2}{2m} + U(\mathbf{r}) - \mu + 2 g_0 \, [n_c(\mathbf{r}) + \tilde{n}(\mathbf{r})] \bigg\} \hat{\eta}(\mathbf{r},t) + g_0 \, [n_c(\mathbf{r}) + \tilde{m}(\mathbf{r})] \hat{\eta}^{\dagger}(\mathbf{r},t),
\label{HFBeta}
\end{equation}
whose solution can be implemented with different levels of approximation, the most convenient being the Popov approximation \cite{griffin1996}, where the anomalous density contribution $\tilde{m}(\mathbf{r})$ is neglected. 
We now decompose $\hat{\eta}(\mathbf{r},t)$ as 
\begin{align}
\begin{split}
\hat{\eta}(\mathbf{r},t) = \sum_j \big[ u_j(\mathbf{r}) e^{-i E_j t/\hbar} \, \hat{a}_j + v_j^{*}(\mathbf{r}) e^{i E_j t/\hbar} \, \hat{a}_j^{\dagger}  \big],
\end{split}
\end{align}
and similarly for $\hat{\eta}^{\dagger}$, where $\hat{a}_j$, $\hat{a}_j^{\dagger}$ are operators satisfying bosonic commutation rules.
Consequently, the functions $u_j(\mathbf{r})$ and $v_j(\mathbf{r})$ satisfy the relation 
\begin{equation}
\int \text{d}\mathbf{r} \, \big[ u_j^{*}(\mathbf{r}) u_k(\mathbf{r}) - v_j^{*}(\mathbf{r})  v_k(\mathbf{r}) \big] = \delta_{jk},
\end{equation}
which ensures their orthonormalization. 
Substituting this decomposition into the previous Eq.~\eqref{HFBeta}, we get the Bogoliubov-de Gennes equations 
\begin{align}
\begin{split}
\hat{\mathfrak{L}} u_j(\mathbf{r}) - g_0 \, [n_c(\mathbf{r}) + \tilde{m}(\mathbf{r})] v_j(\mathbf{r}) &= E_j u_j(\mathbf{r}),
\\
\hat{\mathfrak{L}} v_j(\mathbf{r}) - g_0 \, [n_c(\mathbf{r}) + \tilde{m}^{*}(\mathbf{r})] u_j(\mathbf{r}) &= - E_j v_j(\mathbf{r}),
\end{split}
\end{align}
where we define the differential operator 
\begin{equation}
\hat{\mathfrak{L}} = - \frac{\hbar^2 \nabla^2}{2m} + U(\mathbf{r}) - \mu + 2 g_0 \, [n_c(\mathbf{r}) + \tilde{n}(\mathbf{r})]. 
\end{equation}
The solution of the Bogoliubov-de Gennes equations, for a given external potential $U(\mathbf{r})$, gives the full set of functions $\{ u_j(\mathbf{r}) , v_j(\mathbf{r}) \}$. The total density is then given by \cite{griffin1996}
\begin{equation}
n(\mathbf{r}) = n_{\text{c}}(\mathbf{r}) + \tilde{n}(\mathbf{r}) + \tilde{m}(\mathbf{r}), 
\end{equation}
with the noncondensed density written as 
\begin{equation}
\tilde{n}(\mathbf{r}) = \sum_j \big\{ \big[|u_j(\mathbf{r})|^2+|v_j(\mathbf{r})|^2 \big] \braket{\hat{a}_j^{\dagger}\hat{a}_j} + |v_j(\mathbf{r})|^2 \big\},
\end{equation}
and where the anomalous density reads
\begin{equation}
\tilde{m}(\mathbf{r}) = - \sum_j u_j(\mathbf{r}) v_j^{*}(\mathbf{r}) \big( 2 \braket{\hat{a}_j^{\dagger}\hat{a}_j} + 1  \big), 
\end{equation}
in which 
\begin{equation}
\braket{\hat{a}_j^{\dagger}\hat{a}_j} = \frac{1}{e^{\beta E_j}-1}
\end{equation}
is the Bose-Einstein distribution.

We now describe how to calculate the critical temperature of an ellipsoidal bubble-trapped condensate by {implementing the previous formalism with a few approximations.} 
Within the standard Popov approximation, we neglect the anomalous density $\tilde{m}(\mathbf{r})$ in the previous relations, and we solve the Bogoliubov-de Gennes equations by considering the semiclassical approximation $\nabla = i \mathbf{k}$, where $\mathbf{k}$ is a continuous wave vector for which $\sum_j \to \int \text{d}\mathbf{k}/(2\pi)^{3}$ \cite{salasnich2018}. 
{In this case, the Bogoliubov-de Gennes equations yield the spectrum \cite{giorgini1997} 
\begin{equation}
E(k,\mathbf{r}) = \bigg[ \bigg( \frac{\hbar^2 k^2}{2m} + U(\mathbf{r}) - \mu + 2 g_0 \, [n_c(\mathbf{r}) + \tilde{n}(\mathbf{r})] \bigg)^2 - g_0^2 n_c^2(\mathbf{r})  \bigg]^{1/2},
\end{equation}
and the functions:
\begin{align}
\begin{split}
u^2(k,\mathbf{r}) &= \frac{\hbar^2 k^2 / (2m) + U(\mathbf{r}) - \mu + 2 g_0 \, [n_c(\mathbf{r}) + \tilde{n}(\mathbf{r})] + E(k,\mathbf{r})}{2 E(k,\mathbf{r})},
\\
v^2(k,\mathbf{r}) &= \frac{\hbar^2 k^2 / (2m) + U(\mathbf{r}) - \mu + 2 g_0 \, [n_c(\mathbf{r}) + \tilde{n}(\mathbf{r})] - E(k,\mathbf{r})}{2 E(k,\mathbf{r})},
\\
u(k,\mathbf{r})v(k,\mathbf{r}) &= - \frac{g_0 n_c(\mathbf{r})}{2 E(k,\mathbf{r})}.
\end{split}
\end{align}
An additional approximation, valid when the semiclassical energy spectrum $E(k,\mathbf{r})$ is much larger than the chemical potential $\mu$, consists in considering the Hartree-Fock spectrum \cite{giorgini1997} 
\begin{equation}
E_{\text{HF}}(k,\mathbf{r}) =  \frac{\hbar^2 k^2}{2m} + U(\mathbf{r}) - \mu + 2 g_0 \, [n_c(\mathbf{r}) + \tilde{n}(\mathbf{r})].
\end{equation}
}
At the critical temperature of Bose-Einstein condensation, i.~e.~$T_{\text{BEC}}$, the condensate density $n_c(\mathbf{r})$ is zero and the thermal density $\tilde{n}(\mathbf{r})$ coincides with the total density $n(\mathbf{r})$. In this specific case, the density reads 
\begin{equation}
n(\mathbf{r}) = \int \frac{\text{d}\mathbf{k}}{(2\pi)^{3}} \, \frac{1}{e^{E_{\text{HF}}(k,\mathbf{r})/(k_{\text{B}} T_{\text{BEC}})}-1},
\label{nHartreeFock}
\end{equation}
where $E_{\text{HF}}(k,\mathbf{r})$ is evaluated at $n_c(\mathbf{r}) = 0$. 

\begin{figure}
\centering
\includegraphics[width=0.518\columnwidth]{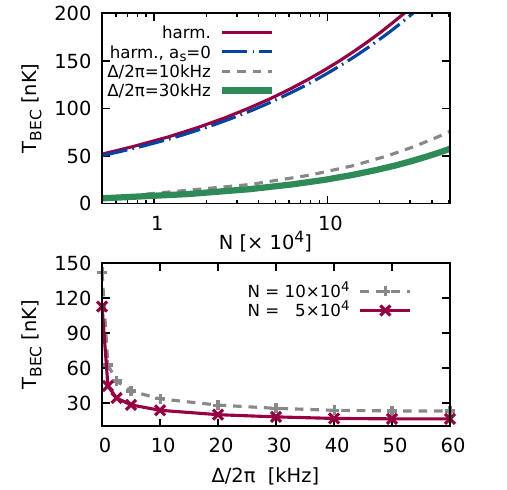}
\caption{
Critical temperature of Bose-Einstein condensation in ellipsoidal bubble-trapped condensates. 
Top: 
critical temperature as a function of the atom number $N$, obtained solving Eq.~\eqref{nHartreeFock} self-consistently. For the same number of atoms, the critical temperature is much smaller for bubble-trapped condensates than for harmonically-trapped ones. 
Bottom:
as the detuning $\Delta$ is increased, the harmonic trap (corresponding to $\Delta=0$) becomes a hollow bubble-trapped condensate, and the critical temperature drops quickly. 
From Ref.~\cite{tononi3}.
}
\label{figcinti1}
\end{figure}

In Fig.~\ref{figcinti1}, top panel, we plot the critical temperature of bubble-trapped ellipsoidal condensates as a function of the total number of particles, which is fixed by the chemical potential.
For this plot, we consider the external potential of Eq.~\eqref{bubble} where $\omega_x/(2\pi) = 30 \, \text{Hz}$, $\omega_y/(2\pi) = \omega_y/(2\pi) = 100 \, \text{Hz}$, and $\Omega_{\text{r}}/(2\pi) = 5 \, \text{kHz}$. 
We emphasize that, to consider the results of the semiclassical approximation to be valid, it is necessary to analyze temperature regimes where the thermal energy $k_{\text{B}} T$ is larger than the typical spacing between the quantum levels. 
As the bottom panel of Fig.~\ref{figcinti1} shows, the critical temperature decreases during the adiabatic expansion of a harmonic trap into a shell-shaped condensate. 
{The gas cools down during this process, but the geometry changes the phase-space density of the system in such a way that an initial condensate can either remain condensate or become a thermal cloud. 
In the recent microgravity experiments \cite{carollo2021} (see also the modeling of Ref. \cite{rhyno2021}) the expansion of a barely-condensed cloud resulted in a thermal shell-shaped gas, and reaching Bose-Einstein condensation in fully-closed shells stands as the next experimental challenge. 
This circumstance shows that the theoretical investigation of finite-temperature properties is therefore particularly relevant for the modeling of current and future experiments. }

\begin{figure}
\centering
\includegraphics[width=0.518\columnwidth]{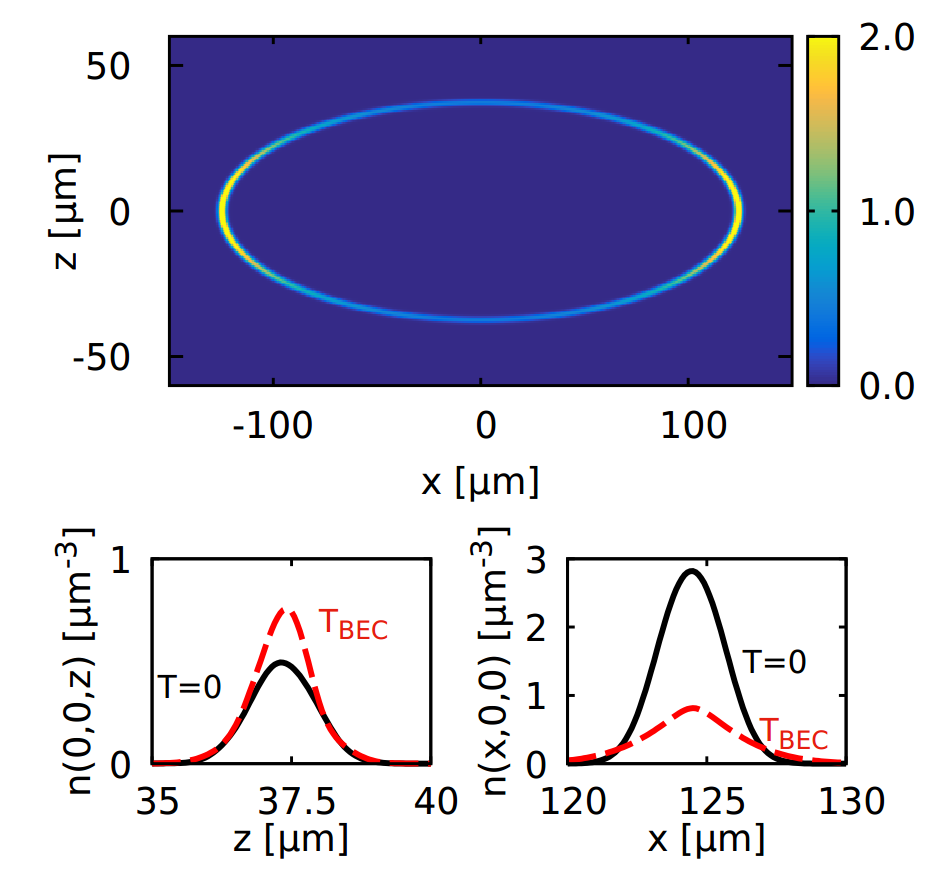}
\caption{
Density distribution of a shell-shaped condensate.
Top: 
atomic density at zero temperature, where the colorbox unit is $\mu\text{m}^{-3}$.
Bottom:
cuts of the condensate density at $T=0$, in comparison with the thermal density at $T_{\text{BEC}}$.
From Ref.~\cite{tononi3}.
}
\label{figcinti2}
\end{figure}

At zero temperature, within the degree of approximation adopted here, all the particles occupy the condensate state. In this case, the condensate density $n_c(\mathbf{r}) = |\Phi(\mathbf{r})|^2$ can be calculated by solving numerically the Gross-Pitaevskii equation \cite{gross1961,pitaevskii1961}
\begin{equation}
\mu \Phi(\mathbf{r}) = \bigg[ - \frac{\hbar^2 \nabla^2}{2m} + U(\mathbf{r})  + g_0 |\Phi(\mathbf{r})|^2  \bigg] \Phi(\mathbf{r}), 
\label{GPE}
\end{equation}
where the chemical potential $\mu$ determines the total number of particles in the system. 
In the top panel of Fig.~\ref{figcinti2} we plot the condensate density in the $xz$ plane of a system of $N=57100$ atoms confined in a trap with the same parameters of Fig.~\ref{figcinti1} and with $\Delta/(2\pi) = 30 \, \text{kHz}$. 
A relevant difference with respect to the spherical case is that the condensate density is not uniform along the shell, but the atoms tend to concentrate on the lobes. 
This distribution is the consequence of a nonuniform local harmonic trapping, since the external potential can be locally approximated as a harmonic trap with an effective trapping frequency that varies across the shell. 
Thus, in the lobes of the ellipsoid at $x \sim \pm 100 \, \mu\text{m}$, the effective frequency is lower than in the region near $x \sim 0$, and the trap accommodates less atoms in the latter region. 

In the bottom panel of Fig.~\ref{figcinti2} we compare the density along the main ellipsoidal axes at zero and at the critical temperature. 
While the zero-temperature condensate is not uniform, the thermal cloud has a constant density across the whole shell. 
In the first experiments with Bose-Einstein condensates, which were held in harmonic traps, this different density distribution of the thermal and of the condensate components was also observed \cite{anderson1995,davis1995,bradley1995}, and it was actually considered a proof of the phenomenon of condensation. 
From the opposite perspective, the uniformity of the density distribution can be used experimentally to estimate the temperature of the shell-shaped condensate. 

\subsubsection{Free expansion}
The total energy a gas of ultracold bosonic atoms can be thought as the sum of three contributions: kinetic, potential, and interaction ones. 
In single-component quantum gases, usually, the density distribution corresponds to the optimal configuration in which the attractive potential energy is balanced by the repulsive kinetic and interaction contributions. 
In the absence of any confinement, thus, the system expands freely in space due to the diffusion and to the repulsive interactions between bosons. 

Connected to these concepts, let us briefly discuss one of the most common destructive experimental techniques: absorption imaging. 
The basic idea of the technique is to suddenly turn off the external potential and, after some time of free expansion, to flash the gas with a resonant laser light. 
The projection of the atomic cloud is then recorded on a CCD camera, which yields a larger signal for the low-density regions and a lower signal for denser ones. 
A precise quantitative understanding of how a trapped quantum gas expands is thus fundamental for the comparison with the experiments. 

In this subsection, we analyze theoretically how a bubble-trapped gas freely expands and self-interferes after the sudden turning off of the trapping potential. 
For this scope, we solve numerically the time-dependent version of the Gross-Pitaevskii equation of Eq.~\eqref{GPE}, namely 
\begin{equation}
i \hbar \frac{\partial \Phi(\mathbf{r},t)}{\partial t} = \bigg[ - \frac{\hbar^2 \nabla^2}{2m} + U(\mathbf{r})  + g_0 |\Phi(\mathbf{r},t)|^2  \bigg] \Phi(\mathbf{r},t), 
\label{TDGPE}
\end{equation}
which models the dynamics of the macroscopic wave function. 
Strictly speaking, since bubble-traps confine the atoms in a dressed state that results from the superposition of different hyperfine states $\ket{F,m_F}$, it would be necessary to solve $2 m_F + 1$ coupled Gross-Pitaevskii equations. 
However, the experiments with shell-shaped gases are carried on with $^{87}$Rb atoms, in which the scattering lengths between the atoms in the different hyperfine states are almost equal, i.~e.~$\sim 100 \, a_0$, with $a_0$ the Bohr radius.
It is therefore a good approximation to solve a single Gross-Pitaevskii equation. 

\begin{figure}
\centering
\includegraphics[width=0.518\columnwidth]{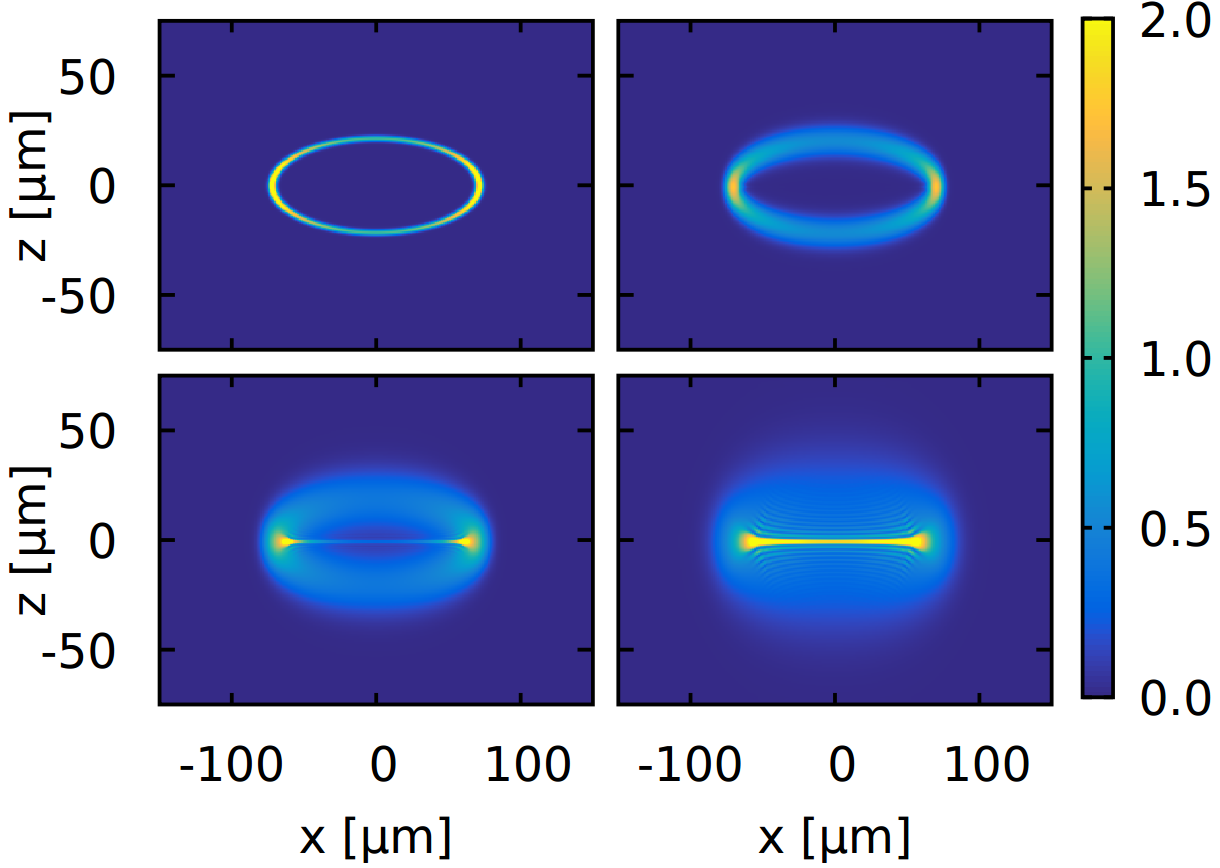}
\caption{
Contour plot of the condensate density distribution $n_0(x,0,z,t)$, cut along the $xz$ plane, of an ellipsoidal shell-shaped condensate. 
We plot the expansion at increasing times: $\{0$, $4.5$, $9$, $18\}$ $\text{ms}$.
The colorbox is in units of $\mu\text{m}^{-3}$. 
From Ref.~\cite{tononi3}.
}
\label{figcinti3}
\end{figure}

The theoretical modeling of the free expansion follows these steps: first, one solves the Gross-Pitaevskii equation of Eq.~\eqref{GPE} in imaginary time, which leads to the ground-state solution for the potential $U(\mathbf{r})$, which we have discussed in the previous subsection. 
Then, by using this ground state distribution as the initial condition, we solve numerically Eq.~\eqref{TDGPE} in which, during the dynamics, we set $U(\mathbf{r})=0$. 
In Fig.~\ref{figcinti3} we show how the condensate density $n_0(\mathbf{r},t)=|\Phi(\mathbf{r},t)|^2$ evolves during the free expansion. 
From left to right and from top to bottom, the simulation times, measured in $\text{ms}$, are given by: $0$, $4.5$, $9$, $18$. 
Due to the hollow shape of the condensate, the cloud expands both inwards and outwards, self-interfering at the center of the trap. 
This feature was qualitatively observed in 2022 in experiments conducted with dual-species repulsive Bose mixtures \cite{jia2022}.

\subsection{BCS-BEC crossover in a spherical superfluid shell}

In a recent paper \cite{chien2022} inspired by the realizations of bubble traps for ultracold atoms in microgravity, it has been discussed the theory of a two-component atomic Fermi gas on a spherical shell. In particular, it has been analyzed the crossover from the Bardeen-Cooper-Schrieffer (BCS) regime of weakly attractive fermions to the Bose-Einstein condensation of dimers. Usually, this crossover is induced by tuning the attractive interaction. Remarkably, the spherical-shell geometry introduces a different way of inducing the crossover. 
It has been shown that a curvature-induced BCS-BEC crossover can be obtained by changing the radius of the sphere at fixed particle number and interaction strength \cite{chien2022}. 
The paper has also investigated the role of the curved geometry on the superfluid fraction of the system as a function of the temperature and it calculated the BKT critical temperature above which the proliferation of quantized vortices breaks up the superfluidity \cite{chien2022}. 

A possible extension of these interesting results \cite{chien2022} could be obtained by including beyond-mean-field Gaussian fluctuations (relevant in the BEC side of the 2D crossover \cite{bighin2016}), {or by deriving the BKT critical temperature within the renormalization group theory procedure described in Section \ref{sectionBKTsphere}, thus going beyond the Nelson-Kosterliz criterion which tends to overestimate the critical temperature. 
Another possibility consists in studying an ellipsoidal shell, where differently from the spherical geometry the local spatial curvature is not constant. In this case, the BKT superfluid transition is complicated by the fact that the superfluid density is a tensorial quantity with an azimuthal component and a polar one, and that the physics of vortices can be affected both globally by the topology of the gas and locally by the geometric potential.}

\section{Hydrodynamic excitations of 2D atomic superfluids}
\label{section4}

Hydrodynamic excitations, i.~e.~long-wavelength collective modes of a quantum liquid at zero and at finite temperature, constitute one of the main probes of superfluid hydrodynamics in atomic condensates. 
{When the mean time between atomic collisions is much larger than the period of the propagating wave, the hydrodynamic excitations of the system need to be modeled through collisionless approaches \cite{ota2018,cappellaro2018}. 
Here we will focus instead on the collisional regime, in which several atomic collisions occur during a full oscillation of the wave. 
In this case, by modeling} the system as a mixture of a normal part and of a superfluid part, two branches of hydrodynamic excitations can coexist.
In flat superfluids, discussed in Sections \ref{sectionsound2Dfermi} and \ref{sectionsound2Dbosons}, these excitations propagate as sound waves, thus with a linear relation between the wave vector and the frequency. 
In spherically-symmetric superfluids, discussed in Section \ref{sectionsoundsphere}, the hydrodynamic excitations are complex surface modes that can be decomposed in the basis of the spherical harmonics. 
Before analyzing these setups in detail, let us introduce the Landau two-fluid model. 

\subsection{Landau two-fluid model} 
\label{sectiontwofluidmodel}
An interacting quantum gas in a regime of quantum degeneracy displays the property of superfluidity. This peculiar phenomenon, which consists in the capability of a quantum liquid to flow through narrow capillaries without friction, was first discovered in experiments with Helium-II by Kapitza \cite{kapitza1938}. In a seminal paper, published in 1941, Landau formulated a complete theory to describe this and other experimental properties of Helium-II \cite{landau1941}, developing the concepts previously introduced by Tisza \cite{tisza1938,tisza1938bis,tisza1938tris}. 

The behavior of a superfluid, which is intrinsically quantum mechanical, can be explained with a semiclassical two-fluid model. 
The main idea is that the fluid can be thought as composed by two parts: 
the superfluid part, which is not viscous and has zero entropy, 
and the normal part, which behaves like a viscous fluid. 
At sufficiently low temperatures, therefore, the hydrodynamic behavior of a many-body system consists in the superposition of the superfluid motion (i.~e.~a portion of the fluid in the same coherent state), and of the motion of the collective excitations, constituting the normal fluid. 
Notably, the two fluids do not exchange momentum with each other: 
there is no internal friction between the normal fluid and the superfluid 
one, neither within the superfluid itself.
It is also important to stress that the fluids are not separable, 
but must be thought as the simultaneous capability of a quantum 
system to display two different and independent motions.

To develop an hydrodynamic description, we suppose that the thermodynamic quantities fluctuate around the average value that they assume at thermodynamic equilibrium. 
Thus, we require that the system is always in local equilibrium during the dynamics, so that the thermodynamic functions acquire an additional dependence on space and on time.
Under these assumptions, let us briefly introduce and comment the equations of Landau-Tisza two-fluid model \cite{landau1941}. 
We consider a system of mass density $\rho$, such that
\begin{equation} 
\rho = \rho_n + \rho_s,
\label{density}
\end{equation}
where $\rho_s$ is the mass density of the superfluid part of the fluid and $\rho_n$ is the mass density of the normal part of the fluid. 
The total mass current results from the superposition of the normal flow and of the superflow, namely 
\begin{equation} 
\mathbf{j} = \rho_n \mathbf{v}_n + \rho_s \mathbf{v}_s,
\label{current}
\end{equation}
where $\mathbf{v}_n$ is the normal fluid velocity and $\mathbf{v}_s$ is the superfluid velocity. 
The conservation of mass is expressed by the continuity equation 
\begin{equation}
\frac{\partial \rho}{\partial t} + {\nabla} \cdot \mathbf{j} = 0, 
\end{equation}
and the conservation of momentum is described by 
\begin{equation}
\frac{\partial j_i}{\partial t} + \frac{\partial \Pi_{ik}}{\partial x_k} = 0, 
\end{equation}
with we define the tensor $\Pi_{ik} = P \, \delta_{ik} + \rho_n {v}_{n,i} {v}_{n,k} 
+ \rho_s {v}_{s,i} {v}_{s,k}$, with $P$ the pressure of the fluid. 
As a consequence of the irrotational nature of the superfluid velocity, i.~e.~$\nabla \times \mathbf{v}_s = 0$, the superfluid motion is reversible, and entropy is carried only by the normal part of the fluid. 
Therefore, the continuity equation for the entropy density $\rho \tilde{s}$ can be written as 
\begin{equation}
\frac{\partial \rho \tilde{s}}{\partial t} + \rho \tilde{s} \, {\nabla} \cdot \mathbf{v}_n = 0,
\label{entropyeq}
\end{equation}
where $\tilde{s} = S/M$ is the entropy per unit mass of fluid $M$ described in this model. 

To close the system of equations of the two-fluid model, another relation is needed. 
It is the Newton equation of the superfluid, obtained by equating the acceleration of the superfluid, i.~e.~$\text{d} \mathbf{v}_s /\text{d} t$, to the force per unit of mass necessary to accelerate it.
The latter quantity is equal to $- \nabla (\partial U / \partial M)_{S,V}$, where $U$ is the internal energy and $(\partial U / \partial M)_{S,V}$ is the potential energy per unit of mass associated to the motion of the sole superfluid. 
Now, considering the thermodynamic relation $(\partial U / \partial M)_{S,V}=(\partial G / \partial M)_{P,T}$, we conveniently discuss the derivative of $G$, the Gibbs free energy, rather than that of $U$. 
We write $G$ as $G_0 + \mathbf{P}^2/(2 M_n)$, with $G_0$ the free energy of the stationary fluid, $\mathbf{P}=M_n (\mathbf{v}_n-\mathbf{v}_s)$ the momentum of the normal fluid, and $M_n \propto M$ the normal fluid mass. Differentiating $G$ with respect to $M$, and putting all the elements together, Landau obtained the following equation \cite{landau1941}
\begin{equation} 
\frac{\partial \mathbf{v}_s}{\partial t} = - {\nabla} \bigg[ \frac{G_0}{M} + 
\frac{{\mathbf{v_s}}^2}{2} - \frac{1}{2}(\mathbf{v}_n-\mathbf{v}_s)^2 
\frac{\rho_n}{\rho_n + \rho_s} \bigg], 
\label{acceleration1}
\end{equation}
where the Lagrangian derivative has been expressed in terms of the Eulerian one as $\text{d} \mathbf{v}_s /\text{d} t = \partial \mathbf{v}_s /\partial t + \nabla (\mathrm{v}_s^2 /2)$. Note that $G_0$ satisfies the Gibbs-Duhem relation $G_0 = \mu N$, with $\mu$ the chemical potential of the stationary fluid. Thus, interpreting the mass $M$ as $M = m N$, with $m$ the mass of the identical microscopic constituents of the fluid, the term $G_0 / M$ in Eq.~\eqref{acceleration1} is equal to $\mu/m$. 

Having introduced the equations of the two-fluid model, we specify that all the thermodynamic quantities depend on space and time; for simplicity, their explicit dependence is omitted unless it becomes necessary. 

\subsection{Derivation of the sound velocities}
\label{soundflatsection}
Within the phenomenological description of the two-fluid model, a quantum liquid admits two different kinds of hydrodynamic excitations. 
If the system is uniformly confined in a $D$-dimensional box $V=L^{D}$, these excitations are labelled by a wave vector and correspond to sound waves: the \textit{first} and the \textit{second} sound. 
In this case, the velocity with which these waves propagate is simply defined as the constant of proportionality between the frequency of these modes and the wave vector itself. 

Starting from the equations of the two-fluid model, we derive here the Landau biquadratic equation for the calculation of the sound velocities. 
In particular, we will limit our derivation to the case in which the superfluid and the 
normal fluid velocities are small, thus neglecting all the $\text{o}(v^2)$ terms, and we consider the linearized equations \cite{landau1941}
\begin{align}
\frac{\partial \rho}{\partial t} + \nabla \cdot \mathbf{j} = 0, \label{eq1}\\
\frac{\partial \rho \tilde{s}}{\partial t} + \rho \tilde{s} \, \nabla \cdot \mathbf{v}_n = 0, \label{eq2}\\
\frac{\partial \mathbf{j}}{\partial t} + \nabla P = 0, \label{eq3}\\
\frac{\partial \mathbf{v}_s}{\partial t} + \nabla \bigg( \frac{G}{M} \bigg) = 0, \label{eq4}
\end{align}
where, in the last equation, we omit the subindex $0$ in the stationary Gibbs free energy. 
Deriving Eq.~(\ref{eq1}) with respect to $t$ and substituting it into Eq.~(\ref{eq3}) we get 
\begin{equation} 
\frac{\partial^2 \rho}{\partial t^2} = \nabla^2 P,
\label{eqrho}
\end{equation}
which is the familiar wave equation of a single classical fluid. Indeed, if the system would be composed only by one species, Eq.~(\ref{eqrho}) would describe a wave where pressure fluctuations induce density fluctuations and vice versa. 

However, a system composed by two fluids hosts two sounds, and this additional degree of freedom is encoded in the equation
\begin{equation} 
\frac{\partial^2 \tilde{s}}{\partial t^2} = \tilde{s}^2 \frac{\rho_s}{\rho_n} \, \nabla^2 T,
\label{eqS}
\end{equation}
which, if it were decoupled from the previous equation,
would describe the propagation of temperature fluctuations that induce entropy fluctuations. 

We now briefly discuss how to obtain Eq.~\eqref{eqS}, by reviewing the original derivation of Landau \cite{landau1941}.
We substitute $\partial \rho/\partial t$ of Eq.~(\ref{eq1}) into Eq.~(\ref{eq2}) and, employing the relations (\ref{density}) and (\ref{current}), we get 
\begin{equation}
\frac{\partial \tilde{s}}{\partial t} = 
\tilde{s} \, \frac{\rho_s}{\rho} \, {\nabla} \cdot (\mathbf{v}_s - \mathbf{v}_n).
\label{alaa}
\end{equation}
Then, differentiating with respect to time, we rewrite this differential identity as 
\begin{equation}
\frac{\partial^2 \tilde{s}}{\partial t^2} = 
\tilde{s} \, \frac{\rho_s}{\rho} \, {\nabla} \cdot \bigg[ \frac{\partial(\mathbf{v}_s - \mathbf{v}_n)}{\partial t} \bigg],
\end{equation}
where we neglect higher order terms in the velocities.
As can be seen differentiating the Gibbs-Duhem relation $G=\mu N$ and considering $M=m N$, we have $\text{d}(G/M)= m^{-1} \text{d}\mu = - \tilde{s} \, \text{d}T + \rho^{-1} \, \text{d}P$. From this relation it follows that 
\begin{equation}
{\nabla} P = \rho \tilde{s} \, {\nabla} T + \rho \, {\nabla} (G/M),
\end{equation}
and, inserting ${\nabla} (G/M)$ from Eq.~(\ref{eq4}), $\nabla P$ from Eq.~(\ref{eq3}) and using again 
Eqs.~(\ref{density}) and (\ref{current}), we obtain 
\begin{equation} 
\rho_n \, \frac{\partial (\mathbf{v}_s - \mathbf{v}_n)}{\partial t} = \rho \tilde{s} \, \nabla T.
\label{alab}
\end{equation}
Finally, the wave equation \eqref{eqS} can be obtained calculating the divergence of Eq.~(\ref{alab}) and inserting it into Eq.~(\ref{alaa}). 

The sound velocities can be calculated by expanding Eqs.~(\ref{eqrho}) 
and (\ref{eqS}) for small perturbations around the equilibrium configuration, namely 
\begin{equation}
\tilde{s} = \tilde{s}_0 + \tilde{s}', \qquad \rho = \rho_0 + \rho', \qquad P = P_0 + P', 
\qquad T = T_0 + T',
\label{elongationfields}
\end{equation}
where the primed quantities represent small fluctuations with 
respect to the uniform and constant equilibrium values and, therefore, depend on the coordinates $(\mathbf{r},t)$. 
In particular, the fluctuation fields are not independent from each other, and, knowing the equation of state, it is possible to determine two of the thermodynamic variables by fixing two of the others. For instance, we write
\begin{equation}
\rho ' = \bigg( \frac{\partial \rho}{\partial T} \bigg)_P T' +  
\bigg( \frac{\partial \rho}{\partial P} \bigg)_T P' \qquad 
\tilde{s} ' = \bigg( \frac{\partial \tilde{s}}{\partial T} \bigg)_P T' +  
\bigg( \frac{\partial \tilde{s}}{\partial P} \bigg)_T P',
\label{expansionrhos}
\end{equation}
where the derivatives are calculated using the equilibrium values. 
Employing these equations to eliminate $\rho'$ and $\tilde{s}'$ and decomposing each 
fluctuation field $P'(\mathbf{r},t)$ and $T'(\mathbf{r},t)$ in the basis of plane waves $\exp[i \omega (t - x/c)]$, 
where $c$ is the sound wave velocity and $\omega$ its frequency, the equations of sound become
\begin{equation}
\begin{cases} 
{P}'(\omega) \big[ - c^2 \big( \frac{\partial \rho}{\partial P} \big)_T +1 \big] 
+ {T}'(\omega) \big[ - c^2 \big( \frac{\partial \rho}{\partial T} \big)_P  \big] = 0, 
\\
{P}'(\omega) \big[ - c^2 \big( \frac{\partial \tilde{s}}{\partial P} \big)_T  \big] 
+ {T}'(\omega) \big[ - c^2 \big( \frac{\partial \tilde{s}}{\partial T} \big)_P + \tilde{s}^2 \frac{\rho_s}{\rho_n} \big] = 0,
\label{system}
\end{cases}
\end{equation}
where ${P}'(\omega)$ and ${T}'(\omega)$ are the amplitudes of the Fourier components of the pressure and temperature fluctuation fields. 
To find a nontrivial solution of the system, we impose that its determinant is zero, obtaining an equation for the velocity of sound, namely
\begin{align} 
\begin{split}
c^4 \bigg[\bigg( \frac{\partial \rho}{\partial P} \bigg)_T \bigg( \frac{\partial \tilde{s}}{\partial T} \bigg)_P 
- \bigg( \frac{\partial \rho}{\partial T} \bigg)_P \bigg( \frac{\partial \tilde{s}}{\partial P} \bigg)_T   \bigg] 
- c^2 \bigg[ \bigg( \frac{\partial \tilde{s}}{\partial T} \bigg)_P 
+ \tilde{s}^2 \frac{\rho_s}{\rho_n} \bigg( \frac{\partial \rho}{\partial P} \bigg)_T \bigg] + \tilde{s}^2 \frac{\rho_s}{\rho_n} = 0,
\label{almosteqsound}
\end{split}
\end{align}
which is a biquadratic equation for the velocity $c$ of the sound waves. 
To put it in a standard form we need to invert the coefficient of the $c^4$ 
term, which, using the Jacobian notation \cite{landaustat}, can be rewritten in the following equivalent ways: either as $\partial (\rho,\tilde{s})/\partial (P,T) = \tilde{c}_V/[T (\partial P/\partial \rho)_T]$,
where $\tilde{c}_{V}$ is the specific heat at constant volume per unit of mass $M$, or as $\partial (\rho,\tilde{s})/\partial (P,T) = ( \partial \tilde{s}/\partial T )_P ( \partial \rho/\partial P )_{\tilde{s}}$.
From these relations, we finally obtain the Landau equation of sound, namely
\begin{equation} 
c^4 - c^2 \bigg[ \bigg(\frac{\partial P}{\partial \rho}\bigg)_{\tilde{s}}  
+ \frac{T \tilde{s}^2 \rho_s}{\tilde{c}_V \rho_n}  \bigg] + \frac{\rho_s T \tilde{s}^2 }{\rho_n \tilde{c}_V } 
\bigg(\frac{\partial P}{\partial \rho}\bigg)_T = 0,
\label{eqsound}
\end{equation}
which relates the sound velocity $c$ to the equilibrium thermodynamics of a quantum liquid. Note that this biquadratic equation admits only two linearly independent solutions, and the degeneracy is associated to the positive or negative speed with respect to the direction of propagation. 

To simplify the notation, in the coefficients of the equation of sound we introduce the definition of the following velocities: 
\begin{equation}
v_{{A}} = \sqrt{\bigg(\frac{\partial P}{\partial \rho}\bigg)_{\tilde{s}}} , 
\quad
v_{{T}} = \sqrt{\bigg(\frac{\partial P}{\partial \rho}\bigg)_{T}} , 
\quad
v_{{L}}=\sqrt{\frac{\rho_s T \tilde{s}^2 }{\rho_n \tilde{c}_V }  } , 
\label{vTAL}
\end{equation}
which are, respectively, the adiabatic velocity $v_{A}$, the isothermal velocity $v_{T}$ and the Landau velocity $v_{L}$. Depending on the thermodynamics of the physical system considered and on the temperature regime, it may occur that the general solutions of Eq.~\eqref{eqsound} are well approximated by these expressions. Before analyzing specific systems, we write the Landau equation as
\begin{equation}
c^4 - (v_A^2 + v_L^2) \, c^2 + v_L^2 v_T^2 = 0,
\end{equation}
and we explicitly calculate its general solutions, namely
\begin{align}
\begin{split}
c_1 = \bigg[ \frac{v_{A}^2 + v_{L}^2}{2} + \sqrt{\bigg(\frac{v_{A}^2 + v_{L}^2}{2}\bigg)^2 -v_{L}^2 v_T^2} \bigg]^{1/2},
\qquad
c_2 = \bigg[ \frac{v_{A}^2 + v_{L}^2}{2} - \sqrt{\bigg(\frac{v_{A}^2 + v_{L}^2}{2}\bigg)^2 -v_{L}^2 v_T^2} \bigg]^{1/2},
\label{c1c2}
\end{split}
\end{align}
which are the velocities of the first and second sound. Note that, by definition, the first sound propagates with a higher velocity with respect to the second one. 

At the critical temperature at which the superfluid density vanishes, the Landau velocity $v_L$ and, consequently, the second sound velocity $c_2$, become zero. 
Therefore, only the first sound can propagate in this high-temperature regime, and its velocity $c_1$ coincides with the usual adiabatic sound velocity $v_A$.
Actually, $v_A$ is usually derived as the solution of Eq.~\eqref{eqrho}, which encodes the interplay of pressure and density fluctuations. 
In the standard discussion of sound waves, thus, there is not an analogous of the sound equation \eqref{eqS}, and only Eq.~\eqref{eqrho} is adopted. 
Indeed, in a classical fluid, heat propagates in a diffusive manner rather than as a wave. 
On the contrary, in a superfluid system, heat can propagate as a wave, as a consequence of the zero entropy of the superfluid component. 
In general, the pressure-density oscillations of a superfluid and entropy-temperature ones are not mutually independent: $c_1$ and $c_2$ are the result of these combined phenomena and, indeed, are obtained by putting Eqs.~\eqref{eqrho} and \eqref{eqS} together. 

In this review we will discuss the propagation of sound in different physical systems, whose thermodynamic and superfluid properties produce a different qualitative and quantitative behavior of the sound modes. 
To facilitate that analysis, it is historically and scientifically relevant to discuss the propagation of sound in superfluid $^{4}$He. 
In this system, the adiabatic compressibility $\kappa_{\tilde{s}} = \rho^{-1} (\partial \rho / \partial P)_{\tilde{s}}$ and the isothermal one $\kappa_{T} = \rho^{-1} (\partial \rho / \partial P)_T$ are approximately the same. As a consequence, also the adiabatic and the isothermal velocities, as can be seen from their definitions, are approximately equal: $v_A \approx v_T$. In this case, the velocities of the sound modes of Eq.~\eqref{c1c2} can be approximated as $c_1 \approx v_A$, $c_2 \approx v_L$,
which explains the notation of the ``Landau'' velocity, for the additional (second) sound velocity that he originally identified. 

In liquid helium, the first sound propagates with $v_A$ and is, therefore, a pure density wave in which the normal fluid and the superfluid propagate in phase. 
The second sound, instead, propagates with $v_L$ and constitutes a pure heat wave, in which the normal fluid and the superfluid oscillate with opposite phase. 
Thus, the pressure-density oscillations of $^{4}$He are essentially decoupled from the temperature-entropy ones. 
This statement can be expressed mathematically as $(\partial \tilde{s} / \partial P)_{T} = (\partial \rho / \partial T)_{P} = 0$, where the equality follows from known Maxwell relations. 
It is also easy to verify that when these partial derivatives are zero, the equations of the system \eqref{system} are decoupled and, the same expressions for $c_1$ and $c_2$ obtained for $^{4}$He can be obtained from this complementary perspective. 

\subsection{Sound propagation in 2D Fermi gases}
\label{sectionsound2Dfermi}
In this section, we discuss sound propagation in systems of uniform two-dimensional fermions. 
In particular we review the calculation of the velocities of first and second sound by implementing the Landau two-fluid model, which is a general phenomenological description of a quantum fluid, independent of the underlying quantum statistics. 
In particular, we will derive the thermodynamics of two-dimensional uniform fermions across the BCS-BEC crossover and, calculating also the superfluid density, we will obtain the sound speeds and compare them with a recent experiment \cite{bohlen2020}. 

\subsubsection{Thermodynamics along the 2D BCS-BEC crossover}
To obtain the thermodynamics of the system, we need to calculate first the grand canonical partition function $\mathcal{Z}$. 
Our starting point is Eq.~\eqref{Zfermions}, in which $\mathcal{Z}$ is expressed as a functional integral over two fields: the fermionic Grassmann field $\psi_{\sigma}(\mathbf{r},\tau)$, and the bosonic pairing field $\Delta(\mathbf{r},\tau)$. 
At a mean-field level, the gap can be assumed to be uniform and constant, namely $\Delta(\mathbf{r},\tau) = \Delta_0$,
and, substituting this expression into Eq.~\eqref{Zfermions}, we need to perform the functional integration over the fermionic field only. 
Explicitly, to diagonalize the fermionic inverse propagator, and to calculate the Gaussian integral, we first need to expand the fermionic field in the basis of plane waves, labelled by the two-dimensional wave vector $\mathbf{k} = (2\pi/L)(n_x,n_y)$, with $n_x,n_y \in \mathbb{Z}$. 
After doing so, we calculate the grand canonical partition function at the mean-field level $\mathcal{Z}_{\text{mf}}$, which also yields the mean-field grand potential \cite{bighin2016}
\begin{equation}
\Omega_{\text{mf}}=-L^2 \frac{\Delta_{0}^{2}}{g}-\sum_{\mathbf{k}}(E_{k}^{\text{sp}}-\xi_{k})-\frac{2}{\beta} \sum_{\mathbf{k}} \ln (1+e^{-\beta E_{k}^{\text{sp}}}),
\label{omegamffermions}
\end{equation}
where we define the single particle excitation spectrum $E_{k}^{\text{sp}}$ as $E_{k}^{\text{sp}} = (\xi_{k}^2+\Delta_0^2)^{1/2}$,
and where $\xi_{k} = \hbar^2k^2/(2m) - \mu$. 
Note that, to obtain the thermodynamic potential, we have also performed the sum over the fermionic Matsubara frequencies \cite{bighin2016}. 
The mean-field grand potential can be calculated analytically in the regime of zero temperature, in which the third term at the right-hand side of Eq.~\eqref{omegamffermions} is zero. 
In particular, we regularize the divergent sum through the equation \cite{salasnich2016}
\begin{equation}
- \frac{1}{g} = \frac{1}{2L^2}\sum_{\mathbf{k}}\frac{1}{\hbar^2 k^2/(2m)+\epsilon_{\text{B}}/2}, 
\end{equation}
where $\epsilon_{\text{B}}$ is the two-body binding energy, and, performing the sum as an integral, we obtain 
\begin{equation}
\frac{\Omega_{\text{mf}}(T=0)}{L^2} = - \frac{m}{4 \pi \hbar^2} \bigg[ \frac{\Delta_0^2}{2} +\mu^2 +\mu \sqrt{\mu^2+\Delta_0^2} + \Delta_0^2 \, \ln \bigg(\frac{\sqrt{\mu^2+\Delta_0^2}-\mu}{\epsilon_{\text{B}}}   \bigg) \bigg], 
\label{omegamfT0fermions}
\end{equation}
which is the mean-field zero-temperature grand potential per unit of area. 

To implement a beyond-mean-field Gaussian description, we include higher order terms in the expansion of the pairing field around the uniform configuration, by writing 
\begin{equation}
\Delta(\mathbf{r},\tau) = \Delta_0 + \eta(\mathbf{r},\tau), 
\end{equation}
where $\eta(\mathbf{r},\tau)$ represents the small fluctuation of the pairing gap with respect to its uniform configuration. 
Substituting this expansion into Eq. \eqref{Zfermions}, we neglect the terms in the field $\eta$ whose order is higher than quadratic, and again, we move to Fourier space and we integrate the fermionic field. 
While the lowest-order terms reproduce the previous results obtained with the simple mean-field approximation, the beyond-mean-field terms can be rearranged to write the partition function as 
\begin{equation}
\mathcal{Z}=\mathcal{Z}_{\text{mf}} \int \mathcal{D} [\bar{\eta},\eta] \, \text{e}^{-\frac{S_{\text{g}}(\bar{\eta},\eta)}{\hbar}},
\end{equation}
where the Gaussian action is written as
\begin{equation}
S_{g}(\eta, \bar{\eta})=\frac{1}{2} \sum_{\Omega_n} \sum_{\mathbf{q}}\begin{bmatrix}
\eta(\Omega_n,\mathbf{q}) \ \
\bar{\eta}(-\Omega_n,-\mathbf{q})
\end{bmatrix}
\mathbf{M}(\Omega_n,\mathbf{q})
\begin{bmatrix}
\eta(\Omega_n,\mathbf{q}) \\
\bar{\eta}(-\Omega_n,-\mathbf{q})
\end{bmatrix},
\end{equation}
where $\mathbf{q}$ is the bosonic wave vector, and $\Omega_n$ are the bosonic Matsubara frequencies of the pairing fluctuation field. 
The elements of the matrix $\mathbf{M}(\Omega_n,\mathbf{q})$ have involved expressions that are reported in detail in Refs.~\cite{diener2008,klimin2012}.
Here we simply show the final results for the grand potential at a beyond-mean-field level, which is given by 
\begin{equation}
\Omega(\mu, T,L^2, \Delta_0)  = 
\Omega_\text{mf} (\mu, T,L^2, \Delta_0) + \Omega_\text{g} (\mu, T,L^2, \Delta_0)
\label{grandpotentialfermions}
\end{equation}
where $\Omega_\text{mf}$ is defined in Eq.~\eqref{omegamffermions}, while
\begin{equation}
\Omega_\text{g} =\frac{1}{2 \beta} \sum_{\Omega_n} \sum_{\mathbf{q}} \ln \operatorname{det} \mathbf{M}(\Omega_n,\mathbf{q})
\end{equation}
is the beyond-mean-field contribution to the grand potential. 
Clearly, the explicit evaluation of the last term requires the determination of the spectrum of the bosonic collective excitations, $E^\text{col}_{\mathbf{q}} = \hbar \omega^\text{col}_{\mathbf{q}}$,
where the frequencies $\omega^\text{col}_{\mathbf{q}}$ correspond to the poles of the inverse of the bosonic propagator, and are determined by the equation $\operatorname{det} \mathbf{M}(\omega,\mathbf{q}) = 0$ for the variable $\omega$. 

Since we are considering a uniform system, the grand potential of Eq.~\eqref{grandpotentialfermions} is an extensive quantity and, thus, scales linearly with the area $L^2$. 
Therefore, neglecting any finite-size correction, the grand potential per unit of area reads
\begin{equation}
\begin{split}
\frac{\Omega}{L^2}  = 
\frac{\Omega_{\text{mf}}(T=0)}{L^2}
-\frac{2}{\beta} \int \frac{\text{d}\mathbf{k}}{(2\pi)^2} \ln (1+e^{-\beta E_{k}^{\text{sp}}}) 
+ \frac{1}{2 \beta} \sum_{\Omega_n} \int \frac{\text{d}\mathbf{q}}{(2\pi)^2} \ln \operatorname{det} \mathbf{M}(\Omega_n,\mathbf{q}),
\label{omegadensityfermions}
\end{split}
\end{equation}
where we have substituted the summations over the wave vectors with integrals, and where the right-hand side is a function of intensive thermodynamics quantities only: $\mu$, $T$, and $\Delta_0$. 
But to compare the theoretical results with the experiments, $\mu$ and $\Delta_0$ should be expressed as functions of controllable parameters as the number density $n$ and the temperature $T$. This is achieved by solving simultaneously the number equation
\begin{equation}
n = - \frac{1}{L^2}
\bigg( \frac{\partial \Omega}{\partial \mu}\bigg)_{T}, 
\label{numbereq}
\end{equation} 
and the gap equation 
\begin{equation}
\bigg( \frac{\partial \Omega_\text{mf}}{\partial \Delta_0} \bigg)_{\mu,T} = 0. 
\label{gapeq}
\end{equation}
At zero temperature, the gap equation yields the simple analytical result $\Delta_0 = \sqrt{\epsilon_{\text{B}}^2 + 2 \mu \epsilon_{\text{B}}}$,
which can be verified by deriving the mean-field grand potential $\Omega_\text{mf}$ of Eq.~\eqref{omegamfT0fermions}. 
Concerning the number equation, the situation is more complicated. 
In Ref.~\cite{bighinthesis} the Eqs.~\eqref{numbereq} and \eqref{gapeq} were solved at zero temperature and at a beyond-mean-field level, obtaining $\mu/\epsilon_{\text{F}}$ and $\Delta_0/\epsilon_{\text{F}}$ as a function of the crossover parameter $\ln(\epsilon_{\text{B}}/\epsilon_{\text{F}})$, where $\epsilon_{\text{F}}=\hbar^2 \pi n/m$ is the Fermi energy. 
Given these functions, the single-particle excitation spectrum $E_{k}^{\text{sp}}$ can be immediately calculated as a function of $\ln(\epsilon_{\text{B}}/\epsilon_{\text{F}})$, and also the zero-temperature spectrum of the bosonic excitations $E^\text{col}_{\mathbf{q}}$ \cite{bighinthesis}. 
This modeling provides, therefore, the knowledge of the beyond-mean-field grand potential at zero temperature as a function of the crossover parameter $\ln(\epsilon_{\text{B}}/\epsilon_{\text{F}})$. 
The extension at finite-temperature is not trivial, as it requires the evaluation and the regularization of multiple wave vector and Matsubara frequency summations for solving the gap and the number equations. 
Since the temperature dependence of $\mu$ and of $\Delta_0$ is expected to be weak, it is possible to include the finite-temperature effects only through the factor $\beta$. 
This simplified description works well and is justified in the cases in which the equation of state depends marginally on temperature. 

\begin{figure}
\centering
\includegraphics[width=0.518\columnwidth]{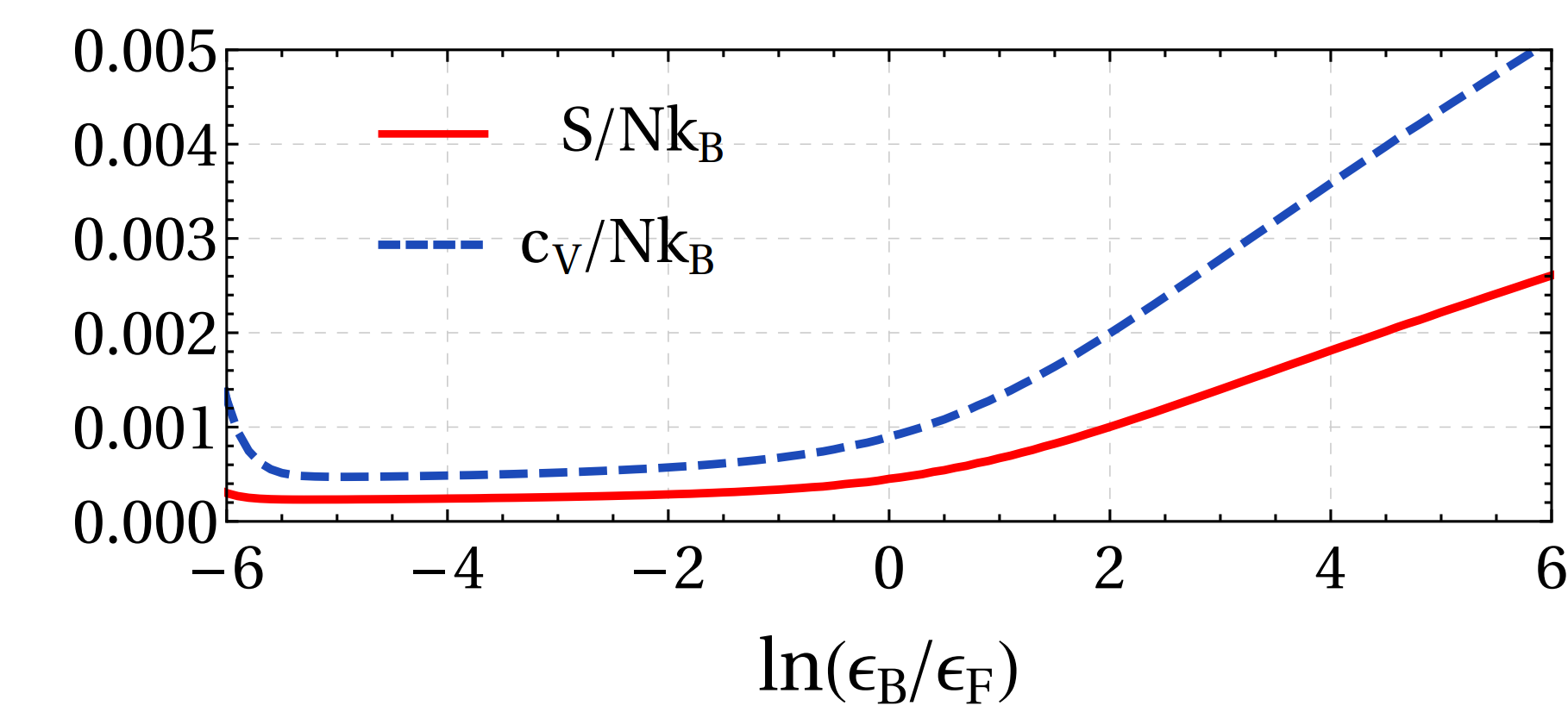}
\caption{
Dimensionless entropy, $S/(Nk_{\text{B}})$, and dimensionless specific heat at constant volume,  $c_V/(Nk_{\text{B}})$, of a two-dimensional Fermi gas  along the BCS-BEC crossover. From Ref.~\cite{tononi9}.
}
\label{fig2bcsbecsounds}
\end{figure}

The effective low-temperature free energy of the system, which includes both fermionic single-particle excitations and bosonic collective ones, reads \cite{salasnich2010}
\begin{equation}
F = \mu N + \Omega_{\text{mf}}(T=0)  -\frac{2}{\beta} \sum_{\mathbf{k}} \ln (1+e^{-\beta E_{k}^{\text{sp}}}) +\frac{1}{\beta} \sum_{\mathbf{q}} \ln (1-e^{-\beta E_{\mathbf{q}}^{\text{col}}}), 
\label{freeenergyfermions}
\end{equation}
where $\Omega_{\text{mf}}(T=0)$ is given by Eq.~\eqref{omegamfT0fermions}. 
This formulation is particularly convenient since the ratio $F/\epsilon_{\text{F}}$ depends only on the reduced temperature $T/T_{\text{F}}$, where $T_{\text{F}}=\epsilon_{\text{F}}/k_{\text{B}}$ is the Fermi temperature. 
Given $F$, one can evaluate numerically all the other thermodynamic functions. 
For instance, in Fig.~\ref{fig2bcsbecsounds}, we plot the entropy and the specific heat, which we rescale with $N k_{\text{B}}$, and which are shown as a function of the crossover parameter $\ln(\epsilon_{\text{B}}/\epsilon_{\text{F}})$. 
We also derive the pressure and the derivatives of the pressure with respect to density, both at constant temperature, and, at constant entropy, using the relation \cite{verney2015}
\begin{equation}
\bigg(\frac{\partial P}{\partial \rho}\bigg)_{S}=\bigg(\frac{\partial P}{\partial \rho}\bigg)_{T}+\frac{m N T}{\rho^{2} c_{V}}\bigg[\bigg(\frac{\partial P}{\partial T}\bigg)_{\rho}\bigg]^{2},
\end{equation}
which completes the set of thermodynamic functions needed to calculate the velocities of the first and second sound. 

\subsubsection{Sound modes and comparison with the experiments}
\label{sectionsound2Dfermions}
To calculate the sound modes, we need two different theoretical inputs: on the one hand the thermodynamic functions that we have obtained, as the entropy, the specific heat, and the derivatives of the pressure; on the other hand the superfluid density of the system. 
In a two-dimensional Fermi gas, the Berezinskii-Kosterlitz-Thouless transition drives the superfluid density to zero at the BKT temperature $T_{\text{BKT}}$, over which the proliferation of vortices is energetically convenient. 
Thus, before discussing sound propagation, we briefly model this transition, adopting the same procedure developed in bubble-trapped condensates. 
Specifically, we calculate the renormalized superfluid density by solving the renormalization group equations, whose initial conditions are expressed in terms of the bare superfluid density
\begin{equation}
n_{s}^{(0)}=n -\beta \int \frac{\mathrm{d} \mathbf{k}}{(2 \pi)^{2}} \, \frac{\hbar^{2} k^{2}}{m} \frac{e^{\beta E_{k}^{\text{sp}}}}{(e^{\beta E_{k}^{\text{sp}}}+1)^{2}} \\
-\frac{\beta}{2} \int \frac{\mathrm{d} \mathbf{q}}{(2 \pi)^{2}} \, \frac{ \hbar^{2} q^{2}}{m} \frac{e^{\beta E_{\mathbf{q}}^{\text{col}}}}{(e^{\beta E_{\mathbf{q}}^{\text{col}}}-1)^{2}},
\end{equation}
which can be derived from the momentum current of the elementary excitations, following the classic Landau argument \cite{landau1941}. 
The Kosterlitz-Nelson renormalization group equations, written in terms of the adimensional parameters $K(\ell)$ and $y(\ell)$, read 
\begin{align}
\begin{split}
\frac{\mathrm{d}K^{-1}(\ell)}{\mathrm{d} \ell} &= 4 \pi^{3} y(\ell)^{2}, \\
\frac{\mathrm{d}y(\ell)}{\mathrm{d} \ell} 
&=[2-\pi K(\ell)] \, y(\ell),
\label{RGeqsfermions}
\end{split}
\end{align}
where, in the context of two-dimensional superfluid fermions, we define $K(\ell) = \hbar^2 n_s/(4 m k_{\text{B}} T)$, with $n_s$ the renormalized superfluid density we aim to calculate, while $y(\ell)$ is the renormalized fugacity. 
{Note that, with respect to bosonic superfluids, the definition of $K$ of a superfluid of Cooper pairs contains a factor $1/4$ to account for the halved superfluid density and for the mass of the pair being double the atomic mass.}
As initial conditions of Eq.~\eqref{RGeqsfermions}, we adopt here 
\begin{equation}
K(\ell=0) = \frac{\hbar^2 n_s^{(0)}}{4 m k_{\text{B}} T}, \qquad y(\ell=0) = e^{-\beta \mu_{\text{v}}}, 
\end{equation}
and, as the chemical potential of the vortices, {we set $\mu_{\text{v}} = (\pi^2/4) \, \hbar^2 n_s^{(0)}/(4 m k_{\text{B}} T)$ \cite{bighin2017}. 
This choice needs a detailed discussion. While the value of the chemical potential does not affect the BKT transition in the infinite-size (universal) case, it can change the critical temperature and the renormalized superfluid density when stopping the renormalization group flow to a finite system size \cite{mondal2011,maccari2020}. Moreover, while the universal BKT physics is system independent, in the nonuniversal case the chemical potential depends on the specific system. While in the spherical case (see Section \ref{sectionvortices}) we assumed the value of Kosterlitz \cite{kosterlitz1973} as a valid approximation in the large-radius regime, in the case of fermions our choice follows Refs. \cite{minnhagen1985,khawaja2002,zhang2008} (see \cite{bighin2017} for a detailed discussion). The current experiments on BKT physics in quantum gases \cite{christodoulou2021} are unfortunately unable to resolve finite-size effects and cannot help us with discriminating the different choices of the chemical potential. }

\begin{figure}
\centering \includegraphics[width=0.518\columnwidth]{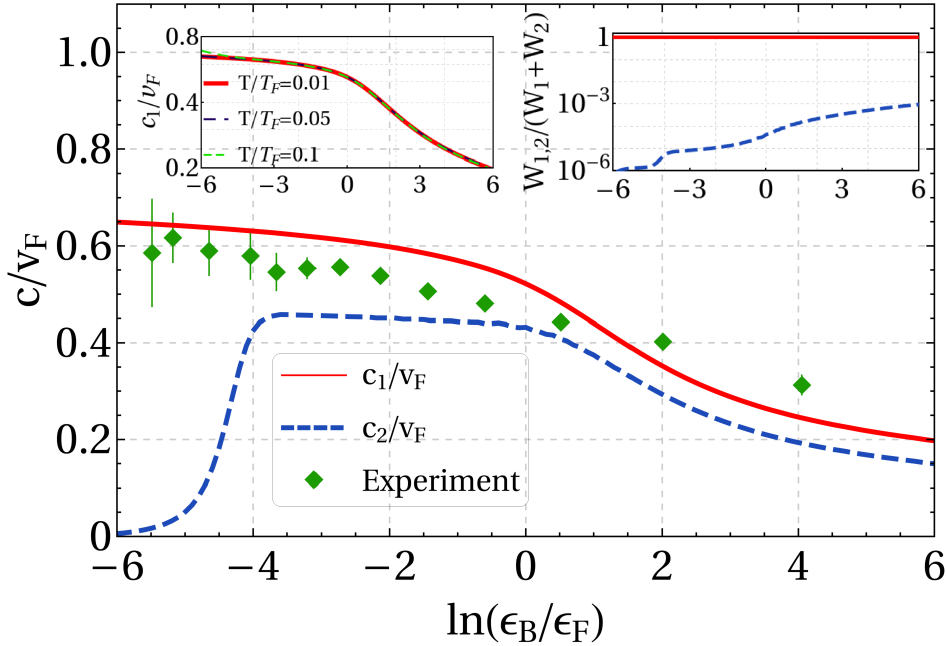}
\caption{
Velocities of the first sound $c_1$ and of the second sound $c_2$, rescaled with the Fermi velocity $v_{\text{F}}=\sqrt{2\epsilon_{\text{F}}/m}$, and plotted as a function of $\ln(\epsilon_B/\epsilon_F)$, which parametrizes the crossover from the BCS regime ($\ln(\epsilon_B/\epsilon_F)\ll 0$) to the BEC regime ($\ln(\epsilon_B/\epsilon_F)\gg 0$).
In this plot, we fix the temperature at $T/T_{\text{F}} = 0.01$ and, since the first sound has a very weak temperature dependence (see the left inset), we can compare the theoretical with the first-sound measurements of Ref.~\cite{bohlen2020}, obtained at $T/T_{\text{F}} \lesssim 0.1$. 
From Ref.~\cite{tononi9}.
} 
\label{fig1bcsbecsounds}
\end{figure}

Let us calculate and analyze the sound modes. 
First, we obtain the thermodynamics of the system, by finding numerically the necessary derivatives (with respect to temperature, volume, density) of Eq.~\eqref{freeenergyfermions}, and then, we solve Eq.~\eqref{RGeqsfermions} to get the renormalized superfluid density. 
Given these quantities as a function of the crossover parameter $\ln (\epsilon_{\text{B}}/\epsilon_{\text{F}})$, we calculate the first and second sound velocities through their general expression of Eq.~\eqref{c1c2}. 
The sound modes across the whole BCS-BEC crossover are shown in Fig.~\ref{fig1bcsbecsounds}, and are obtained at the fixed temperature of $T/T_{\text{F}} = 0.01$. 
The green diamonds represent the experimental measurements of Ref.~\cite{bohlen2020}, where a single sound wave was observed by exciting the Fermi gas with a density perturbation. 
As discussed below, these data points can be identified with the first sound.

A density perturbation $\delta \rho (\mathbf{r},t)$, induced in the experiment via a phase shift in a fermionic Josephson junction \cite{luick2020}, can excite both sound velocities, and can be expressed as \cite{arahata2009}
\begin{equation}
\delta \rho (\mathbf{r},t) = W_1 \, \delta \rho_1 (\mathbf{r} \pm c_1 t,t) + W_2 \, \delta \rho_2 (\mathbf{r} \pm c_2 t,t),
\end{equation}
where $W_1$ weights the density response  of the first sound wavepacket, $\delta \rho_1 (\mathbf{r} \pm c_1 t,t)$, and $W_2$ weights the response of the second one $\delta \rho_2 (\mathbf{r} \pm c_2 t,t)$. 
These amplitudes are actually related to the velocities themselves, since \cite{nozieres1999,ozawa2014}
\begin{equation}
\frac{W_{1}}{W_1+W_2} = \frac{(c_1^2 - v_L^2)\,c_2^2}{(c_1^2 -c_2^2)\,v_L^2},\; \qquad \frac{W_{2}}{W_1+W_2} = \frac{(v_L^2 - c_2^2)\,u_1^2}{(c_1^2 -c_2^2)\,v_L^2}
\end{equation}
are their relative weights. 
\begin{figure}
\centering \includegraphics[width=0.518\columnwidth]{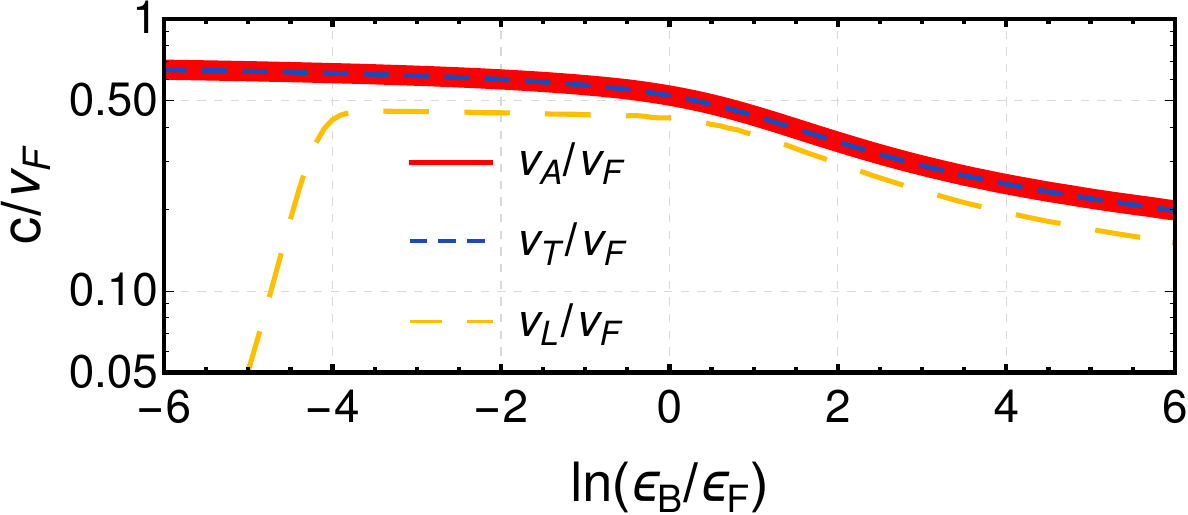}
\caption{
Adiabatic velocity $v_A$, isothermal one $v_T$, and Landau velocity $v_L$ as a function of the crossover parameter at a fixed temperature of $T/T_{\text{F}} = 0.01$. 
From Ref.~\cite{tononi9}.
} 
\label{fig2bbcsbecsounds}
\end{figure}
In the temperature and interaction regimes considered in Fig.~\ref{fig1bcsbecsounds}, we see that $c_2 \approx v_L$ (see also Fig.~\ref{fig2bbcsbecsounds}), and that, therefore, the density perturbation adopted in Ref.~\cite{bohlen2020} excites mainly the first sound. 
This consideration is quantitatively verified in the right panel of Fig.~\ref{fig1bcsbecsounds}, where we plot the relative weights of the first and second sound. 
Conversely, we expect that a heat perturbation excites mainly the second sound \cite{sidorenkov2013}, which is still unobserved in uniform 2D Fermi gases. 

\begin{figure}
\centering
\includegraphics[width=0.518\textwidth]{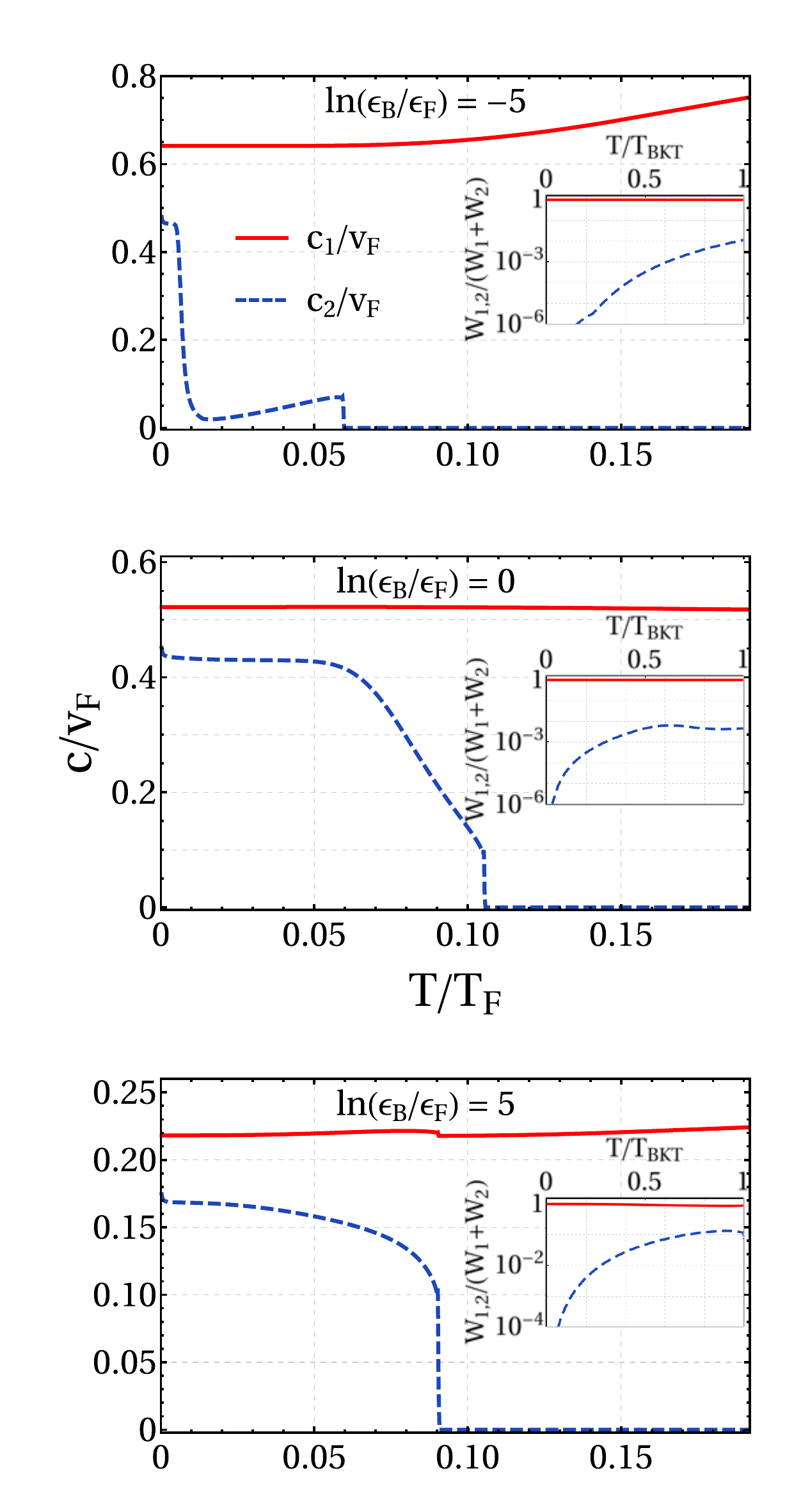}
\caption{
First and second sound velocities shown, for fixed values of the crossover parameter, as a function of the adimensional temperature $T/T_{\text{F}}$.
From Ref.~\cite{tononi9}.
}
\label{fig3bcsbecsounds}
\end{figure}

While the temperature influences weakly the behavior of the first sound, the second sound must show a discontinuity at the BKT transition temperature and across the whole BCS-BEC crossover. 
Indeed, the temperature behavior of the sound modes, analyzed in Fig.~\ref{fig3bcsbecsounds}, highlights the stronger temperature dependence of the second sound velocity with respect to the first one. 
In particular, we predict the jump of the second sound velocity to be an excellent probe of the superfluid BKT transition in uniform 2D Fermi gases. 
We also evaluate the weights $W_{1,2}$, and, in a similar way with respect to what happens at low temperatures, we find that $W_1 \gg W_2$, besides in the deep BEC regime and near $T_{\text{BKT}}$. 
As before, this observation confirms that a slight heating of the superfluid will mainly excite the second sound. 

\subsection{Sound propagation in 2D Bose gases} 
\label{sectionsound2Dbosons}
Let us analyze the propagation of sound waves in uniform two-dimensional Bose gases. 
As in the previous Section \ref{sectionsound2Dfermi}, we need on the one hand to derive the thermodynamics of the system, on the other hand to describe the superfluid BKT transition. 
This analysis will be carried on in the next subsections \ref{subsectionthermodynamics2Dbosons} and \ref{subsectionsound2Dbosons}. 

\subsubsection{Thermodynamics of box-trapped 2D bosons}
\label{subsectionthermodynamics2Dbosons}
To analyze the thermodynamics of the system, we implement a beyond-mean-field Gaussian calculation of the grand canonical partition function, see Eq.~\eqref{partfunction}. 
We suppose, in particular, that the bosons are confined in a square box of size $L$, and interact with the contact interaction $V(\mathbf{r}-\mathbf{r}') = g_0 \,  \delta^{(2)}(\mathbf{r}-\mathbf{r}')$. 

We shift the bosonic field as $\psi(\mathbf{r},\tau) = \psi_0 + \eta(\mathbf{r},\tau)$, where $\psi_0$ is the uniform mean-field configuration of Bose gas, and $\eta(\mathbf{r},\tau)$ is the complex fluctuation field around $\psi_0$. 
Substituting this decomposition into the Lagrangian of Eq.~\eqref{lagr}, and neglecting the terms of third and fourth order in $\eta$, the partition function can be calculated as a Gaussian functional integral in the fluctuation field $\eta$. 
Since we are describing a box-trapped gas, confined by repulsive walls into a square of side $L$, we impose open boundary conditions {(i.e. zero derivative at the domain boundaries) on $\eta(\mathbf{r},\tau)$, thus} expanding it in a series of cosines as 
\begin{equation}
\eta(\mathbf{r},\tau) = \frac{2}{L} \sum_{\omega_{n}} \sum_{k_{x,y} >0} \, e^{-i\omega_{n}\tau} \, \cos (k_x x) \cos (k_y y) \, \eta(\mathbf{k},\omega_{n}), 
\label{etacosines}
\end{equation}
and similarly for its complex conjugate $\bar{\eta}(\mathbf{r},\tau)$. 
Due to the boundary conditions, the two-dimensional wave vector can assume only the discrete values $\mathbf{k}=(k_x,k_y) = (\pi/L) (n_x,n_y) $, with $n_{x,y}$ positive integers. 
Expanding the fluctuation field as in Eq.~\eqref{etacosines}, the action of Eq.~\eqref{action} can be written, at a Gaussian level, as 
\begin{equation}
S=S_{0}+\frac{\beta\hbar}{2}\sum_{\omega_{n}} \sum_{k_{x,y} >0} \, 
\begin{bmatrix}
\bar{\eta}(\mathbf{k},\omega_n) & \eta(\mathbf{k},-\omega_n)
\end{bmatrix}\textbf{M}
\begin{bmatrix}
\eta(\mathbf{k},\omega_n) \\
\bar{\eta}(\mathbf{k},-\omega_n)
\end{bmatrix}
\end{equation}
where the mean-field action reads $S_0 = \beta \hbar L^{D}(-\mu \psi_0^2+g\psi_0^4/2)$, and where the matrix elements of $\textbf{M}$ are given by 
\begin{align}
\begin{split}
\textbf{M}_{jj} =& (-1)^{j} \,i\hbar\omega_{n} +\epsilon_k - \mu + 2 g_{0} \psi_{0}^2, \\
\textbf{M}_{12} =& \, \textbf{M}_{21} = g_0 \psi_{0}^2, 
\end{split}
\end{align}
with $\epsilon_k = \hbar^2 k^2/(2m)$. 
Performing the Gaussian functional integral and calculating the sum over the Matsubara frequencies, we obtain the grand canonical partition function and the grand potential $\Omega$. In particular, we get 
\begin{align}
\begin{split}
\Omega(\mu,\psi_0^2) = L^2 \big( -\mu \psi_{0}^2 +g_0 \psi_{0}^4 /2 \big) 
+  \frac{1}{2} \sum_{k_{x,y} >0} \, 
[E_{k}(\mu,\psi_0^2)-\epsilon_k - \mu] 
+ \frac{1}{\beta}  \sum_{k_{x,y} >0} \, \ln\big[1-e^{-\beta E_{k}(\mu,\psi_0^2)}\big],
\end{split}
\end{align}
where the counterterm $-\epsilon_k-\mu$ in the zero point energy appears with the convergence factor regularization \cite{salasnich2016}, and where we define the quasiparticle energies as 
\begin{equation}
E_k(\mu,\psi_0^2) = \sqrt{(\epsilon_k - \mu + 2 g_0 \psi_0^2)^2-(g_0 \psi_0^2)^2}. 
\end{equation}
As in subsection \ref{subsectionfunctionalintegralsphere}, we impose the saddle-point condition, i.~e.~$\partial\Omega/\partial\psi_0 = 0$, which determines the condensate density $\psi_0^2 = n_0(\mu)$, and substituting it into the grand potential, we write 
\begin{equation}
\Omega[\mu,n_0(\mu)]  = - L^2 \, \frac{\mu^2}{2 g_0} + \frac{1}{2} \sum_{k_{x,y} >0} \, (E_{k}^{\text{B}} -\epsilon_k - \mu) + \frac{1}{\beta}  
\sum_{k_{x,y} >0} \, \ln(1-e^{-\beta E_{k}^{\text{B}}}), 
\label{grandpotential2Dflat}
\end{equation}
where we obtain perturbatively the Bogoliubov-Popov excitation spectrum 
\begin{equation}
E_{k}^{\text{B}}=\sqrt[]{\epsilon_k 
(\epsilon_k +  2 \mu ) },
\label{Bogoliubovspectrum2Dflat}
\end{equation}
and where the Gaussian corrections in $n_0(\mu)$ are treated as perturbations with respect to the mean-field term. 

The zero-point energy in the grand potential of Eq.~\eqref{grandpotential2Dflat} is ultraviolet divergent, and needs to be regularized. 
We follow the same procedure implemented in subsection \ref{sectionscatteringth}, by including a {high-}momentum cutoff $\Lambda$ in the {first} sum of Eq.~\eqref{grandpotential2Dflat}, which we calculate as an integral over wave vectors {and expand up to logarithmic terms in $\Lambda$}. 
Then, we express the contact interaction strength as a function of $\Lambda$ and of the two-dimensional $s$-wave scattering length $a_{\text{2D}}$ as \cite{mora2003}
\begin{equation}
g_0 = - \frac{2 \pi \hbar^2 }{m} \frac{1}{\ln\big(a_{\text{2D}} \, \Lambda  \, e^{\gamma}/2\big)}, 
\label{g0}
\end{equation}
{and note that the dependence of the grand potential on the ultraviolet cutoff $\Lambda$ cancels.}
As a final result, we obtain 
\begin{equation}
\frac{\Omega}{L^2} =-\frac{m\mu^2}{8\pi\hbar^2} \ln \bigg[ \frac{4\hbar^2}{m \mu \, a_{\text{2D}}^2 e^{2\gamma+1/2}}  \bigg] 
+ \frac{1}{\beta}  
\int \frac{\text{d}\mathbf{k}}{(2\pi)^2} \, \ln(1-e^{-\beta E_{k}^{\text{B}}}), 
\label{grandpotential2dflat}
\end{equation}
where we have neglected the finite-size corrections which appear due to the lower bound of the integrals over the wave vectors. 
These finite-size effects in $\Omega$ are actually important for small condensates in weakly-interacting regimes, but are not quantitatively relevant for the following analysis. 

We emphasize that the grand potential obtained in Eq.~\eqref{grandpotential2dflat} is a function of $\mu$, $T$, and $L^2$, and that the description of a system with a fixed number of particles is more convenient in terms of the Helmholtz free energy $F$. 
The latter quantity is obtained from $\Omega$ with a Legendre transformation, i.~e.~$F=\mu N + \Omega$, and requires the determination of the function $\mu = \mu(T,L^2,N)$. 
In Ref.~\cite{tononi5}, we have actually calculated $\mu(T,L^2,N)$ numerically to describe precisely the superfluid properties of the system, while the thermodynamics is obtained from an effective simplified calculation of the free energy $F$. 
Thus, to obtain analytical results in the low-temperature regime, we have considered the following expression of the free energy  
\begin{equation}
\frac{F}{L^2} = \frac{1}{2} \, g_0 n^2 -\frac{m\mu^2}{8\pi\hbar^2} \ln \bigg[ \frac{4\hbar^2}{m g_0 n \, a_{\text{2D}}^2 e^{2\gamma+1/2}}  \bigg] 
+ \frac{1}{\beta}  
\int \frac{\text{d}\mathbf{k}}{(2\pi)^2} \, \ln(1-e^{-\beta E_{k}}),
\label{freeenergy2Dflat}
\end{equation}
where $E_k = \sqrt{\epsilon_k(\epsilon_k+2g_0 n)}$, and which can be obtained from the microscopic expression of the grand potential by setting $\mu=g_0 n$. 
For a fixed area $L^2$, for a fixed number of particles in the system $N$, and of the scattering length $a_{\text{2D}}$, the free energy is a known function of the temperature. 
Given $F$, implementing basic thermodynamic relations, see for instance the end of Section \ref{sectionBEC}, one can calculate numerically the wave vector integrals and obtain all the inputs for the determination of the sound velocities of Eq.~\eqref{c1c2}. 

\subsubsection{Sound modes and comparison with the experiments}
\label{subsectionsound2Dbosons}
In parallel with the analysis implemented in the fermionic case, we briefly describe the calculation of the renormalized superfluid density of a box-trapped bosonic gas, and we then obtain the sound velocities. 

Implementing in two dimensions a standard result of Landau \cite{landau1941}, we calculate the bare superfluid density as 
\begin{equation}
n_s^{(0)} = n 
-\beta \int \frac{\mathrm{d} \mathbf{k}}{(2 \pi)^{2}} \, \frac{ \hbar^{2} k^{2}}{2m} \frac{e^{\beta E_{k}^{\text{B}}}}{(e^{\beta E_{k}^{\text{B}}}-1)^{2}},
\label{barens2dflat}
\end{equation}
where the Bogoliubov spectrum is given by Eq.~\eqref{Bogoliubovspectrum2Dflat}.
We then solve the Kosterlitz-Nelson renormalization group equations, which are formally the same of those at Eq.~\eqref{RGeqsfermions}, in the interval $[0,\ell_{\text{max}}]$ of the renormalization group scale $\ell$. 
As the initial conditions at the scale $\ell=0$, we consider the bare parameters $K_0 = K(\ell=0) = \hbar^2 n_s^{(0)}/(m k_{\text{B}} T)$, and $y_0=y(\ell=0) = \exp(-\beta \mu_{\text{v}})$, with the chemical potential of the vortices given by $\beta \mu_{\text{v}} = (\pi^2/2) K_0$ \cite{nagaosa1999}. 
The cutoff $\ell_{\text{max}}$, at which we interrupt the flow of the renormalization group equations, is given by $\ell_{\text{max}} = \ln (L^2/\xi^2)$, with $\xi=\sqrt{\hbar^2/(m g_0 n)}$ the healing length of the superfluid. 
The renormalized superfluid density $n_s$ is then calculated as $n_s= (m k_{\text{B}} T/\hbar^2) K(\ell_{\text{max}})$, since $K(\ell) = \hbar^2 n_s(\ell)/(m k_{\text{B}} T)$. 

\begin{figure}
\centering
\includegraphics[width=0.518\columnwidth]{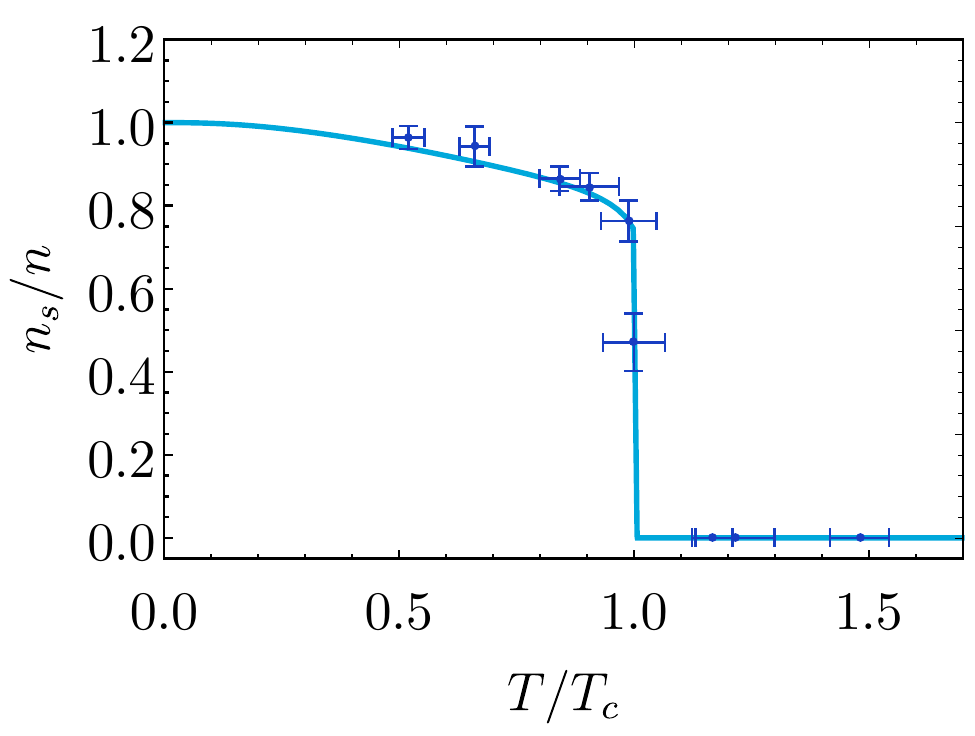}
\caption{
Superfluid density of a box-trapped two-dimensional Bose gas, obtained by solving the renormalization group Eqs.~\eqref{RGeqsfermions}. For their solution, we use {as} input the bare superfluid density of Eq.~\eqref{barens2dflat}, and consider the following parameters of the experiment in Ref.~\cite{christodoulou2021}: density $n=3 \, \mu\text{m}^{-2}$, area $L^2 \approx 33 \times 22 \, \mu\text{m}^{2}$, scattering length $a_{\text{3D}}=522 a_0$, with $a_0$ the Bohr radius, frequency of the transverse confinement $\omega_{\perp} = 2\pi \times 5500 \, \text{Hz}$. Here $T_c$ denotes the Berezinskii-Kosterlitz-Thouless critical temperature.
}
\label{fignsovernflat}
\end{figure}

In Fig.~\ref{fignsovernflat}, we compare the superfluid fraction $n_s/n$ with the experimental data of Refs.~\cite{christodoulou2021,stringari2021}. 
The theory analyzed here, which is crucially based on the calculation of the Bogoliubov-Popov excitation spectrum of Eq.~\eqref{Bogoliubovspectrum2Dflat}, agrees with all the experimental data points. 
This suggests that, concerning at least the determination of the superfluid properties, the Bogoliubov-Popov scheme provides a reliable description of the superfluid density up to the critical temperature of the BKT transition. 
Note that, to implement the calculation of the bare superfluid density, we have determined the equation of state $\mu(T,L^2,N)$ numerically by deriving the grand potential of Eq.~\eqref{grandpotential2dflat}. 
In particular, we model the two-dimensional scattering length as $a_{\text{2D}} = 2.092 \, l_{\perp} \ln[-\sqrt{\pi/2}\, (l_{\perp}/a_{\text{3D}})]$ \cite{dalibard2016}, with $a_{\text{3D}}$ the three-dimensional $s$-wave scattering length, and $l_{\perp} = \sqrt{\hbar/(m \omega_{\perp})}$ the characteristic length of the transverse harmonic confinement of frequency $\omega_{\perp}$. 

\begin{figure}
\centering
\includegraphics[width=0.518\columnwidth]{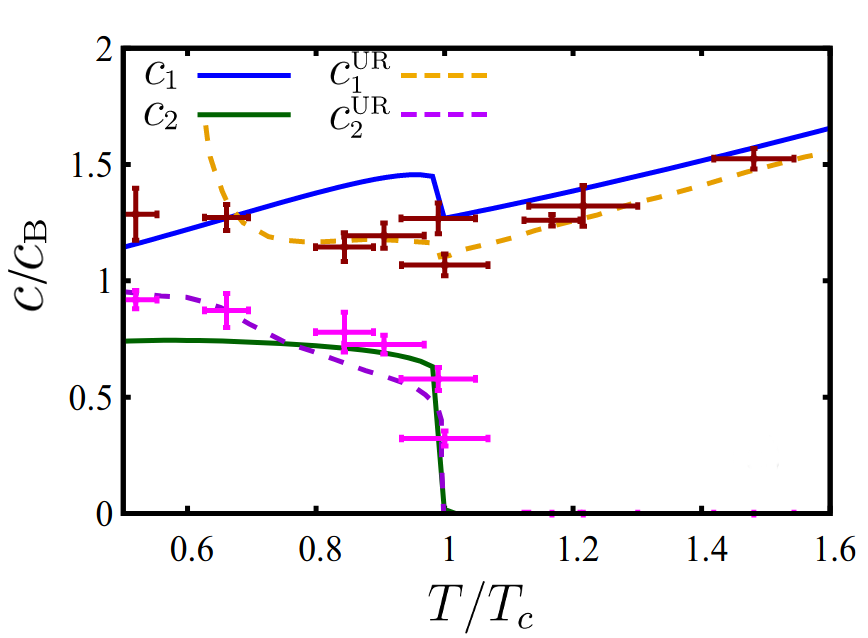}
\caption{
Sound velocities of box-trapped uniform bosons across the Berezinskii-Kosterlitz-Thouless transition. 
The theory reviewed here, plotted as continuous lines, has a good agreement with the experimental data. The dashed lines represent the sound velocities obtained from the scale-invariant theory of Ref.~\cite{prokofev2002}, which works around the superfluid transition. From Ref.~\cite{tononi5}. 
}
\label{figsoundbosons}
\end{figure}

Let us now compare the sound velocities, obtained from the solutions of Eq.~\eqref{c1c2} of the Landau biquadratic equation, with the experimental results of Ref.~\cite{christodoulou2021}. 
As can be seen from Fig.~\ref{figsoundbosons}, there is a good qualitative agreement between the theory and the experimental results. 
To assess the origin of the quantitative differences, we first remark that this theoretical scheme yields a solid estimate of the superfluid fraction, which is obtained in Ref.~\cite{stringari2021} from the measurements of the sound velocities and from the thermodynamics based on the theory of Ref.~\cite{prokofev2002}. 
This agreement suggests that, within the experimental error, the superfluid density is not strongly sensitive to the specific finite-temperature behavior of the equation of state. 
Thus, we can mainly attribute the quantitative discrepancies to a simplified dependence on the temperature, density, and interactions of the thermodynamic functions. 
In the transition region, indeed, the sound modes are very sensitive to the values of the grand potential and of its derivatives, and their precise determination, beyond the effective calculation based on Eq.~\eqref{freeenergy2Dflat}, can lead to a better quantitative agreement. 

\subsection{Hydrodynamic excitations in 2D superfluid shells} 
\label{sectionsoundsphere}
We now discuss how the two-fluid model derived in Section \ref{sectiontwofluidmodel} can be extended to describe the geometry of a two-dimensional spherical superfluid. 
It turns out that most of the previous equations can be rederived in a straightforward manner, and their analysis will elucidate how the hydrodynamic modes propagate in a spherically symmetric superfluid. 
The linearized Eqs.~\eqref{eq1}-\eqref{eq4} of a flat two-fluid system can be written in the spherical case as 
\begin{align}
\frac{\partial \rho}{\partial t} + \nabla_{R} \cdot \mathbf{j} &= 0, \label{eq1sphere}\\
\frac{\partial \rho \tilde{s}}{\partial t} + \rho \tilde{s} \, \nabla_{R} \cdot \mathbf{v}_n &= 0, \label{eq2sphere}\\
\frac{\partial \mathbf{j}}{\partial t} + \nabla_{R} P &= 0, \label{eq3sphere}\\
\frac{\partial \mathbf{v}_s}{\partial t} + \nabla_{R} \bigg( \frac{G_0}{M} \bigg) &= 0, \label{eq4sphere}
\end{align}
where we define $\tilde{\nabla}_{R} = R^{-1}[\mathbf{e}_{\theta} \, \partial_{\theta} +  (\sin\theta)^{-1}\, \mathbf{e}_{\varphi} \,\partial_{\varphi}]$
as the gradient in spherical coordinates for a system with a radius $R$ fixed. 
With this expression of $\nabla_{R}$, we are supposing that the vectorial quantities of the model do not have a radial component, so that the dynamics is constrained to occur only along the surface of the sphere. 

Starting from Eqs.~\eqref{eq1sphere}-\eqref{eq4sphere}, we perform the same calculations that we have implemented in the flat case, see in particular Eqs.~\eqref{eqrho}-\eqref{alab}. In this context, we obtain the following ``wave'' equations 
\begin{align} 
\frac{\partial^2 \rho}{\partial t^2} &= -\frac{\hat{L}^2}{\hbar^2 R^2} P,
\label{eqrhosphere}
\\ 
\frac{\partial^2 \tilde{s}}{\partial t^2} &= - \tilde{s}^2 \frac{n_s}{n_n} \, \frac{\hat{L}^2}{\hbar^2 R^2} T,
\label{eqSsphere}
\end{align} 
where the Laplacian is simply substituted by the angular momentum operator in spherical coordinates, i.~e.~$\hat{L}^2$. 
As in subsection \ref{soundflatsection}, it is possible to expand the thermodynamic functions around their equilibrium configuration and to express the density and entropy fluctuations in terms of temperature and pressure ones [see Eqs.~\eqref{elongationfields}, \eqref{expansionrhos}].
However, in the spherical geometry, the standard sound waves are not the correct basis to analyze Eqs.~\eqref{eqrhosphere} and \eqref{eqSsphere}, which are instead solved in the basis of spherical harmonics. 
Therefore, we expand the fluctuation fields as 
\begin{align}
P'(t,\theta, \varphi) =  \int_{-\infty}^{+\infty} \text{d} \omega \sum_{l=1}^{\infty} \sum_{m_l}^l \, e^{i \omega t} \, \mathcal{Y}_l^{m_l} (\theta, \varphi)\,  P(\omega,l,m_l), 
\\
T'(t,\theta, \varphi) =  \int_{-\infty}^{+\infty} \text{d} \omega \sum_{l=1}^{\infty} \sum_{m_l}^l \, e^{i \omega t} \, \mathcal{Y}_l^{m_l} (\theta, \varphi)\,  T(\omega,l,m_l), 
\end{align}
where $\omega$ is the frequency of these hydrodynamic low-energy modes and $l$, $m_l$ are their quantum numbers. 
With the same previous steps, we obtain 
\begin{equation} 
\omega^4 - \omega^2 \bigg[ \bigg(\frac{\partial P}{\partial \rho}\bigg)_{\tilde{s}}  
+ \frac{T \tilde{s}^2 \rho_s}{\tilde{c}_V \rho_n}  \bigg] \bigg[\frac{l(l+1)}{R^2} \bigg] + \frac{\rho_s T \tilde{s}^2 }{\rho_n \tilde{c}_V } 
\bigg(\frac{\partial P}{\partial \rho}\bigg)_T \bigg[\frac{l(l+1)}{R^2} \bigg]^2 = 0,
\label{eqsoundsphere}
\end{equation}
which is the Landau biquadratic equation extended to a spherical superfluid. 
As this equation shows, there is not a linear relation between the frequency of the wave and the quantum number $l$ which labels the different excitations. 
Therefore, we cannot actually define and calculate a sound speed, but only the frequencies $\omega$ of the hydrodynamic excitations for a given quantum number $l$. 
We also stress that the flat-case velocity is obtained only in the limit of a very large radius $R$, in which $l(l+1)/R^2 \approx l^2/R^2$ and a sound velocity can be approximately defined as $c \approx R\omega/l$. 

\begin{figure}
\centering
\includegraphics[width=0.518\columnwidth]{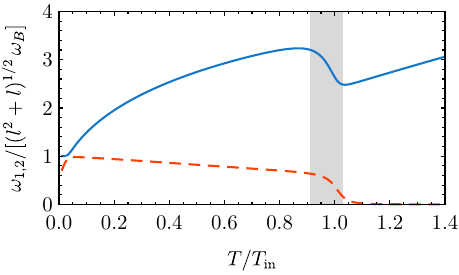}
\caption{
Frequencies of the hydrodynamic excitations rescaled with the Bogoliubov frequency $\omega_B=\sqrt{\mu/(mR^2)}$, and where $l$ is the main quantum number of the angular momentum. 
The gray region, where the hydrodynamic modes are non monotonic, is where the BKT transition occurs. 
For this figure, we use the same parameters adopted to calculate the thermodynamic functions and the renormalized superfluid density in Fig.~\ref{figpelsterfig2}. 
From Ref.~\cite{tononi4}.
}
\label{figpelster3p1}
\end{figure}

The previous considerations do not prevent to express, at least formally, the frequencies of the hydrodynamic modes in terms of the adiabatic, isothermal and Landau velocities defined in Eq.~\eqref{vTAL}. 
In this way, we write the solutions of Eq.~\eqref{eqsoundsphere} as 
\begin{equation}
\omega_{1,2}^2 = \bigg[ \frac{l(l+1)}{R^2} \bigg] \bigg[ \frac{v_{A}^2 + v_{L}^2}{2} \pm \sqrt{\bigg(\frac{v_{A}^2 + v_{L}^2}
	{2}\bigg)^2 -v_{L}^2 v_T^2} \bigg], 
\label{surfaceomegas}
\end{equation}
which are the frequencies of the ``first'' and ``second'' hydrodynamic excitations of a spherical superfluid. 
We plot the hydrodynamic frequencies of a spherical bubble-trapped gas in Fig.~\ref{figpelster3p1}, showing their behavior as a function of temperature at a fixed number of particles and for fixed interactions and trap parameters. 
To calculate $\omega_{1,2}$, we have adopted the thermodynamic description and BKT analysis for shell-shaped condensates developed in Section \ref{chapter1}. 
It is important to highlight that measuring $\omega_{1,2}$, and in particular their discontinuity as a function of temperature, would prove that the BKT transition occurs also in bubble-trapped superfluids. 
Indeed, considering that other curved and compact surfaces, as cylinders and large tori, do not display the renormalization of the superfluid density associated to BKT physics \cite{machta1989}, the experimental investigation of this phenomenon is of fundamental importance. 

\section{Conclusions and outlook}
The study of ultracold atomic gases in reduced dimensionalities and, particularly, the analysis of novel geometries, has often led to new discoveries and applications. 
In this review, we have discussed theoretically the quantum statistics and the superfluidity of two-dimensional bubble-trapped condensates in spherical and in ellipsoidal configurations. 

Shell-shaped condensates are an experimentally-relevant prototype of a curved quantum gas, and display rich equilibrium and nonequilibrium properties. 
Throughout their study, we have analyzed how finite size and curvature influence the transition of Bose-Einstein condensation, and we have derived a renormalization group description of the Berezinskii-Kosterlitz-Thouless transition in the spherical case. 
In this geometry, the results on the interplay of Bose-Einstein condensation and superfluidity and the study of finite-temperature hydrodynamic modes, can offer a theoretical insight to benchmark future and ongoing experiments. 
Most of the results reviewed here are tailored on the specific parameters of microgravity experiments \cite{carollo2021}, but they can also be extended to model other experimental realizations of two-dimensional shell-shaped gases that may be developed in the future (for instance based on the setup of Ref.~\cite{jia2022}).  

We have also reviewed the sound propagation in two-dimensional uniform superfluids. 
To consider the Landau two-fluid model as a valid description of a weakly-interacting gas, it is necessary to detect experimentally both sound modes, which should agree with the predictions of the model. 
This proof has been recently obtained in two-dimensional box-trapped bosons \cite{christodoulou2021}, but, in two-dimensional uniform fermions, the second sound velocity and its vanishing at the BKT transition is still undetected. 
This evidence, for a textbook configuration as that of a quantum gas in a box, would demonstrate the validity of the two-fluid picture in fermions across the whole BCS-BEC crossover.
We also hope to trigger future experimental investigations for observing the hybridization of first and second sound in box-trapped 2D bosons. 
In particular, we discussed the theoretical predictions for the hybridization temperature in two-dimensional weakly-interacting bosons, and for the behavior of the sound modes at low temperatures. 

In the last 20 years, the field of ultracold quantum gases has been stimulated by the experimental implementation of long-range interatomic interactions \cite{lahaye2009,chomaz2023} and by the engineering of artificial gauge fields \cite{dalibard2011}. 
Most of the results reviewed here, both on shell-shaped atomic gases and on flat configurations, are actually obtained for zero-range interatomic interactions. 
Including long-range interactions or interatomic scattering in higher partial waves in the analysis of the quantum statistical properties could lead to a richer phenomenology. 

The research field of low-dimensional quantum gases in curved geometries is not only limited to shell-shaped configurations, but includes also other two-dimensional geometries and, notably, ring-shaped superfluids. This active field of research led to many interesting findings. 
Among them, for instance, a few non-exhaustive examples involve studying the analogy between the phononic excitations in expanding rings and cosmological phenomena \cite{eckel2018}, the dynamical formation of a ring superfluid on a rapidly-rotating curved surface \cite{guo2020}, the role of the locally-varying curvature for Bose-Einstein condensates confined in an elliptical waveguide \cite{salasnich2022}, and the imprinting of persistent currents with high winding numbers in a tunable fermionic ring across the BCS-BEC crossover \cite{roati2022}. 
The rapidly evolving research on ring-shaped superfluids is promising both for its fundamental applications and for the technological ones \cite{amico2022}.

In conclusion, we believe that bubble-trapped condensates deserve further investigations, and that new physics could stem from the study of quantum statistics of interacting particles in nontrivial topologies, possibly at finite temperature, and considering various curved geometries. 

\section*{Acknowledgements} 
A.T. acknowledges support from {ANR} grant ``Droplets'' No.~{ANR-19-CE30-0003-02} and from the {EU Quantum Flagship} {PASQuanS2.1, 101113690}.
L.S. is partially supported by the {BIRD} grant ``Ultracold atoms in curved geometries'' of the University of Padova, by the ``Iniziativa Specifica Quantum'' of {INFN}, by the {European Quantum Flagship project PASQuanS2}, and by the {European Union-NextGenerationEU} within the National Center for HPC, {Big Data and Quantum Computing} (Project No. {CN00000013}, CN1 Spoke 10: ``Quantum Computing'').
ICFO group acknowledges support from:
{European Research Council} AdG NOQIA;
MCIN/AEI ({PGC2018-0910.13039/ 501100011033}, {CEX2019-000910-S/10.13039/501100011033}, Plan National FIDEUA PID2019-106901GB-I00, Plan National STAMEENA PID2022-139099NB-I00, project funded by {MCIN/AEI/10.13039/501100011033} and by the ``{European Union NextGenerationEU/PRTR}'' ({PRTR-C17.I1}), FPI); QUANTERA MAQS PCI2019-111828-2); QUANTERA DYNAMITE PCI2022-132919, QuantERA II Programme co-funded by {European Union's Horizon 2020 program} under Grant Agreement No {101017733};
{Ministry for Digital Transformation and of Civil Service of the Spanish Government} through the QUANTUM ENIA project call - Quantum Spain project, and by the {European Union} through the Recovery, Transformation and Resilience Plan - NextGenerationEU within the framework of the Digital Spain 2026 Agenda;
Fundaci\'o Cellex;
Fundaci\'o Mir-Puig;
Generalitat de Catalunya ({European Social Fund FEDER} and {CERCA program}, {AGAUR} Grant No. {2021 SGR 01452}, {QuantumCAT} \ {U16-011424}, co-funded by {ERDF Operational Program of Catalonia} 2014--2020);
{Barcelona Supercomputing Center MareNostrum} ({FI-2023-1-0013});
Funded by the European Union. Views and opinions expressed are however those of the author(s) only and do not necessarily reflect those of the European Union, European Commission, European Climate, Infrastructure and Environment Executive Agency (CINEA), or any other granting authority.  Neither the European Union nor any granting authority can be held responsible for them ({EU Quantum Flagship PASQuanS2.1}, {101113690}, {EU Horizon 2020 FET-OPEN OPTOlogic}, Grant No {899794}),  {EU Horizon Europe Program} (This project has received funding from the {European Union's Horizon Europe research and innovation program} under grant agreement No {101080086} NeQSTGrant Agreement {101080086} --- NeQST);
ICFO Internal ``QuantumGaudi'' project;
{European Union's Horizon 2020 program} under the Marie Sklodowska-Curie grant agreement No {847648};  
``La Caixa'' {Junior Leaders fellowships, La Caixa'' Foundation} (ID {100010434}): {CF/BQ/PR23/11980043}.

\appendix

\section{Laplace equation in spherical coordinates}
\label{appendixgreen}
In this appendix, we derive the Green's function of Laplace equation in spherical coordinates. 
The Green's function $G(\theta,\varphi,\theta',\varphi')$ satisfies the following Poisson equation 
\begin{equation}
-\frac{\hat{L}^2}{\hbar^2} \, G(\theta,\varphi,\theta',\varphi')=  q \, \bigg[ \delta (\cos\theta-\cos\theta') \, \delta(\varphi-\varphi') -\frac{1}{4\pi} \bigg],
\label{poissoneqappendix}
\end{equation}
where $\hat{L}^2$ is the angular momentum operator in spherical coordinates, and where the additional term $-1/(4\pi)$ eliminates the divergences in the following steps. 
We expand $G(\theta,\varphi,\theta',\varphi')$ and the delta functions in a basis of spherical harmonics as
\begin{align}
G(\theta,\varphi,\theta',\varphi') &= \sum_{l=0}^{\infty} \sum_{m_l=-l}^{l} g_{lm_l} (\theta',\varphi') {\mathcal{Y}_l^{m_l}} (\theta,\varphi), 
\label{greensexpansionappendix}
\\
\delta (\cos\theta-\cos\theta') \, \delta(\varphi-\varphi') &= \sum_{l=0}^{\infty} \sum_{m_l=-l}^{l} {\mathcal{Y}_l^{m_l}}^{*} (\theta',\varphi') {\mathcal{Y}_l^{m_l}} (\theta,\varphi).
\end{align}
where $g_{lm_l} (\theta',\varphi')$ are unknown coefficients.  
To determine them, we substitute these decompositions in Eq.~\eqref{poissoneqappendix}, and using the properties of the spherical harmonics, we find that
\begin{equation}
g_{lm_l} (\theta',\varphi') = \frac{q }{l(l+1)} {\mathcal{Y}_l^{m_l}}^{*} (\theta',\varphi').
\end{equation}
Then, substituting $g_{lm_l} (\theta',\varphi')$ into the Green's function of Eq.~\eqref{greensexpansionappendix}, and summing over $m_l$, we find 
\begin{equation}
G(\theta,\varphi,\theta',\varphi') = \frac{q}{4\pi} \bigg[ \sum_{l=1}^{\infty} \frac{1}{l} P_l(\cos \gamma) + \sum_{l=1}^{\infty} \frac{1}{l+1} P_l(\cos \gamma) \bigg],
\label{eqgreen}
\end{equation}
with $P_l(\cos \gamma)$ the Legendre polynomia, and $\cos \gamma = \cos \theta \cos \theta' +  \sin\theta\sin\theta' \cos(\varphi-\varphi')$. The sums in Eq.~\eqref{eqgreen} can be performed analytically (see Ref.~\cite{gradshteyn1996}), obtaining the Green's function as
\begin{equation}
G(\theta,\varphi,\theta',\varphi') = \frac{q}{4\pi} + \frac{q}{2\pi} \,\ln \bigg( \sqrt{\frac{1-\cos \gamma}{2}} \bigg), 
\end{equation}
which allows to determine the stream function of a spherical superfluid film, as discussed in Section \ref{sectionvortices}.

\end{document}